\documentclass[11pt,a4paper]{article}
\usepackage{jheppub}
\pdfoutput=1
\usepackage{amsmath}
\usepackage{graphicx}
\usepackage{hyperref}
\usepackage{xspace}
\usepackage{xcolor}
\usepackage{placeins}

\newcommand{\eq}[1]{eq.~\eqref{eq:#1}}
\renewcommand{\sec}[1]{sec.~\ref{sec:#1}}

\newcommand{\eqs}[2]{eqs.~\eqref{eq:#1} and \eqref{eq:#2}}
\newcommand{\fig}[1]{fig.~\ref{fig:#1}}

\newcommand{\app}[1]{appendix~\ref{app:#1}}
\newcommand{\tab}[1]{table~\ref{table:#1}}

\providecommand{\refcite}[1]{ref.~\cite{#1}}
\providecommand{\refscite}[1]{refs.~\cite{#1}}

\newcommand{\Tau}{\mathcal{T}}

\newcommand{\cusp}{\mathrm{cusp}}

\newcommand{\DeltaFO}{\Delta_{\mu}}

\newcommand{\nn}{\notag}

\newcommand{\eg}{e.g.\ }

\allowdisplaybreaks

\newcommand{\abs}[1]{\lvert#1\rvert}

\newcommand{\ord}[1]{\mathcal{O}(#1)}
\newcommand{\ORd}[1]{{\mathcal O}\Bigl(#1\Bigr)}
\newcommand{\df}{\mathrm{d}}
\newcommand{\img}{\mathrm{i}}
\let\Re\relax
\DeclareMathOperator{\Re}{Re}

\def\lo*{\text{LO}}
\def\nlo*{\text{NLO}}
\def\nnlo*{\text{NNLO}}
\def\nnnlo*{\text{\texorpdfstring{N\textsuperscript{3}LO}{N3LO}}}
\def\nNlo*{\text{\texorpdfstring{N\textsuperscript{$n$}LO}{N$n$LO}}}

\def\llog*{\text{LL}}
\def\nllog*{\text{NLL}}
\def\nnllog*{\text{NNLL}}
\def\nnnllog*{\text{\texorpdfstring{N\textsuperscript{3}LL}{N3LL}}}
\def\nNllog*{\text{\texorpdfstring{N\textsuperscript{$n$}LL}{N$n$LL}}}

\def\llphi*{\texorpdfstring{\ensuremath{\text{LL}_\varphi}}{LLphi}}
\def\nllphi*{\texorpdfstring{\ensuremath{\text{NLL}_\varphi}}{NLLphi}}
\def\nnllphi*{\texorpdfstring{\ensuremath{\text{NNLL}_\varphi}}{NNLLphi}}
\def\nnnllphi*{\texorpdfstring{\ensuremath{\text{N\textsuperscript{3}LL}_\varphi}}{N3LLphi}}
\def\nNllphi*{\texorpdfstring{\ensuremath{\text{N\textsuperscript{$n$}LL}_\varphi}}{N$n$LLphi}}

\def\llpphi*{\texorpdfstring{\ensuremath{\text{LL}^\prime_\varphi}}{LL'phi}}
\def\nllpphi*{\texorpdfstring{\ensuremath{\text{NLL}^\prime_\varphi}}{NLL'phi}}
\def\nnllpphi*{\texorpdfstring{\ensuremath{\text{NNLL}^\prime_\varphi}}{NNLL'phi}}
\def\nnnllpphi*{\texorpdfstring{\ensuremath{\text{N\textsuperscript{3}LL}^\prime_\varphi}}{N3LLphi}}
\def\nNllpphi*{\texorpdfstring{\ensuremath{\text{N\textsuperscript{$n$}LL}^\prime_\varphi}}{N$n$LLphi}}

\def\llmatched*{\ensuremath{\lo*$+$\llpphi*}}
\def\nllmatched*{\ensuremath{\nlo*$+$\nllpphi*}}
\def\nnllmatched*{\ensuremath{\nnlo*$+$\nnllpphi*}}
\def\nnnllmatched*{\ensuremath{\nnnlo*$+$\nnnllpphi*}}
\def\nNllmatched*{\ensuremath{\nNlo*$+$\nNllpphi*}}

\def\ggH*{\texorpdfstring{\ensuremath{gg \rightarrow H}}{gg to H}}
\def\bbH*{\texorpdfstring{\ensuremath{b\bar{b} \rightarrow H}}{bb to H}}
\def\ggX*{\texorpdfstring{\ensuremath{gg \rightarrow X}}{gg to X}}

\newcommand{\GeV}{\,\mathrm{GeV}}
\newcommand{\TeV}{\,\mathrm{TeV}}

\newcommand{\pb}{\,\mathrm{pb}}

\newcommand{\as}{\alpha_s}
\newcommand{\FO}{\mathrm{FO}}
\newcommand{\res}{\mathrm{res}}

\newcommand{\MSbar}{$\overline{\text{MS}}$\xspace}

\newcommand{\sqrts}{E_\mathrm{cm}}
\newcommand{\mufo}{\mu_\mathrm{FO}}

\newcommand{\rap}{Y}

\newcommand{\rEFT}{\mathrm{rEFT}}
\newcommand{\EFT}{\mathrm{EFT}}

\newcommand{\WidthTwoSubfigs}{0.5\textwidth}
\newcommand{\WidthThreeSubfigs}{0.32\textwidth}

\renewcommand{\tabcolsep}{1ex}

\graphicspath{ {./plots/} }

\title{Resummation Improved Rapidity Spectrum for Gluon Fusion Higgs Production}

\author[a]{Markus A.~Ebert,}
\author[a,b]{Johannes K.\,L.~Michel,}
\author[a]{and Frank J.~Tackmann}
\affiliation[a]{Theory Group, Deutsches Elektronen-Synchrotron (DESY),\\ Notkestra\ss{}e 85, D-22607 Hamburg, Germany}
\affiliation[b]{Institute for Theoretical Physics, WWU M\"unster,\\ Wilhelm-Klemm-Stra\ss{}e 9, D-48149 M\"unster, Germany} 
\emailAdd{markus.ebert@desy.de}
\emailAdd{johannes.michel@desy.de}
\emailAdd{frank.tackmann@desy.de}

\abstract{
Gluon-induced processes such as Higgs production typically exhibit large perturbative corrections.
These partially arise from large virtual corrections to the gluon form factor, which at timelike momentum transfer
contains Sudakov logarithms evaluated at negative arguments $\ln^2(-1) = -\pi^2$.
It has been observed that resumming these terms in the timelike form factor
leads to a much improved perturbative convergence for the total cross section.
We discuss how to consistently incorporate the resummed form factor into the perturbative predictions
for generic cross sections differential in the Born kinematics, including in particular
the Higgs rapidity spectrum. We verify that this indeed improves the perturbative
convergence, leading to smaller and more reliable perturbative uncertainties,
and that this is not affected by cancellations between resummed and unresummed contributions.
Combining both fixed-order and resummation uncertainties, the perturbative uncertainty
for the total cross section at \nnnllmatched* is about a factor of two smaller than at \nnnlo*.
The perturbative uncertainty of the rapidity spectrum at \nnllmatched*
is similarly reduced compared to NNLO.
We also study the analogous resummation for quark-induced processes,
namely Higgs production through bottom quark annihilation and the Drell-Yan rapidity spectrum.
For the former the resummation leads to a small improvement, while for the latter
it confirms the already small uncertainties of the fixed-order predictions.
}

\preprint{\vbox{
\hbox{DESY 16-216}
\hbox{MS-TP-16-30}
\hbox{February 02, 2017}
}}

\arxivnumber{}

\begin{document}
\maketitle

\section{Introduction}
\label{sec:Intro}

After the discovery of the Higgs boson~\cite{Aad:2012tfa, Chatrchyan:2012xdj}, the LHC has entered an era of precision Higgs measurements.
One important goal is the precise determination of the Higgs couplings in order to test the Standard Model and search for
evidence of physics beyond it. Other important color-singlet processes like Drell-Yan production serve as standard candles that are used, for example, to constrain parton distribution functions (PDFs).

In order to match the ever increasing level of experimental precision, precise theoretical predictions for the measured cross sections are needed. An important example is the dominant Higgs production via gluon fusion, which receives large perturbative corrections. This has led to the calculation of the total production cross section up to \nnnlo* \cite{Dawson:1990zj, Djouadi:1991tka, Spira:1995rr, Harlander:2002wh, Anastasiou:2002yz, Ravindran:2003um, Anastasiou:2015ema, Anastasiou:2016cez},
and including the resummation of threshold logarithms up to N$^3$LL$^\prime$ \cite{Ahrens:2008nc, Bonvini:2014joa, Li:2014afw, Bonvini:2014tea, Schmidt:2015cea, Bonvini:2016frm}.
However, due to the limited detector acceptance the experimental measurements cannot measure the cross section fully inclusively but only in a restricted kinematic range, in particular in a restricted range of Higgs rapidities. The interpretation of the experimental measurements thus fundamentally requires theoretical predictions differential in the Higgs kinematics. The essential nontrivial ingredient is the Higgs rapidity spectrum (or equivalently the cross section with a rapidity cut), which is so far known to NNLO~\cite{Anastasiou:2004xq, Anastasiou:2005qj, Anastasiou:2007mz, Catani:2007vq, Grazzini:2008tf}.

A specific class of perturbative corrections to Drell-Yan-like color-singlet production
arises from the associated quark and gluon form factors, which contain Sudakov logarithms $\ln^2(-q^2/\mu^2)$, where $q^\mu$
is the transferred hard momentum. For spacelike momentum transfer, $q^2 = - Q^2 < 0$ as in deep-inelastic scattering,
these logarithms vanish with the standard choice $\mu^2 = Q^2$.
For timelike production processes, the form factor enters the production cross section evaluated at timelike momentum
transfer $q^2 = Q^2 > 0$. With the ordinary scale choice $\mu^2 = Q^2$, the form
factor contains leftover Sudakov logarithms $\ln^2(-1) = -\pi^2$, inducing large corrections at each order in the perturbative series.
For simplicity, we will henceforth refer to these as ``timelike'' logarithms or contributions, as they arise in the ratio
of the timelike and spacelike form factors.%
\footnote{Since the resummed logarithms $\ln^{2n}(-1)$ happen to give factors of $(-\pi^2)^{n}$,
their resummation has been referred to as ``$\pi^2$-resummation''.
Since factors of $\pi^2$ from other (unrelated) sources are typical
to appear in the perturbative coefficients as well,
we will always refer to the resummed logarithms as ``timelike logarithms'', to avoid
any possible confusion as to what is being resummed.
}
This effect was first observed long ago in Drell-Yan production in \refcite{Altarelli:1979ub},
where it was realized that the coefficients of these terms are directly related to infrared (IR) singularities.
Due to the universal structure of IR singularities, these terms arise to all orders and their resummation
is well known~\cite{Parisi:1979xd, Sterman:1986aj, Magnea:1990zb, Eynck:2003fn}.
As discussed in \refcite{Bakulev:2000uh}, the
timelike logarithms are also present in the soft contributions to the pion electromagnetic form factor
providing an enhancement compared to the spacelike case in agreement with the measured enhancement.
The resummation of the timelike logarithms for gluon-fusion Higgs production was carried out in
\refscite{Ahrens:2008qu, Ahrens:2008nc} in the context of soft-gluon (threshold) resummation,
where it was shown that it substantially reduces the large
perturbative corrections to the total $gg\to H$ cross section.

The resummation of the timelike logarithms originating in the form factors has since been included in the resummation of various other exclusive color-singlet cross sections (see e.g.\ \refscite{Stewart:2010pd, Berger:2010xi, Becher:2012yn, Becher:2013xia, Stewart:2013faa, Jaiswal:2014yba, Gangal:2014qda, Neill:2015roa, Ebert:2016idf}), leading to improvements in the perturbative uncertainties. In these contexts, the use of the resummed form factor is unambiguous,
as it explicitly appears as an ingredient in the corresponding factorized cross section.

In this paper, we study in detail the utility of the resummed timelike form factors for predictions of inclusive color-singlet production cross sections. In the case of inclusive cross sections the benefit of the resummation is a priori not obvious, and its applicability has occasionally been called into question. For this reason, we discuss in some detail the arguments for it and its consistent application, as well as the potential pitfalls one might worry about. For our numerical analysis, we consider both gluon-induced and quark-induced processes. The cases we consider include a generic scalar resonance $gg\to X$ as a function of $m_X$, $gg\to H$ as a function of the Higgs rapidity, $b\bar b \to H$, and Drell-Yan $q\bar q \to Z$ as a function of the $Z$ rapidity.

We find that in all cases the resummation of the timelike logarithms leads to stable perturbative predictions.
For the gluon-induced cases it leads to a significantly improved convergence compared to the fixed-order predictions,
as first pointed out in \refscite{Ahrens:2008qu, Ahrens:2008nc}.
This results in perturbative uncertainties that are both smaller and more reliable.
In addition to the total cross section studied previously,
we show how the resummation can be easily and consistently applied to
generic inclusive cross sections differential in the Born kinematics.
This allows us in particular to obtain the currently most precise predictions for the Higgs rapidity spectrum, or equivalently the inclusive cross section with a rapidity cut, with perturbative uncertainties that are reduced by almost a factor of two compared to NNLO.
For the quark-induced processes, the improvement is not as dramatic. Here, the resummed and fixed-order results have a similar stability. With an optimal choice of $\mu_F$ the resummation still provides some improvement in the perturbative convergence and uncertainties. This demonstrates that using the resummed form factor is also viable for quark-induced processes and provides additional confidence in the estimated perturbative uncertainties.

The remainder of the paper is structured as follows:
The basic setup how to consistently incorporate the resummed form factors into the
inclusive cross section is discussed in \sec{Calculation}.
The application to gluon-fusion processes is then discussed in \sec{GluonFusion},
to Higgs production through bottom quark annihilation in \sec{bbH}, and to
Drell-Yan production in \sec{DY}. We conclude in \sec{Conclusion}.
For completeness all required perturbative ingredients for the resummed form factors
are collected in \app{ingredients}.

\section{Calculational setup}
\label{sec:Calculation}

\subsection{Resummation framework}
\label{sec:ResummationScheme}

We consider the hadronic production $gg\to L$ or $q\bar q\to L$ of a color-singlet
final state $L$ with total invariant mass $Q^2 = q^2 > 0$.
The hard virtual corrections to these processes are described by the corresponding
QCD form factors. The full form factors contain infrared divergences, which when combined
into the full cross section cancel against the infrared divergences in the real corrections.
Hence, what enters in the final cross section are the IR-finite parts of the form factor.
In the context of soft-collinear effective theory (SCET)~\cite{Bauer:2000ew, Bauer:2000yr, Bauer:2001ct, Bauer:2001yt},
these are equivalent to the Wilson coefficients from matching the QCD currents defining the
form factors onto the corresponding SCET currents~\cite{Bauer:2002nz, Manohar:2003vb, Bauer:2003di}.
For the cases we consider, these are the gluon, quark vector, and quark scalar form factors.
The corresponding matching conditions read schematically
\begin{align} \label{eq:currents}
G_{\mu\nu} G^{\mu\nu} &\to C_{gg}\, Q^2 \mathcal{B}_{n \perp} \mathcal{B}_{\bar n \perp}
\,, \nn \\
\bar q \gamma^\mu q &\to C_{q\bar q}^V\, \bar \chi_{n} \gamma^\mu \chi_{\bar n}
\,, \nn \\
\bar q q &\to C_{q\bar q}^S\, \bar \chi_{n} \chi_{\bar n}
\,,\end{align}
where the $\mathcal{B}_{n \perp}$ and $\chi_n$ are collinear gluon and quark fields in SCET.
(The exact matching conditions for the currents can be found e.g.\ in \refscite{Stewart:2009yx, Berger:2010xi}.)
The IR divergences in the full QCD form factors, given by the quark and gluon matrix elements of the left-hand side, are exactly reproduced by the corresponding matrix elements of the SCET operators on the right-hand side, such that the hard Wilson coefficients $C_{ij}$ are given in terms of the IR-finite parts of the form factors.

The relevant object entering the cross section is the hard function given by the square of the Wilson coefficient, which we write as
\begin{equation}
H(q^2,\mu) = |C(q^2,\mu)|^2 = 1 + H^{(1)}(q^2, \mu) + H^{(2)}(q^2, \mu) + \dotsb
\,,\end{equation}
where by default we normalize $H$ to unity at leading order, and $H^{(n)}$ denotes the $\ord{\as^n}$ term.
To all orders in perturbation theory, $C$ and $H$ depend on the hard momentum transfer $q^\mu$
through logarithms $L \equiv \ln[(-q^2 - \img0)/\mu^2]$.
For spacelike processes, $q^2 = -Q^2 < 0$ such that $L = \ln(Q^2/\mu^2) = 2\ln(Q/\mu)$, while for timelike processes $q^2 = Q^2 > 0$ such that $L = 2\ln(-\img Q/\mu)$.

The Wilson coefficients in SCET obey the renormalization group equation (RGE)
\begin{align} \label{eq:hard RGE}
\mu \frac{\df C(q^2,\mu)}{\df\mu} &= \gamma_H(q^2,\mu) \, C(q^2, \mu)
\,,\nn \\
\gamma_H(q^2,\mu) &= \Gamma_\cusp[\as(\mu)]\,\ln\frac{-q^2 -\img0}{\mu^2} + \gamma_H[\as(\mu)]
\,,\end{align}
where $\Gamma_\cusp(\as)$ is the cusp anomalous dimension and $\gamma_H(\as)$ the noncusp term.
Integrating \eq{hard RGE} yields the solution
\begin{align} \label{eq:hard resummed}
H(q^2,\mu) &= H(q^2,\mu_H) \, U_H(\mu_H,\mu)
\,, \\
\label{eq:def UH}
U_H(\mu_H,\mu) &= \biggl\lvert\exp\biggl[ \int_{\mu_H}^\mu \frac{\df\mu'}{\mu'} \gamma_H(q^2,\mu')\biggr] \biggr\rvert^2
\,.\end{align}
The explicit result for the evolution kernel $U_H$ is given in \app{RGE}.
By choosing the imaginary-valued scale $\mu_H = -\img Q$, the hard function $H(Q,\mu_H)$ is free of logarithms and can be calculated in fixed-order perturbation theory, while the evolution kernel $U_H$ resums
all logarithms $\ln(\mu_H/\mu) = \ln(-\img Q/\mu)$.

The hard function explicitly appears in calculations of exclusive cross sections as
\begin{equation} \label{eq:exclusive}
\frac{\df\sigma}{\df\Tau} = \sigma_B \times H(Q^2, \mu_\Tau) \times SC(\Tau, \mu_\Tau) \times [1 + \ord{\Tau/Q}]
\,.\end{equation}
Here $\Tau$ denotes a resolution variable, which resolves additional emissions, such that in the limit $\Tau \ll Q$ the cross section is restricted to the soft-collinear regime. In this limit it is dominated by hard virtual corrections contained in $H$, and soft and collinear contributions (both real and virtual) at lower scales $\mu_\Tau\sim \Tau$ contained in $SC$, while hard real emissions are forbidden. At the partonic level, an example for $\Tau$ is the partonic threshold variable $(1 - z)Q$.
More physical examples of $\Tau$ are beam thrust or the $p_T$ of the leading jet.
The precise form of the soft-collinear contribution $SC$ depends on the definition of $\Tau$ but is irrelevant for our discussion. For a given process always the same hard function appears independently of the precise choice of $\Tau$. The factorization in \eq{exclusive} implies that in the $\Tau\ll Q$ limit $H$ appears as a well-defined perturbative object (namely as a hard matching coefficient), which is fully factorized from the rest of the cross section. In particular, the only dependence on the hard timelike momentum transfer $Q^2$ resides in $H$, while $SC$ only depends on parametrically smaller soft and collinear scales proportional to $\Tau$. In practice, \eq{exclusive} can be used to perform the resummation of logarithms of $\Tau$ in $\df\sigma/\df\Tau$, which involves using \eq{hard resummed} to evolve $H$ from its natural scale $\mu_H = -\img Q$ to the relevant lower scale $\mu_\Tau \sim \Tau$.

We want to apply the resummed form factor to the inclusive cross section for color-singlet production. Here, inclusive refers to the fact that the cross section is fully integrated over any additional QCD emissions, but it can still be differential in or contain cuts on any kinematic variables that are present at Born level and describe the produced color-singlet system, such as its total rapidity $Y$ or total invariant mass $Q$.
To do so, we can factor out the hard function from the inclusive cross section
\begin{equation} \label{eq:Rdef}
 \sigma(X) = H(Q^2, \mufo) \times R(X, \mufo)
\,,\end{equation}
which \emph{defines} the remainder $R(X, \mufo)$.
Here, $X$ denotes any dependence on Born variables or cuts.
By definition, $H$ only depends on the Born kinematics via $Q$, while the remainder $R$ can depend on $X$.

We write the perturbative expansion of the remainder as
\begin{equation} \label{eq:Rexpansion}
R(X, \mufo) = \sigma^{(0)}(X, \mufo) \bigl[ 1 + R^{(1)}(X, \mufo) + R^{(2)}(X, \mufo) + \cdots \bigr]
\,,\end{equation}
where for convenience we pulled out the leading-order cross section $\sigma^{(0)}(X, \mufo)$.
The dependence on the factorization scale $\mu_F$ related to the PDFs
entirely cancels within $R$, and we will mostly suppress it. The $\mufo$ scale in \eqs{Rdef}{Rexpansion}
is equivalent to the renormalization scale $\mu_R$ in the fixed-order prediction, and its
dependence explicitly cancels between $H$ and $R$.
The $R^{(n)}$ coefficients depend primarily only on the total color-singlet invariant mass and
rapidity, while any dependence on additional Born kinematics or cuts resides primarily in $\sigma^{(0)}$.
(This becomes exact for a scalar resonance in the narrow-width approximation like the Higgs.)
In the following we will for simplicity suppress the dependence on $X$ and $Q^2$.

We also define the $K$ factor
\begin{equation} \label{eq:Kfactor}
K(\mu) = \frac{\sigma}{\sigma^{(0)}(\mu)} = 1 + K^{(1)}(\mu) + K^{(2)}(\mu) + \cdots
\,,\end{equation}
which captures the total perturbative correction relative to the leading-order result.
Expanding \eq{Rdef} order by order in $\alpha_s(\mu)$, it is straightforward to obtain the
fixed-order coefficients of $R$ from those of $K$ and $H$. Up to \nnnlo* we have,
\begin{align}
R^{(1)}(\mu) &= K^{(1)}(\mu) - H^{(1)}(\mu)
\,, \nn \\
R^{(2)}(\mu) &= K^{(2)}(\mu) - H^{(2)}(\mu) - R^{(1)}(\mu)\, H^{(1)}(\mu)
\,, \nn \\
R^{(3)}(\mu) &= K^{(3)}(\mu) - H^{(3)}(\mu) - R^{(2)}(\mu)\, H^{(1)}(\mu) - R^{(1)}(\mu)\, H^{(2)}(\mu)
\,.\end{align}

To resum the timelike logarithms from the form factor in the cross section
we can simply take the resummed result for the hard function \eq{hard resummed} and use it in \eq{Rdef},
\begin{align} \label{eq:sigmares}
\sigma_\res
&=  H(\mu_H)\, U_H(\mu_H,\mufo) \, R(\mufo)
\\\nn
&= U_H(\mu_H,\mufo)\, \sigma^{(0)}\, \bigl[1 + H^{(1)}(\mu_H) + R^{(1)}(\mufo)
\nn \\ & \qquad\qquad\qquad\qquad\qquad
+ H^{(2)}(\mu_H) + R^{(2)}(\mufo) + H^{(1)}(\mu_H)\, R^{(1)}(\mufo) + \dots \bigr]
\nn\,.\end{align}
As indicated, the fixed-order expansions for $H(\mu_H)$ and $R(\mufo)$ are
reexpanded against each other (but without reexpanding the $\as(\mu_H)$ inside
the coefficients $H^{(n)}(\mu_H)$ in terms of $\as(\mufo)$).
This is analogous to the standard treatment in resummed predictions as would be used
for example in \eq{exclusive}. This ensures that in the limit $\mu_H = \mufo$
we exactly recover the usual fixed-order result without inducing any higher-order cross terms
between $H$ and $R$.
Using the definition of $R$ in \eq{Rdef}, the resummed cross section in \eq{sigmares} can equivalently be written as
\begin{equation} \label{eq:sigmaresAlt}
 \sigma_\res = U_H(\mu_H,\mufo)  \biggl[ \frac{H(\mu_H)}{H(\mufo)}\, \sigma_\FO \biggr]_\FO
\,,\end{equation}
where the brackets $[\dots]_\FO$ indicate the fixed-order reexpansion in powers of $\as(\mufo)$ and $\as(\mu_H)$,
with $\sigma_\FO$ the usual fixed-order cross section expanded in $\as(\mufo)$.
Written in this way, the ratio of timelike to spacelike form factors is manifest.

Equation~\eqref{eq:sigmares} will be the basis of all our results. For consistency with
the fixed-order limit, we always include $H(\mu_H)$ and $R(\mufo)$ to the same fixed order.
Furthermore, we always combine the \nNlo* fixed-order
contributions with the \nNllog* resummation for $H$, which corresponds to the
primed resummation counting and ensures consistency with the exclusive
resummations~\cite{Berger:2010xi, Stewart:2013faa} based on \eq{exclusive}.
We will denote the perturbative accuracy by N$^n$LO$+$N$^n$LL$'_\varphi$, where
the subscript indicates that the resummed logarithms correspond to the complex phase $\varphi$
of the hard scale in the form factor.

While the remainder $R$ is uniquely defined by \eq{Rdef}, one should of course ask the question to what extent
it is justified or meaningful to ``brute-force'' factorize the perturbative series for the inclusive
cross section into those for $H$ and $R$.

First, one might be worried by the fact that the remaining nonlogarithmic constant terms in the fixed-order
expansion of $H(\mu_H)$ are scheme-dependent, i.e.\ they depend on the fact that $H$ is renormalized in the \MSbar scheme
and using a different scheme would result in different constant terms.
However, this fixed-order scheme dependence is canceled by $R$ up to higher orders, and this
cancellation is explicitly ensured in our implementation in \eq{sigmares}
by the fact that we always reproduce the exact fixed-order result, as discussed above.
The cancellation can also be seen explicitly from \eq{sigmaresAlt}.
Expanding the ratio $H(\mu_H)/H(\mufo)$, including expanding $\as(\mu_H)$ in terms of $\as(\mufo)$, the constant terms in $H$
explicitly drop out. In particular, the nonlogarithmic constant terms at $\ord{\as^n}$
cancel up to $\ord{\as^{n+2}}$ since
\begin{equation} \label{eq:finite terms cancel}
\Re\bigl[\as^n(-\img\mufo) - \as^n(\mufo)\bigr] = \ord{\as^{n+2}}
\,.\end{equation}

Therefore, the relevant question is whether the series of timelike Sudakov logarithms present in $H$ can be considered to be independent from the perturbative series in $R$. This would not be the case if (and only if) $R$ were to contain contributions at each order correlated with the timelike Sudakov series in $H$ and of opposite sign, which would then lead to large cancellations between $H$ and $R$ at each order in perturbation theory. These cancellations would then be spoiled by resumming the timelike logarithms in $H$ while keeping the corresponding pieces in $R$ at fixed order. This would imply that the perturbative corrections for $R$ would be noticeably \emph{larger} than for the cross section itself, and since the resummation of $H$ eliminates its large corrections, the larger perturbative corrections of $R$ would result in the resummed cross section being worse behaved. In other words, the absence or presence of sizeable cancellations between the resummed terms and the unresummed fixed-order terms, is mathematically equivalent to whether the resummation improves the perturbative convergence of the cross section or not. This is of course easy to check up to the available order, and in all our applications we have verified that there are indeed no large cancellations that are being spoiled by the resummation.

The primary reason one could be worried about such cancellations is that this is actually what happens in the reverse timelike process, namely color-singlet decays such as $H\to gg$, $H\to b\bar b$, $e^+e^-\to Z\to q\bar q$, or hadronic $\tau$ decays. These processes involve the same timelike form factor, but their perturbative series is known to not contain timelike Sudakov logarithms. The relation to these processes was already discussed in some detail in \refcite{Ahrens:2008nc}.
In these processes, timelike logarithms only appear as single logarithms (and thus only at higher orders) through the running of $\as$, for which analogous analytic continuation methods have been considered, \eg for $e^+ e^- \rightarrow \text{hadrons}$ in \refscite{Pennington:1981cw,Radyushkin:1982kg,Krasnikov:1982fx,Bakulev:2000uh,Bakulev:2010gm} and hadronic $\tau$-decays \eg in \refscite{Pivovarov:1991rh, LeDiberder:1992jjr, Neubert:1995gd} (see also \refscite{Broadhurst:2000yc,Stefanis:2009kv} and references therein).

However, the situation is fundamentally different when the hard partons appear in the initial vs.\ the final state.
An explicit discussion how the timelike Sudakov logarithms cancel in the final-state case but not in the initial-state
case can be found in \refcite{Bakulev:2000uh}.
In the final-state case, the process can be written as the imaginary part of forward matrix elements summed over all possible cuts, in which case the whole calculation can be deformed into the Euclidean domain where the timelike logarithms never appear. That is, the timelike Sudakov logarithms fully cancel between all cuts, or equivalently between the virtual corrections to the form factor and the real corrections to the corresponding remainder.
The same does not happen if the partons appear in the initial state, which simply cannot be obtained from cutting a diagram, i.e.\ the process with incoming partons is intrinsically more exclusive, which exposes the timelike Sudakov logarithms in the form factor.
Note also that if the same cancellations as in the final-state case were present in the initial-state case, they would have to be present at each order starting at NLO. The fact that we do not observe this even in the first several orders of the perturbative series provides clear evidence that this is indeed not the case.

It is also easy to understand why one finds a substantial numerical improvement for
inclusive Higgs production.
Comparing \eq{Rdef} with the exclusive cross section in \eq{exclusive}, in the soft-collinear
limit the remainder $R$ reduces to the soft-collinear contributions times power corrections,
\begin{equation}
R \to \sigma_B\, SC(\Tau)\,[1 + \ord{\Tau/Q}]
\,.\end{equation}
Therefore, the factorization in \eq{Rdef} also becomes formally justified when
the inclusive cross section is numerically dominated by soft-collinear contributions.
It is well known that a large portion of the Higgs cross section comes from
the partonic threshold limit,
in which the hard function factors out of the cross section as in \eq{exclusive}.
One can also take the more physical limit and simply veto additional hard radiation
(which is also a weaker limit as it allows both soft and collinear radiation).
Going from this exclusive $0$-jet region, to which \eq{exclusive} strictly applies,
to the inclusive cross section amounts to factoring out
the form factor also from the nonsingular power corrections.
As pointed out in \refscite{Berger:2010xi, Stewart:2013faa},
using either beam thrust or the $p_T$ of the leading jet to veto hard radiation,
one finds that utilizing the resummed form factor
for both singular and nonsingular corrections, and hence for the full
inclusive cross section, is actually important, since not doing so can easily
lead to unphysical results with the inclusive cross section being smaller than the $0$-jet cross section.

Finally, we note that it has been argued in \refcite{Anastasiou:2016cez} on the basis of the coefficient of the $\delta(1-z)$ term in the partonic cross section that the timelike logarithms are not a dominant source of higher-order corrections
and in particular that their resummation fails to improve the results beyond NNLO.
We need to disagree with this assessment, because this coefficient is strongly scheme dependent and not a very well-defined quantity. Rather the impact or improvement should be judged at the level of the physical cross section.
A more detailed discussion on this is given in \app{estimates}.

\subsection{Perturbative uncertainties and numerical inputs}
\label{sec:Uncertainties}

For our numerical predictions we consider the LHC at $\sqrts=13\TeV$.
We use the \texttt{PDF4LHC\_nnlo\_100} \cite{Butterworth:2015oua, Dulat:2015mca, Harland-Lang:2014zoa, Ball:2014uwa, Gao:2013bia, Carrazza:2015aoa}
NNLO PDFs with $\as(m_Z)= 0.118$. Since we are interested in the size of the coefficients in
the perturbative series we always use this same PDF independent of the perturbative order in consideration.
The numerical value of $\alpha_s(\mu_R)$ is obtained with the corresponding three-loop running,
except for the total gluon-fusion cross section known at \nnnlo*, where we use four-loop running
(though the numerical differences are negligible).
For bottom-quark annihilation we use the PDF sets from
\refscite{Bonvini:2015pxa, Bonvini:2016fgf}, which are reevolved from \texttt{PDF4LHC\_nnlo\_mc}
in order to allow varying the $b$-quark matching scale separately from the $b$-quark mass.
The relevant masses entering our predictions are $m_H = 125\GeV$, $m_t = 172.5 \GeV$,
$\overline{m}_b(\overline{m}_b) = 4.18\GeV$, and $m_Z = 91.1876 \GeV$.

Since we are primarily interested in investigating the perturbative structure, we do not
consider parametric uncertainties due to PDFs and the value of $\as(m_Z)$, which
are straightforward to evaluate. They are essentially unaffected by the resummation of the form factor,
since all PDF dependence, as well as the dominant overall dependence on $\alpha_s(m_Z)$
in case of Higgs production, resides in the remainder $R$.

An important aspect of precision predictions is a reliable assessment of
the theory uncertainties due to missing higher-order corrections.
Our predictions in principle involve three scales that we can vary as a means to estimate the size of higher-order corrections:
the factorization scale $\mu_F$ probing collinear logarithms in the PDFs, the renormalization scale $\mu_R$ probing higher orders in the fixed-order series, and the hard resummation scale $\mu_H$ probing higher orders in
the series of timelike Sudakov logarithms.
We like to stress that these scales are unphysical parameters whose variations simply provide a convenient way to probe the ``typical'' size of the associated missing higher-order terms. The resulting variations in the cross section
must be interpreted as such. In particular, we do not assign any meaning to accidentally small one-sided scale variations
that yield asymmetric uncertainties, which are just artifacts of a nonlinear scale dependence, which is frequently encountered
in predictions at higher orders or involving resummation. We therefore always consider the maximum absolute deviation from
the central result at the chosen central scale as the (symmetric) uncertainty. To be explicit, an observed scale variation of
$+|x|$ and $-|y|$ in the cross section is interpreted as a perturbative uncertainty of $\pm \max\{|x|, |y|\}$.

We parametrize the three scales as
\begin{equation}
\mu_H = \mufo \exp(-\img \varphi)
\,,\qquad
\mu_R = \mufo
\,,\qquad
\mu_F = \kappa_F\,\mufo
\,.\end{equation}
The choices for $\mufo$ and $\kappa_F$ for the central value depend on the process we consider.
For the resummed predictions we use the central choice $\varphi = \pi/2$ , while the
fixed-order predictions correspond to taking $\varphi = 0$, which turns off the resummation.

\begin{figure*}[t!]
\centering
\includegraphics[width=\WidthTwoSubfigs]{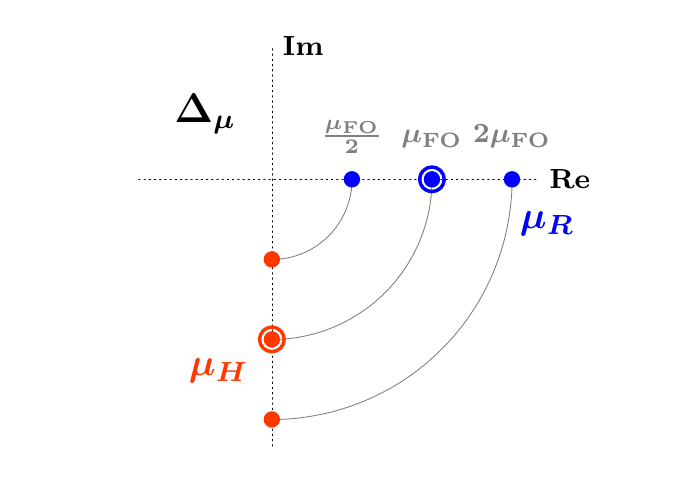}%
\hfill%
\includegraphics[width=\WidthTwoSubfigs]{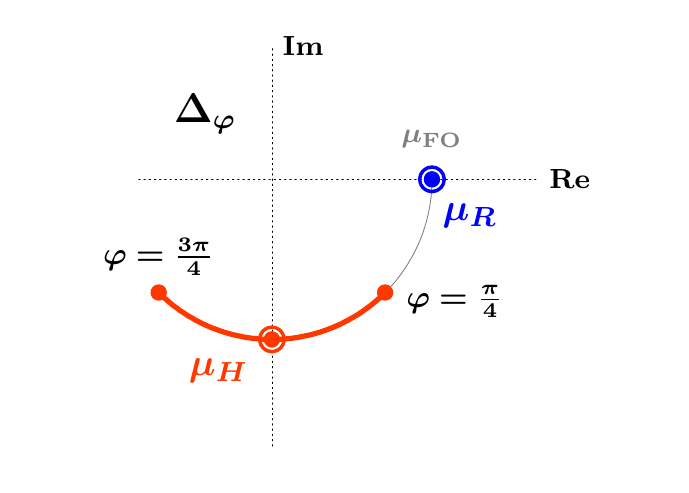}%
\caption{
Illustration of the scale variations used to estimate the perturbative uncertainties.
Left: The overall variations of $\mufo$, which determines $\DeltaFO$ (in
conjunction with the variation of $\kappa_F$, which is not shown).
Right: The phase variation for $\mu_H$ for fixed $\mufo$, which determines the resummation
uncertainty $\Delta_\varphi$.}
\label{fig:VariationSchemes}
\end{figure*}

We explicitly distinguish two different sources of perturbative uncertainties,
namely fixed-order and resummation uncertainties, that are associated to the
two independent perturbative series involved.
The fixed-order uncertainty, denoted as $\DeltaFO$, is obtained via the conventional
variations of $\mu_R$ and $\mu_F$. This comprises a collective overall variation
of $\mufo$ by a factor of two around its central value, which is combined with an additional
variation of $\kappa_F$ by a factor of two around its central value, without considering the
extreme variations where both are varied up or down at the same time. That is,
relative to the central values we consider the set of variations
\begin{equation}
V_\FO = \Bigl\{
\frac{\mufo}{2},\, 2\mufo,\, \frac{\kappa_F}{2},\, 2\kappa_F,\, \Bigl(\frac{\mufo}{2}, 2\kappa_F\Bigr),
\Bigl(2\mufo, \frac{\kappa_F}{2}\Bigr)
\Bigr\}
\,,\end{equation}
from which the fixed-order uncertainty $\DeltaFO$ is obtained as the
maximum deviation from the central value
\begin{equation}
\DeltaFO = \max_{v \in V_\FO} \bigl\lvert \sigma_{\rm vary}(v) - \sigma_{\rm central} \bigr\rvert
\,.\end{equation}
In the limit where the resummation is turned off, this reproduces the perturbative uncertainty
in the fixed-order predictions.
For the resummed predictions, the magnitude of the hard scale by construction follows the $\mufo$ variation,
$|\mu_H| = \mufo$, as illustrated in \fig{VariationSchemes} on the left, such that the fixed-order
variations do not change the resummed logarithms $\ln(\mu_H/\mufo)$.

For the resummation uncertainty, we vary the phase $\varphi$ in the interval $[\pi/4, 3\pi/4]$
around the central value of $\varphi = \pi/2$, while keeping $\mufo$ at its central value,
as illustrated in \fig{VariationSchemes} on the right.
This probes the intrinsic size of the higher-order timelike logarithms.
The phase variation by $\pm \pi/4$ is chosen to be roughly equivalent to the usual factor of 2
for conventional logarithms since $\pi/4 \simeq \ln 2$. The uncertainty $\Delta_\varphi$ is then obtained
as the maximum observed deviation from the central value (usually happening at one of the endpoints), such that
\begin{equation}
\Delta_{\varphi} = \max_{\varphi \in [\pi/4,\, 3\pi/4 ]} \bigl\lvert \sigma_{\rm vary}(\varphi) - \sigma_{\rm central} \bigr\rvert
\,.\end{equation}
This additional resummation uncertainty was not considered in earlier treatments, but
has already been included in the resummed 0/1/2-jet-bin results reported in \refcite{deFlorian:2016spz}.

The total perturbative uncertainty is obtained by adding the two independent sources,
\begin{equation}
 \DeltaFO \oplus \Delta_\varphi = \sqrt{\DeltaFO^2 + \Delta_\varphi^2}
\,.\end{equation}
For $b\bar b\to H$ we follow \refcite{Bonvini:2016fgf} and consider the low-scale
matching at $\mu_b$ onto the $b$-quark PDFs as a third independent source of
uncertainty $\Delta_b$, which is estimated by varying $\mu_b$ by a factor of two.

\section{Gluon fusion}
\label{sec:GluonFusion}

Gluon-fusion processes are well-known to contain large perturbative corrections,
which are partially due to the timelike logarithms in the gluon form factor,
as first demonstrated in \refcite{Ahrens:2008qu}. We first consider the total production
cross section up to \nnnllmatched* for a generic scalar final state $gg\to X$ in \sec{ggX}
and for the SM Higgs boson in the rEFT $m_t\to\infty$ limit in \sec{ggHInclusive}.
In \sec{quarkMassEffects} we discuss how to incorporate quark-mass and electroweak effects
into the resummed results. In \sec{ggHRapidity} we then present our results for the
Higgs rapidity spectrum and the cross section with a rapidity cut to \nnllmatched*.

\subsection{Color-singlet production}
\label{sec:ggX}

We first consider the total production cross section from gluon fusion for a generic
color-singlet scalar $X$ with mass $m_X$.
Its coupling to gluons at the scale $\mu \sim m_X$ can be expressed in terms of an effective Lagrangian as
\begin{equation} \label{eq:LagrangeX}
\mathcal{L}_\text{eff}(m_X) \supset -\frac{C_X}{\Lambda}\,\as\, G_{\mu\nu}^a G^{a,\mu\nu} X
\,,\end{equation}
where $\Lambda$ is a suitable high mass scale and $C_X$ is the Wilson coefficient
from integrating out heavy particles that mediate the effective $ggX$ interaction.
This effective operator arises for SM Higgs production in the $m_t\to\infty$ limit,
which we discuss in more detail in \sec{ggHInclusive}. Here, we use it as a simple case
to study the effects of the resummation and its dependence on the mass over a wide
range $m_X \in [100,1000]\GeV$. For this purpose, the precise values of the effective coupling
$C_X(\mu=m_X)/\Lambda$ need not be specified, as it drops out for the $K$-factor
$\sigma/\sigma^{(0)}$ on which the resummation acts.

We obtain the total \ggX* cross section to \nnnlo* from
\texttt{SusHi 1.6.0}~\cite{Harlander:2002wh, Harlander:2005rq, Harlander:2012pb,
Harlander:2016hcx, Chetyrkin:2000yt, Anastasiou:2014lda, Anastasiou:2015yha, Anastasiou:2016cez}.
Our central scale choices are $\mufo = m_X$ and $\kappa_F = 1$, such that $\mu_R = \mu_F = m_X$.
Away from $\mu = m_X$, the perturbative running of $C_X(\mu)$ induces logarithms of
$m_X/\mu$ at \nnlo* and \nnnlo*. Their resummation is irrelevant and can be neglected,
and they are instead included in the fixed-order cross section~\cite{Harlander:2016hcx}.

The gluon form factor is known up to three loops \cite{Harlander:2000mg, Gehrmann:2005pd, Moch:2005tm, Baikov:2009bg, Lee:2010cga, Gehrmann:2010ue}, and the Wilson coefficient $C_{gg}$ is explicitly extracted from it in \refcite{Gehrmann:2010ue} (see also \refscite{Idilbi:2005ni, Idilbi:2006dg}),
\begin{equation} \label{eq:GluonHard}
H_{gg}(m_X^2, \mu) = \bigl| C_{gg}(m_X^2, \mu) \bigr|^2
  = \biggl| 1 + \sum_{n=1}^\infty \biggl[\frac{\as(\mu)}{4\pi}\biggr]^n C_{gg}^{(n)} \biggl( \ln \frac{-m_X^2-\img 0}{\mu^2} \biggr) \biggr|^2
\,,\end{equation}
where now $Q^2 = m_X^2$. The RGE of $C_{gg}$ reads
\begin{align} \label{eq:RGECgg}
\mu \frac{\df}{\df \mu} C_{gg}(m_X^2, \mu)
&= \gamma_{gg}(m_X^2, \mu) \,C_{gg}(m_X^2, \mu)
\,,  \\\nn
\gamma_{gg}(m_X^2, \mu)
&= \Gamma^g_\mathrm{cusp} [\as(\mu)] \ln \frac{-m_X^2-\img 0}{\mu^2}
      + 2 \gamma_C^g[\as(\mu)]
      - \gamma_t [\as(\mu)]
      - \frac{\beta[\as(\mu)]}{\as(\mu)}
\,,\end{align}
where $\Gamma^g_\mathrm{cusp}(\as)$ is the gluon cusp anomalous dimension and the last three terms are the total noncusp contribution. All the relevant ingredients are collected in \app{ingredients}.

\begin{figure*}
\includegraphics[width=\WidthThreeSubfigs]{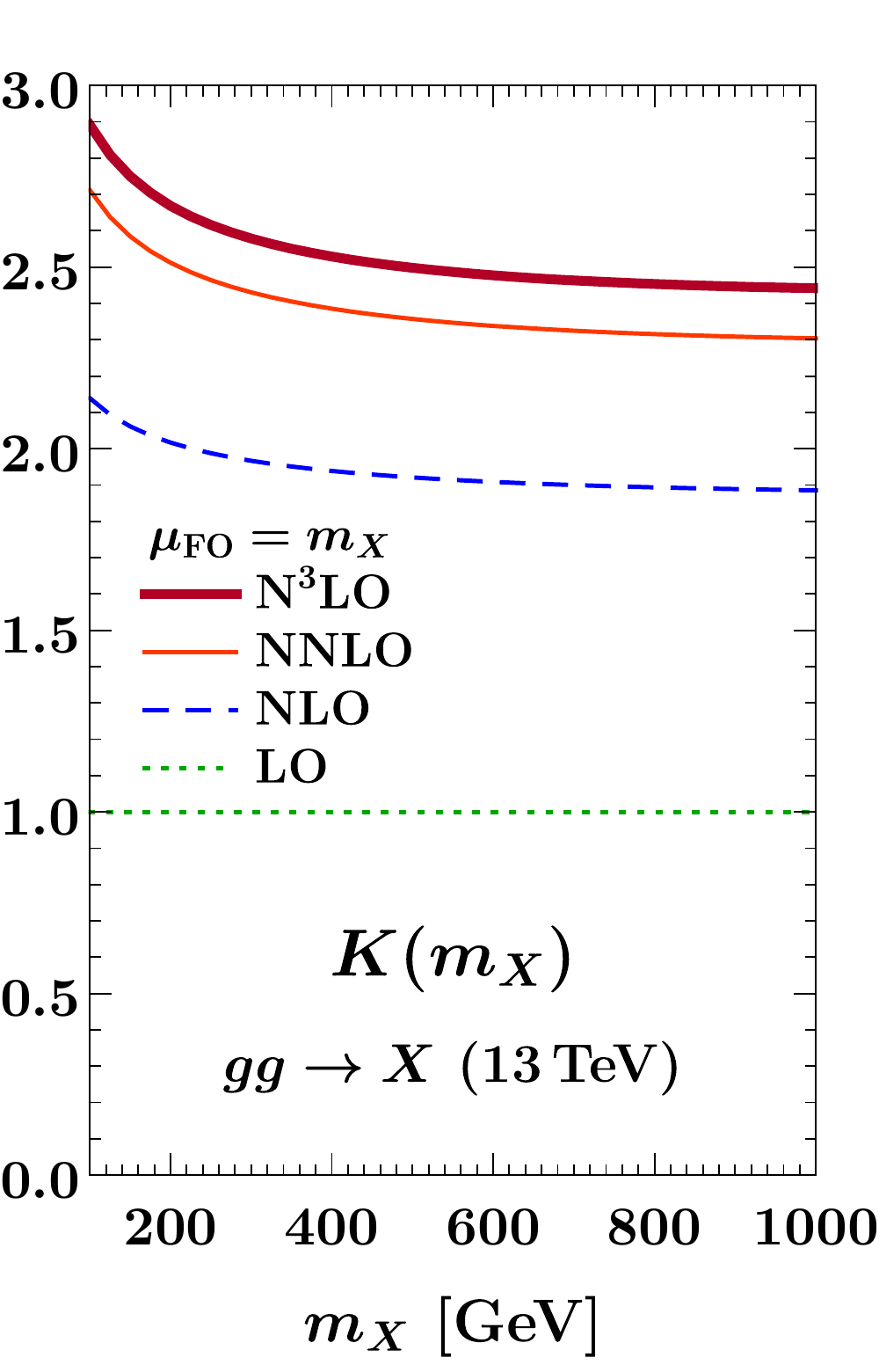}%
\hfill%
\includegraphics[width=\WidthThreeSubfigs]{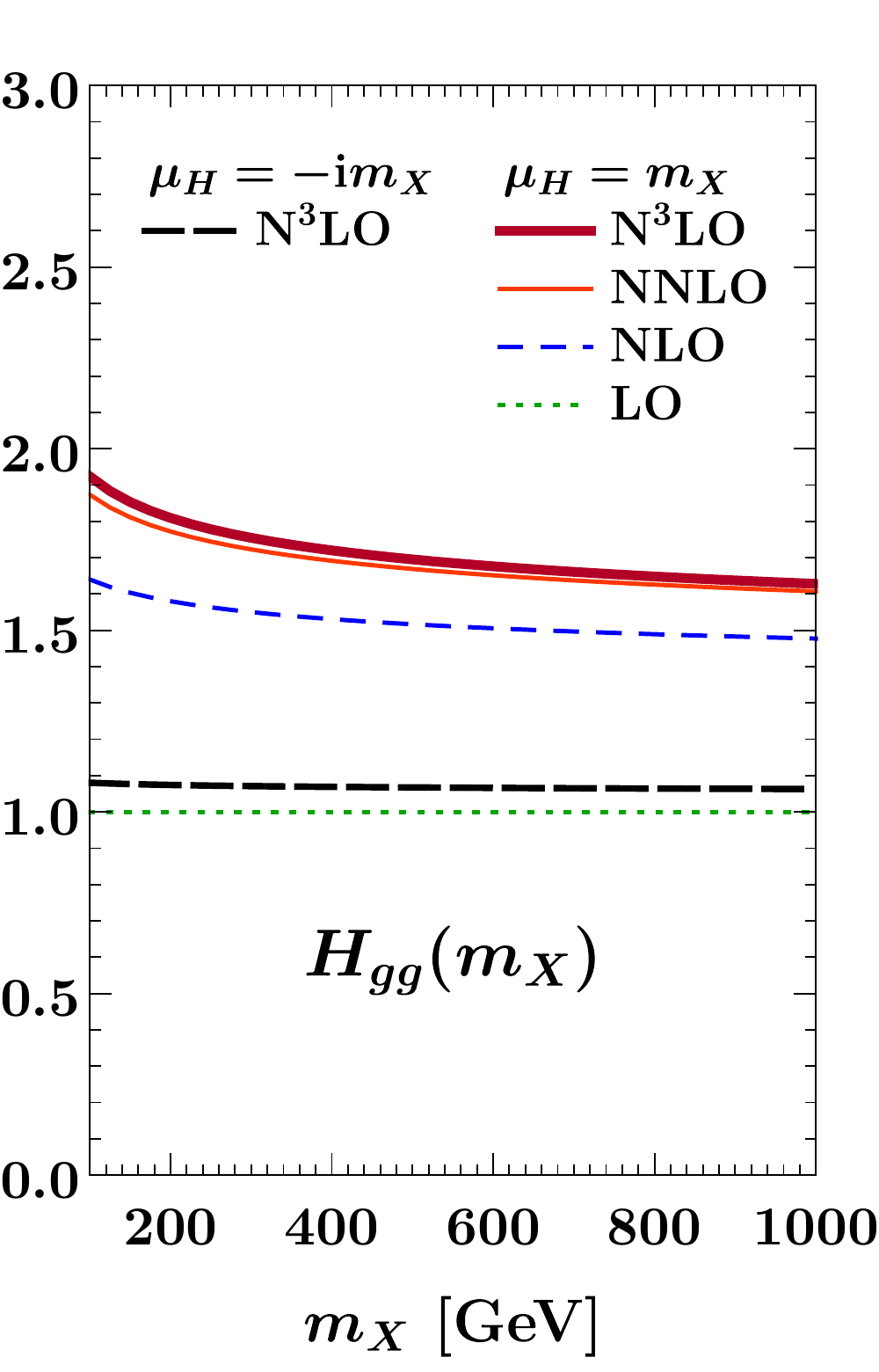}%
\hfill%
\includegraphics[width=\WidthThreeSubfigs]{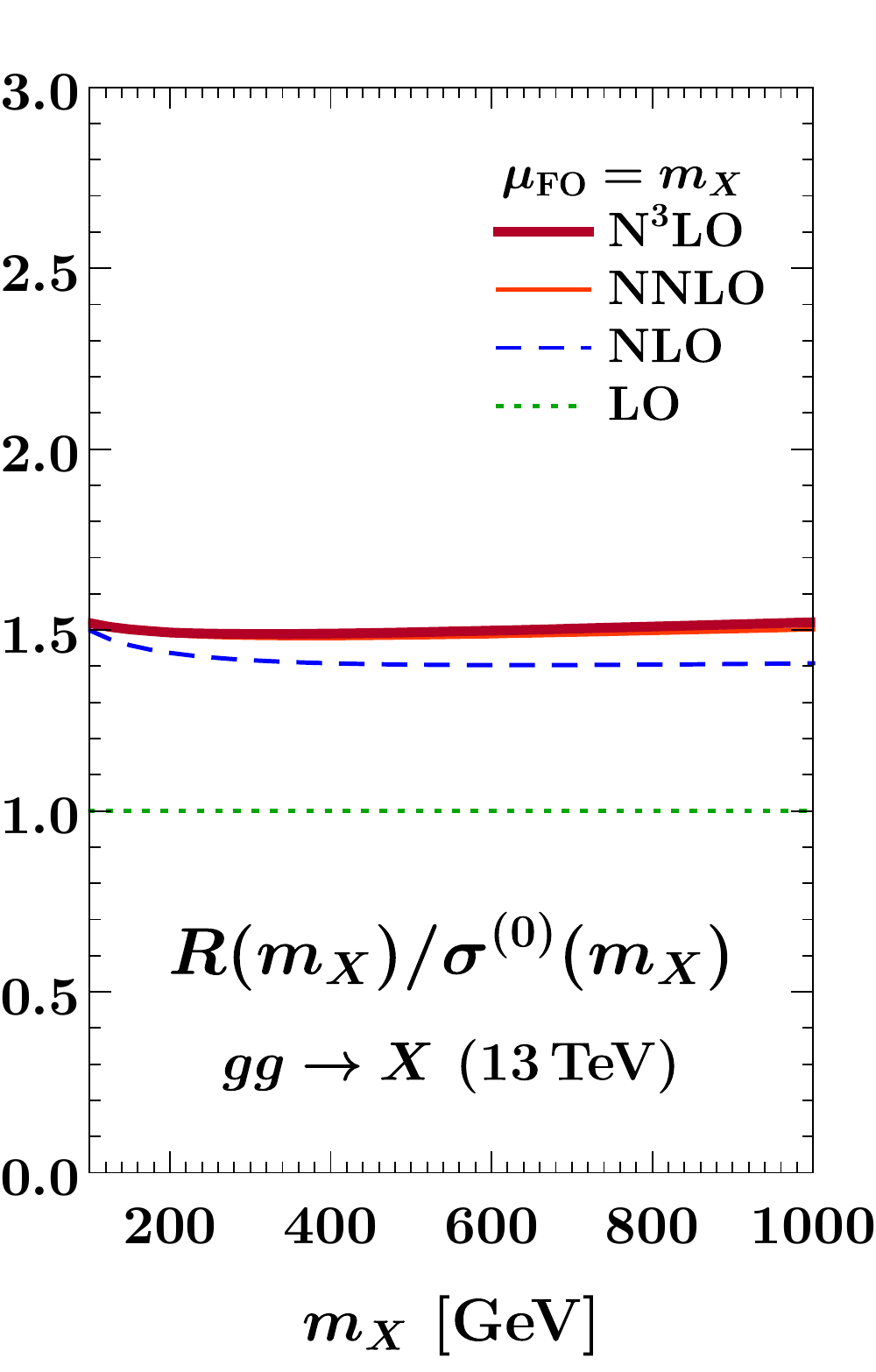}
\caption{
Illustration of the fixed-order perturbative series for \ggX* at $\mufo = m_X$ for the inclusive $K$-factor (left), the hard function $H_{gg}$ at $\mu_H = m_X$ (center), and the normalized remainder $R/\sigma^{(0)}$ (right).
The middle panel also shows the \nnnlo* hard function $H_{gg}$ at $\mu_H = -\img m_X$ (black long dashed),
for which it contains no timelike logarithms.
}
\label{fig:ggX:KvsHvsR}
\end{figure*}

The separation of the perturbative series for the $K$ factor at fixed order into those of $H$ and $R$ is shown
in \fig{ggX:KvsHvsR} as a function of $m_X$. Half of the large NLO $K$ factor comes from $H$ and half from $R$,
while beyond NLO the corrections in $H$ are larger than for $R$.
Hence, the large corrections to the $K$-factor present at each order are driven to a large
extent (but also not entirely) by the corrections from $H$.
In particular, the remainder $R$ by itself has a much better behaved perturbative series than $K$, and there are
clearly no cancellations between $H$ and $R$. (Otherwise, as already explained in \sec{ResummationScheme}, $R$ would
need to have negative corrections that are larger in size than those in $K$.) This pattern holds
independently of $m_X$. The visible increase in the corrections toward smaller $m_X$ is due to the
running of $\alpha_s(m_X)$.

The large perturbative corrections in $H_{gg}$ at the real scale $\mu_H = m_X$ are absent
at the imaginary scale $\mu_H = -\img m_X$, as shown by the long-dashed curve
in the middle panel of \fig{ggX:KvsHvsR}.
To illustrate this more explicitly, the numerical values for an example mass of $m_X = 750 \GeV$ are,%
\footnote{The value is chosen purely for historical reasons.}
\begin{alignat}{3} \label{eq:GluonHardNumbersSevenFifty}
&H_{gg}(m_X^2, \mu_H=m_X)       &&= 1 + 0.49279 + 0.13855 + 0.02288
\,, \nn \\
&H_{gg}(m_X^2, \mu_H=-\img m_X) &&= 1 + 0.06820 - 0.00102 - 0.00251
\,,\end{alignat}
where each term is the contribution from a subsequent order in $\as$ up to \nnnlo*.
Clearly, the large corrections to the gluon form factor at real scales are almost entirely due to the
timelike Sudakov logarithms that are present for $\mu_H = m_X$ and are eliminated by taking
$\mu_H = -\img m_X$.
Since the corrections in $H_{gg}$ at $\mu_H = -\img m_X$ are very small, the perturbative
convergence of the resummed cross section will be essentially determined by that of the remainder $R$.

\begin{figure*}
\includegraphics[width=\WidthTwoSubfigs]{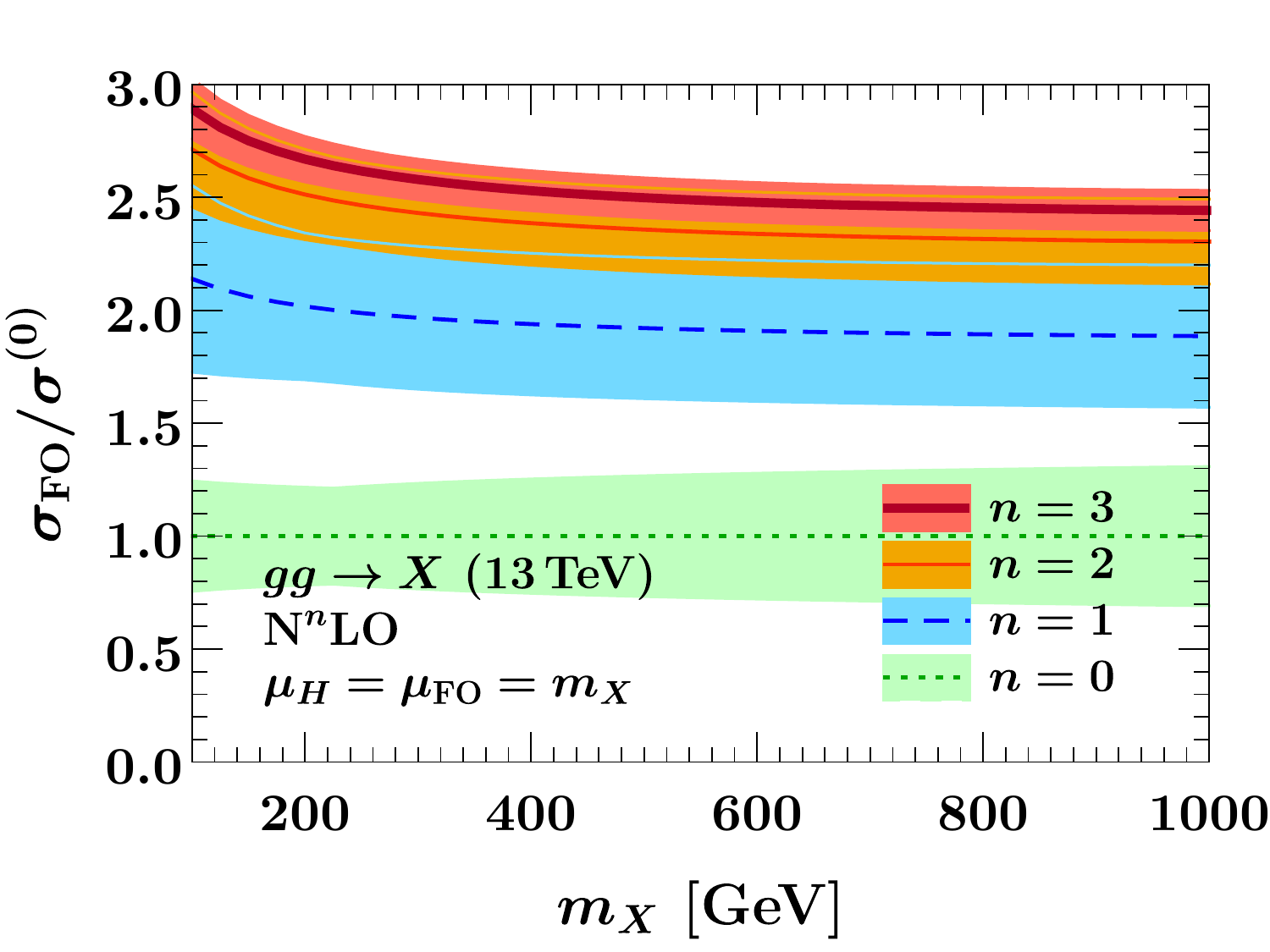}%
\hfill%
\includegraphics[width=\WidthTwoSubfigs]{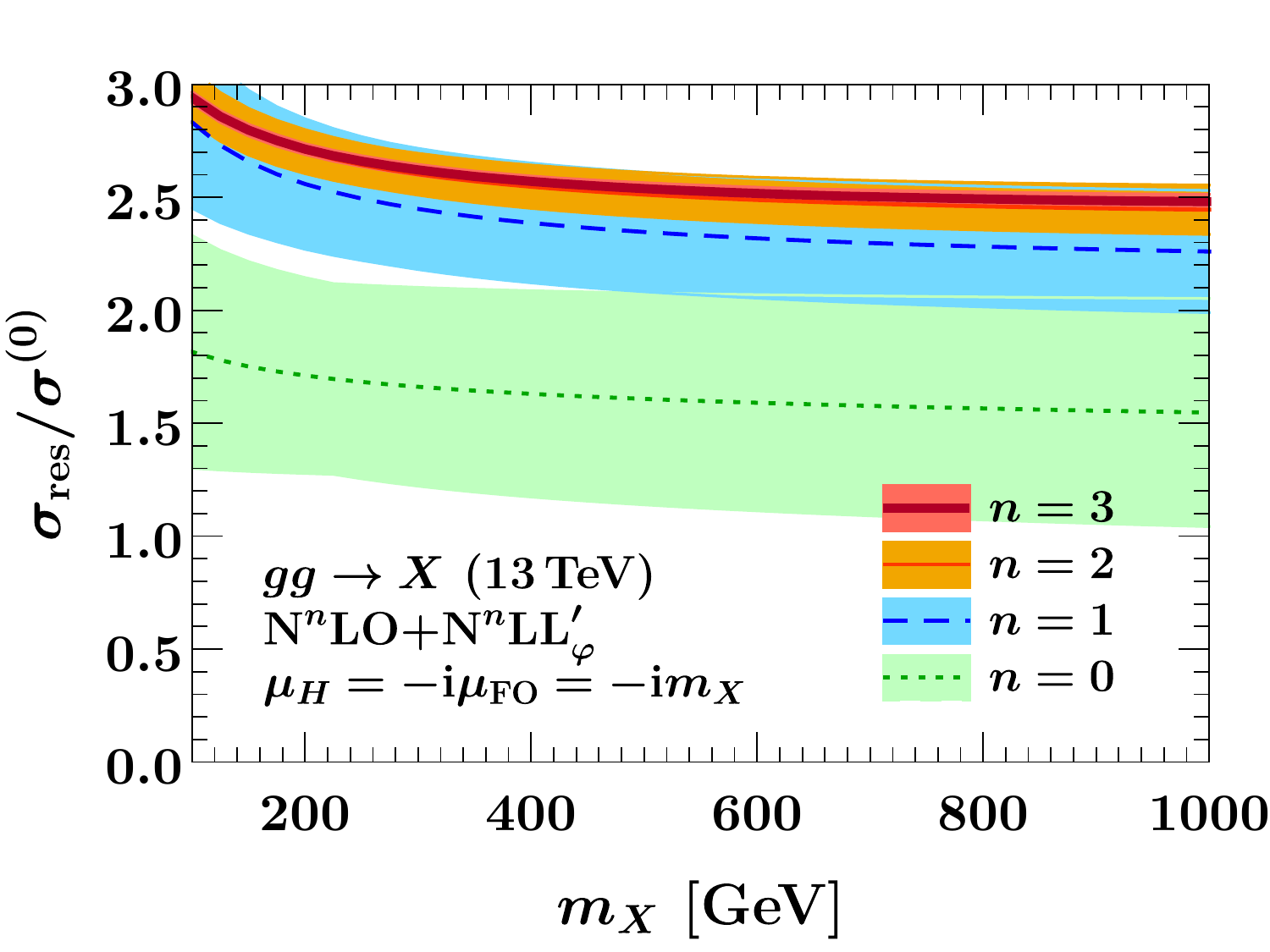}%
\caption{
The total cross section for \ggX* at $\sqrts = 13 \TeV$
at fixed order (left) and including the resummation of timelike logarithms (right).
All results are normalized to the central \lo* prediction at $\mu_\FO = m_X$.
}
\label{fig:ggX:FOvsRes}
\end{figure*}

In \fig{ggX:FOvsRes}, we compare the fixed-order and resummed cross sections as a function of $m_X$,
with the bands showing the total perturbative uncertainties evaluated as discussed in \sec{Uncertainties}.
(Note that in case of \ggX* and \ggH*, the fixed-order uncertainties come from the variation of $\mu_R$ for fixed $\mu_F$.) All results are normalized to the \lo* prediction $\sigma^{(0)}$ at fixed $\mufo = m_X$.
As expected, the absence of large corrections in the resummed hard function directly translates into a much faster convergence of the resummed cross section.
Furthermore, the uncertainties in the resummed predictions at lower orders cover the higher-order bands
much better than at fixed order, while at the same time being substantially reduced at higher orders.
Hence, even at \nnlo* and \nnnlo*, where the fixed-order results start to show convergence,
the resummation noticeably improves the predictions.
Due to their better convergence, the resummed predictions provide substantially improved
uncertainty estimates both in terms of their reliability and their size.
In particular, we can be reasonably confident that the result at the next higher order will lie within the small
\nnnllmatched* uncertainty band.

\subsection{Inclusive Higgs production in the rEFT scheme}
\label{sec:ggHInclusive}

We now turn to the case of Higgs production through gluon fusion as an important
application of the singlet production discussed above. For Higgs masses below the
top threshold, $m_H < 2 m_t$, the gluon-fusion cross section can be well approximated
by an effective theory where the top quark is integrated
out~\cite{Wilczek:1977zn, Shifman:1978zn,Inami:1982xt,Spiridonov:1988md},
giving rise to an effective Lagrangian analogous to \eq{LagrangeX},
\begin{equation} \label{eq:LagrangeH}
 \mathcal{L}(m_H) \supset - \frac{C_t}{12\pi v}\,\as\, G_{\mu\nu}^a G^{a,\mu\nu} H
\,.\end{equation}
In this case, the Wilson coefficient $C_t$
itself receives sizable QCD corrections, which have been calculated to N$\textsuperscript{4}$LO in \refscite{Chetyrkin:1997un,Schroder:2005hy,Chetyrkin:2005ia}.
The effective operator in \eq{LagrangeH} is the same as in \eq{LagrangeX}, giving rise to
the same gluon form factor and hard function $H_{gg}$ in \eq{GluonHard}.

Rescaling the cross section $\sigma^\EFT$ obtained from \eq{LagrangeH} by the LO $m_t$ dependence~\cite{Georgi:1977gs}
\begin{equation} \label{eq:def F0}
F_0(\rho) = \frac{3}{2\rho} - \frac{3}{2\rho} \biggl| 1- \frac{1}{\rho} \biggr| \arcsin^2( \sqrt{\rho} )
\,,\qquad
\rho = \frac{m_H^2}{4 m_t^2} < 1
\,,\end{equation}
one obtains the inclusive cross section in the ``rescaled EFT'' scheme (rEFT),
\begin{equation} \label{eq:def rEFT}
\sigma^\rEFT
= |F_0(\rho)|^2\, \sigma^\EFT
\,.\end{equation}
This rescaling is known to well reproduce the $m_t$-exact result at NLO,
and hence it is believed to be a useful approximation also at higher orders~\cite{Graudenz:1992pv, Dawson:1993qf, Spira:1995rr, Harlander:2009bw, Pak:2009bx, Harlander:2009mq, Pak:2009dg, Harlander:2009my}.
The inclusion of further quark mass and electroweak effects will be discussed in \sec{quarkMassEffects}.

We use \texttt{SusHi 1.6.0}~\cite{Harlander:2002wh, Harlander:2005rq, Harlander:2012pb,
Harlander:2016hcx, Chetyrkin:2000yt}
to compute the total cross section in the rEFT scheme to \nnlo*.
For the \nnnlo* contribution we use the results of \refcite{Anastasiou:2016cez}
as implemented in \texttt{ggHiggs 3.5}~\cite{Bonvini:2016frm}.%
\footnote{%
In \texttt{SusHi 1.6.0}, the $\mu_F$ and $\mu_R$ dependence at \nnnlo* is threshold expanded consistently with
the $\mu$-independent terms, while it is kept exact in \refscite{Anastasiou:2016cez, Bonvini:2016frm}.
There is no clear theoretical preference for either treatment. The resulting numerical differences away
from the canonical values $\mu_R = \mu_F = m_H$ are around 0.3\%, consistent with the level of systematic uncertainties
expected from the threshold expansion~\cite{Anastasiou:2016cez}.
To ease numerical comparisons we use the numerical values corresponding to the exact running here.
}
We use $m_t^\mathrm{OS} = 172.5 \GeV$ and $m_H = 125 \GeV$. To study the perturbative series and
the resummation effects we choose the canonical values $\mufo = m_H$ and $\kappa_F = 1$
(so $\mu_R = \mu_F = m_H$) as central values. With these settings,
the \nnnlo* cross section has the perturbative series
\begin{align} \label{eq:ggHxsecFO}
\sigma_\FO^\rEFT &= (1 + 1.291 + 0.783 + 0.296) \times 13.80 \pb
\,, \nn \\
H_{gg}(m_H^2, \mu_H=m_H)  &= \,\,1 + 0.619 + 0.219 + 0.045
\,, \nn \\
R(\mufo = m_H) &= (1 + 0.672 + 0.148 + 0.012) \times 13.80 \pb
\,,\end{align}
where again each term gives the contribution from a subsequent order in $\alpha_s$.
The remainder $R$ now includes the corrections to $|C_t|^2$.
As before, its perturbative series is much better behaved than that of the
cross section, whose large perturbative corrections are thus driven by the
large corrections from timelike logarithms in $H_{gg}$.

To illustrate the improved convergence of the resummed form factor,
we consider the hard function $H_{gg}(m_H, \mu_H)$ at various scales $\mu_H$,
\begin{alignat}{3} \label{eq:GluonHardNumbersHiggs}
&H_{gg}(m_H^2, \mu_H=m_H) 	&&= 1 + 0.61925 + 0.21878 + 0.04539 \,, \nn \\
&H_{gg}(m_H^2, \mu_H=-\img m_H) 	&&= 1 + 0.08408 - 0.00145 - 0.00441 \,, \nn \\
&H_{gg}(m_H^2, \mu_H= m_H/2) 	&&= 1 + 0.57325 + 0.12361 - 0.00839 \,, \nn \\
&H_{gg}(m_H^2, \mu_H=-\img m_H/2) &&= 1 - 0.01553 - 0.01544 - 0.00247 \,, \nn \\
&H_{gg}(m_H^2, \mu_H=m_H/5) 	&&= 1 + 0.08090 - 0.16424 - 0.00552
\,.\end{alignat}
For both imaginary-valued scales $\mu_H = -\img m_H$ and $\mu_H = -\img m_H/2$,
the corrections are drastically reduced compare to the real scale choice.
For comparison, choosing a real value $\mu_H = m_H/5$ that yields the same reduced NLO correction as
$\mu_H = -\img m_H$ still leads to much larger NNLO corrections.

\begin{figure*}
\includegraphics[width=\WidthTwoSubfigs]{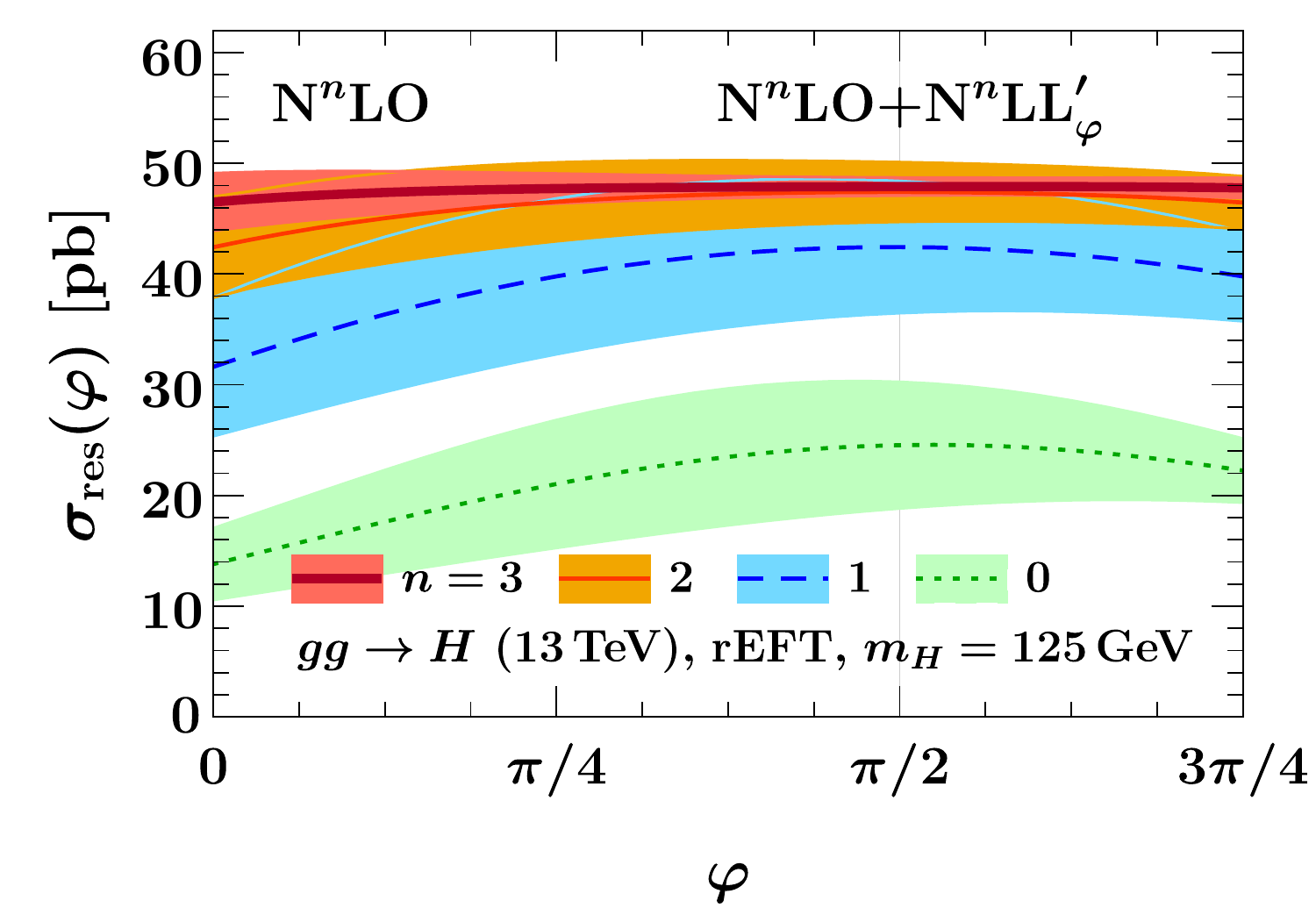}%
\hfill%
\includegraphics[width=\WidthTwoSubfigs]{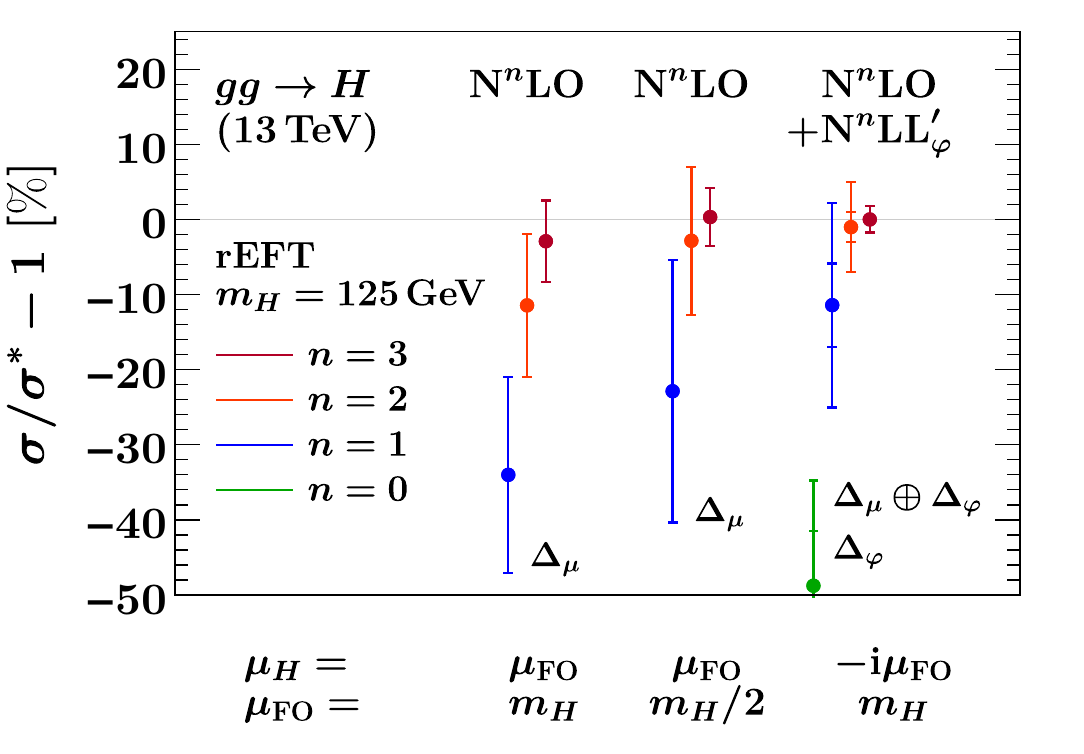}%
\caption{
The \ggH* cross section at $\sqrts = 13 \TeV$ and $m_H = 125\GeV$ in the rEFT scheme.
Left: The cross section as a function of the resummation phase $\varphi$ of the
hard scale $\mu_H = \mu_\FO \exp(-\img \varphi)$, with the uncertainty bands corresponding
to $\DeltaFO$ only.
Right: Comparison of the fixed-order results for $\mufo = m_H$ and
$\mufo = m_H/2$, and the resummed results with $\mu_\FO = \img \mu_H = m_H$.
All results are given as the percent difference from the \nnnllmatched* central value.
The uncertainty bars show $\DeltaFO$ for the fixed-order results and
$\DeltaFO \oplus \Delta_\varphi$ for the resummed results (with the inner bars
visible at the lower orders showing $\Delta_\varphi$ only).
The fixed \lo* results are out of range.
}
\label{fig:ggH:PhaseVariation}
\end{figure*}

To examine the dependence on the resummation phase $\varphi$ of the hard scale, $\mu_H = \mufo \exp(-\img \varphi)$,
we show in the left panel of \fig{ggH:PhaseVariation} the resummed cross section as a function of $\varphi$.
Here, the uncertainty bands only show the fixed-order uncertainty $\DeltaFO$.
At $\varphi = 0$, $\sigma_\res(\varphi)$ is just the fixed-order cross section.
As $\varphi \rightarrow \pi/2$, the timelike resummation is turned on, visibly improving the convergence of the cross section and providing better coverage of the uncertainty bands. The $\varphi$ dependence
becomes stationary at $\varphi = \pi/2$, where the timelike Sudakov logarithms exactly vanish.
Beyond $\varphi = \pi/2$, powers of $\varphi - \pi/2$ start to enter again.

In the right panel of \fig{ggH:PhaseVariation}, we compare the fixed-order results at
the conventional scales of $\mufo = m_H$ and $\mufo = m_H/2$ with the resummed results.
The results are shown as relative corrections to our best prediction at \nnnllmatched*.
For the resummed results, the inner uncertainty bars indicate $\Delta_\varphi$ alone,
while the outer ones show $\DeltaFO \oplus \Delta_\varphi$.
While $\Delta_\varphi$ contributes to obtaining a more realistic uncertainty estimate at \llmatched* (compared to \lo*), its impact is strongly reduced at higher orders.
The overall picture and conclusions from the generic color-singlet case are unaffected
by the presence of the Wilson coefficient $\vert C_t \vert^2$ in the cross section.
The resummation yields again a clear improvement in convergence and uncertainties,
also compared to the fixed-order results at $\mufo = m_H/2$, which are
already better behaved than those at $\mufo = m_H$.
In particular, the \nllmatched* result already fully covers the highest-order result,
which is not the case at fixed NLO, and the precision of the \nnllmatched* result
is roughly comparable to the fixed N$^3$LO results.
This gives us good confidence in the small remaining uncertainty at \nnnllmatched*,
which is reduced by a factor of two compared to N$^3$LO.
The explicit numerical results at the highest order are
\begin{alignat}{9} \label{eq:ggHxsecRes}
&\sigma_\FO^\rEFT &&= (46.51\pm 2.60_\mu)\pb~&&(5.59\%)
\,,\quad && (\text{\nnnlo*},\, \mufo = m_H)
\,, \nn \\
&\sigma_\FO^\rEFT &&= (48.06\pm 1.83_\mu)\pb~&&(3.82\%)
\,, \quad && (\text{\nnnlo*},\, \mufo = m_H/2)
\,, \nn \\
&\sigma_\res^\rEFT &&= (47.90\pm 0.82_\mu \pm 0.18_\varphi)\pb~&&(1.75\%)
\,, \quad  && (\text{\nnnllmatched*},\, \mufo=m_H)
\,.\end{alignat}
Note that for the \nnnlo* results in \refcite{Anastasiou:2016cez} the perturbative
uncertainties are estimated by varying $\mufo$ but keeping $\kappa_F = 1$ fixed.
Doing so reduces $\Delta_\mu$ to $2.21\pb$ $(4.76\%)$ at central $\mufo = m_H$ and
$1.54\pb$ $(3.21\%)$ at central $\mufo = m_H/2$.%
\footnote{Ref.~\cite{Anastasiou:2016cez} further utilizes the $\overline{\text{MS}}$
top-quark mass $\overline{m}_t(\mufo)$ in the rescaling factor in \eq{def rEFT},
which participates in the overall $\mufo$ scale variation and further reduces its effect to $2.4\%$.
However, the perturbative series for the $\overline{\text{MS}}$
top-quark mass entering in the rescaling factor has nothing to do with the perturbative
series of the $gg\to H$ cross section in the $m_t\to\infty$ limit arising from the
effective Lagrangian \eq{LagrangeH}. Hence, the fact that their $\mufo$
dependences partially compensate numerically is purely accidental.}
Similarly dropping the $\kappa_F$ variation in the resummed results gives
$\Delta_\mu = 0.67\pb$, which combined with $\Delta_\varphi$ then yields a total
perturbative uncertainty of $1.44\%$. 
Note also that using the threshold-expanded running in $\mu_R$ and $\mu_F$ as implemented
in \texttt{SusHi 1.6.0}, the \nnnlo* result at $\mufo = m_H/2$ increases to
$(48.17\pm 1.99_\mu)\pb~(4.14\%)$, with a corresponding increase in $\Delta_\mu$ since
the result at $\mufo = m_H$ is unaffected.

\subsection{Incorporating quark mass and electroweak effects beyond rEFT}
\label{sec:quarkMassEffects}

While the previous section focused on the QCD corrections to Higgs production in the
$m_t\to\infty$ limit, further corrections arise from finite quark-mass effects
as well as electroweak contributions.
Here we discuss how to consistently combine them with the resummation of timelike logarithms.

The full dependence of the cross section on the heavy quark masses $m_t$, $m_b$, $m_c$ is fully known
at \nlo*~\cite{Graudenz:1992pv, Dawson:1993qf, Spira:1995rr, Harlander:2005rq,
Anastasiou:2006hc, Aglietti:2006tp, Anastasiou:2009kn}.
We define $\delta\sigma^q_\text{(N)LO}$ as the correction of the exact result relative to
the rEFT result,
\begin{align} \label{eq:def delta sigma}
\sigma_\nlo*^{t,b,c} &= \sigma^\rEFT_\nlo* + \delta\sigma_\lo*^{b,c} + \delta\sigma^{t,b,c}_\nlo*
\,.\end{align}
On top of the exact \nlo* corrections, top-quark mass effects are also known in an
asymptotic expansion in $1/m_t$ at \nnlo*~\cite{Harlander:2009bw, Pak:2009bx, Harlander:2009mq, Pak:2009dg, Harlander:2009my}.

In the following we consider the top-mass effects in more detail.
As discussed in \refscite{Harlander:2009my, Harlander:2009mq, Pak:2009dg, Harlander:2016hcx},
the asymptotic $1/m_t$ corrections at NNLO cannot be expected to reliably improve
over the $m_t\to\infty$ limit. Rather, they can serve to estimate the uncertainty due
to the still unknown full NNLO $m_t$ corrections. For this reason we will only take
into account the NLO corrections $\delta\sigma^{t}_\nlo*$. (The inclusion of the NNLO
$m_t$ corrections would be completely analogous.)
This is also consistent with our analysis of the rapidity spectrum in \sec{ggHRapidity},
for which the $m_t$-corrections are only known at \nlo*.
For illustration, the numerical results for $\delta\sigma^{t}_\nlo*$ are
\begin{alignat}{7} 
&\delta\sigma^{t}_\nlo* = -0.210 \pb \,,\qquad &&(\mufo = m_H)
\,, \\
&\delta\sigma^{t}_\nlo* = -0.315 \pb \,, &&(\mufo = m_H/2)
\,. \nn
\end{alignat}

The finite $m_t$ contributions correspond to a correction to the $m_t\to\infty$ limit in \eq{LagrangeH},
from which the gluon form factor arises, and so a priori they do not
involve the same local gluon form factor. Therefore, one option to include them in
the resummed results is to simply add them to the rEFT results in \eq{ggHxsecRes}, which yields
\begin{alignat}{7} \label{eq:ggHxsecRes delta sigma added}
&\sigma_\FO^\rEFT + \delta\sigma^{t}_\nlo* = (46.30\pm 2.55_\mu)\pb~&&(5.50\%)
\quad && (\text{\nnnlo*},\, \mufo = m_H)
\,, \nn \\
&\sigma_\FO^\rEFT + \delta\sigma^{t}_\nlo* = (47.74\pm 1.75_\mu)\pb~&&(3.66\%)
\quad && (\text{\nnnlo*},\, \mufo = m_H/2)
\,, \\ \nn
&\sigma_\res^\rEFT + \delta\sigma^{t}_\nlo* = (47.69\pm 0.78_\mu \pm 0.18_\varphi)\pb~&&(1.68\%)
\,, \quad  && (\text{\nnnllmatched*},\, \mufo=m_H)
\,.
\end{alignat}
The complete results including those at lower orders are collected in \tab{ggH:Inclusive}.

Alternatively, following \refcite{Berger:2010xi} we can perform a one-step matching
of the full Standard Model including the top quark onto SCET, simultaneously integrating
out both the top quark and hard virtual corrections. The resulting hard function $H_{gg}^t = |C_{gg}^t|^2$
corresponds to the full SM $gg\to H$ form factor and includes all virtual finite-$m_t$ effects.
It takes the form~\cite{Berger:2010xi}
\begin{align} \label{eq:GluonHardFull}
H_{gg}^t(m_t, m_H^2, \mu)
= |F_0(\rho)|^2 |\as(\mu)|^2 \,
\Bigl\{
   &\bigl|C_t(m_t, \mu)\, C_{gg}(m_H^2, \mu) \bigr|^2
\nn \\
   &+ 2 \Re \Bigl[ \frac{\as(\mu)}{4\pi} ( F_1(\rho) - F_1(0) ) \Bigr]
   + \ord{\rho \, \as^2}
\Bigr\}
\,,\end{align}
where as before $\rho \equiv m_H^2/(4m_t^2)$.
Compared to $H_{gg} = |C_{gg}|^2$, the LO $m_t$ dependence $F_0(\rho)$ and the contributions
from $C_t$ are now moved from the remainder into the hard function.
The $F_1(\rho)$ contains the full virtual $m_t$ dependence at NLO and the $\ord{\rho \, \as^2}$
terms denote the neglected \nnlo* virtual $m_t$ corrections.%
\footnote{The $-F_1(0)$ here simply removes the leading $m_t\to\infty$ part of $F_1(\rho)$, which
is already included via $C_t$. We drop all cross terms of $F_1(\rho) - F_1(0)$ with $C_{gg}$, which are
of $\ord{\rho \, \as^2}$ and higher, because these terms are also not included in the
fixed-order cross section.}
Although $H_{gg}^t$ is no longer normalized to unity at leading order, we can continue to use \eq{sigmaresAlt}
to obtain the resummed cross section.
Compared to \eq{sigmares}, the result now contains an overall factor $\vert \as(\mu_H) / \as(\mufo) \vert^2$
from the ratio of hard functions, which replaces the $\as^2(\mufo)$ inside the \lo* cross section by $|\as(\mu_H)|^2$.

The RGE for $C_{gg}^t$ is given by
\begin{align} \label{eq:RGECggPrime}
\mu \frac{\df}{\df \mu} C_{gg}^t(m_t, m_H^2, \mu)
&= \gamma_{gg}^t(m_H^2, \mu) \, C_{gg}^t(m_t, m_H^2, \mu)
\,, \nn \\*
\gamma_{gg}^t(m_H^2, \mu)
&= \Gamma^g_\mathrm{cusp}[\as(\mu)] \ln \frac{-m_H^2-\img 0}{\mu^2}
   + 2 \gamma_C^g[\as(\mu)]
\,.\end{align}
The noncusp terms in $\gamma_{gg}^t$ differ from those in $\gamma_{gg}$ in \eq{RGECgg}
due to the additional $\mu$ dependence of $\as(\mu) C_t(\mu)$, which is now included in the
hard Wilson coefficient. The overall $|\as(\mu)|^2 |C_t(m_t,\mu)|^2$
in \eq{GluonHardFull} is now evaluated at $\mu_H = -\img\mufo$ and then evolved
back to $\mufo$. For the overall $\as(\mu)$ this is largely irrelevant since it
is ultimately evolved starting from $\as(m_Z)$. For $C_t(\mu)$, which is treated in fixed order,
this induces different subleading timelike logarithms starting at NNLO compared to $H_{gg}$.
This is reflected in the noncusp terms differing by $\gamma_t$, whose numerical effect
however is not significant. Also, the perturbative convergence of $|C_t(\mu)|^2$ at
$\mu = m_H$ and $\mu = -\img m_H$ (and at its natural scale $\mu = m_t$)
is practically the same.

The perturbative convergence of $H_{gg}^t$ shows the same improvement as seen for $H_{gg}$
when evaluated at $\mu_H = -\img m_H$ rather than $\mu_H = m_H$,
\begin{alignat}{3} \label{eq:HPrimeNumbers}
&H_{gg}^t(m_H, \mu_H=m_H)         &&= \vert \as(m_H) \vert^2 \, \vert F_0 \vert^2
&\times\, \bigl(1 + 0.82152 + 0.36170 + 0.10268 \bigr)
\,, \nn \\
&H_{gg}^t(m_H, \mu_H=-\img m_H)   &&= \vert \as(-\img m_H) \vert^2 \, \vert F_0 \vert^2
&\times\, \bigl(1 + 0.27631 + 0.04244 - 0.00257 \bigr)
\,.\end{alignat}
The main difference compared to $H_{gg}$ are the additional constant terms from $C_t$
that are now included in $H_{gg}^t$. The finite-$m_t$ corrections have a very
small effect on the NLO contribution, contributing a $+0.005$ to the above
$0.82152$ and $0.27631$.

For reference, we first consider the rEFT limit and drop the finite-$m_t$ terms in
$H_{gg}^t$ as well as $\delta\sigma_\nlo*^t$.
The rEFT result based on $H_{gg}^t$ at \nnnllmatched* then reads
\begin{equation} \label{eq:ggHxsecResFullHardFunction}
\sigma_{\res,\,H^t}^\rEFT = (47.98\pm 0.85_\mu \pm 0.24_\varphi)\pb~(1.85\%)
\,.\end{equation}
This is equivalent to the results at $\mu_H = -\img m_H$
reported in \refcite{deFlorian:2016spz}.
Including the full NLO $m_t$ dependence, we obtain
\begin{equation} \label{eq:ggHxsec finite top resummed H prime}
\left( \sigma^\rEFT + \delta\sigma^{t}_\nlo* \right)_{\res,\,H^t}
= (47.84\pm 0.81_\mu \pm 0.25_\varphi)\pb~(1.77\%)
\,.\end{equation}
The full set of results including the lower orders are shown in the last column of
\tab{ggH:Inclusive}.

\begin{table}
\centering
\begin{small}
\begin{tabular}{l|ll|ll}
\hline\hline
\multicolumn{5}{c}{$\sigma\,[\!\pb]$ for $gg\to H$, $\sqrts = 13 \TeV$, $m_H = 125\GeV$}
\\
\hline\hline
& \multicolumn{2}{c|}{$\sigma^\rEFT_\FO \!+ \delta\sigma^t_\nlo*$}
& \multicolumn{1}{c}{$\sigma^\rEFT_\res \!+ \delta\sigma^t_\nlo*$}
& \multicolumn{1}{c}{$(\sigma^\rEFT + \delta\sigma^t_\nlo*)_\res$}
\\
$n$
& \multicolumn{1}{c}{$\nNlo* ,\,\mufo = m_H$} & \multicolumn{1}{c|}{$\nNlo* ,\,\mufo = \frac{m_H}{2}$}
& \multicolumn{1}{c}{$\nNllmatched*\,\,(H_{gg})$} & \multicolumn{1}{c}{$\nNllmatched*\,\, (H^t_{gg})$}
\\
\hline
$0$
&$13.8\!\pm\! 3.2_\mu\,(23\%)$
&$16.0\!\pm\! 4.3_\mu\,(27\%)$
&$24.5\!\pm\! 5.7_\mu\!\pm\! 3.5_\varphi\,(27\%)$
&$23.3\!\pm\! 5.1_\mu\!\pm\! 3.4_\varphi\,(26\%)$
\\
$1$
&$31.4\!\pm\! 6.2_\mu\,(20\%)$
&$36.6\!\pm\! 8.2_\mu\,(23\%)$
&$42.2\!\pm\! 5.9_\mu\!\pm\! 2.7_\varphi\,(15\%)$
&$41.8\!\pm\! 5.7_\mu\!\pm\! 2.8_\varphi\,(15\%)$
\\
$2$
&$42.2\!\pm\! 4.5_\mu\,(11\%)$
&$46.2\!\pm\! 4.6_\mu\,(10\%)$
&$47.2\!\pm\! 2.6_\mu\!\pm\! 1.0_\varphi\,(6.0\%)$
&$47.3\!\pm\! 2.7_\mu\!\pm\! 1.0_\varphi\,(6.1\%)$
\\
$3$
&$46.3\!\pm\! 2.5_\mu\,(5.5\%)$
&$47.7\!\pm\! 1.7_\mu\,(3.7\%)$
&$47.7\!\pm\! 0.8_\mu\!\pm\! 0.18_\varphi\,(1.7\%)$
&$47.8\!\pm\! 0.8_\mu\!\pm\! 0.25_\varphi\,(1.8\%)$
\\
\hline\hline
\end{tabular}
\end{small}
\caption{
Total \ggH* cross section at $\sqrts = 13 \TeV$ and $m_H = 125\GeV$.
All results include the exact $m_t$ dependence $\delta\sigma^t$ at NLO.
The percent uncertainties for the resummed results correspond to the total uncertainty
$\DeltaFO \oplus \Delta_\varphi$.
}
\label{table:ggH:Inclusive}
\end{table}

Comparing the last two columns of \tab{ggH:Inclusive},
the resummed results using the two different ways to include the top-quark contributions
are perfectly compatible with each other.
The fixed-order uncertainty is essentially unaffected, because it is
insensitive to the precise split of the constant terms into $H$ and $R$
due to the reexpansion of their fixed-order contributions [see \eq{sigmares}].
The resummation uncertainty $\Delta_\varphi$ increases somewhat in the one-step matching,
which reflects the fact that the $C_t$ contributions introduce an additional
residual $\mu$ dependence and that they are evaluated at $\mu = -\img m_H$ rather than
their natural scale $\mu = m_t$.
Overall, the numerical differences are however completely insignificant,
which shows that the results are insensitive to the
precise treatment of the top contributions. This also provides nontrivial verification
that the scheme dependence in how the nonlogarithmic constant terms are split between $H$ and $R$
at each order is much smaller than the perturbative uncertainties and hence irrelevant.

A complete numerical inclusion of all known corrections beyond the rEFT limit is beyond the scope
of this paper. The inclusion of $b$-quark and electroweak effects can proceed
completely analogously to the treatment of the top contributions. Any multiplicative
contributions can be trivially included, while additive corrections such as the
NLO $m_b$-dependent terms can be treated analogously to the finite-$m_t$ corrections.
For example, the dominant known electroweak corrections can be included by replacing~\cite{Anastasiou:2008tj}
\begin{equation}
 C_t \to C_t + \delta_\mathrm{EW} ( 1 + C_{1w} \as + \cdots)
\,,\end{equation}
where $\delta_\text{EW}$ is the pure NLO electroweak correction to the LO cross section~\cite{Aglietti:2004nj, Actis:2008ug}
and $C_{1w}$ contains the mixed $\ord{\alpha \as}$ correction calculated in \refcite{Anastasiou:2008tj}
by integrating out $W$- and $Z$-bosons as an estimate of the full $\ord{\alpha \alpha_s}$
corrections.
These additional contributions will not affect the  benefit of the resummation, in the same
way the inclusion of the top corrections for $gg\to H$ did not affect the conclusions compared
to the generic scalar $gg\to X$ case.

\subsection{Higgs rapidity spectrum}
\label{sec:ggHRapidity}

As discussed in \sec{ResummationScheme}, the resummed form factor can be incorporated
in the same way as for the total production cross section into generic cross sections
that are differential in or contain cuts on the Born kinematics.
Here we consider the primary example of the rapidity spectrum as well as the cross
section with a rapidity cut.
For simplicity we do not consider additional fiducial cuts on the Higgs decay products
here, but stress again that these are straightforward to include.

\begin{figure*}
\centering
\includegraphics[width=\WidthTwoSubfigs]{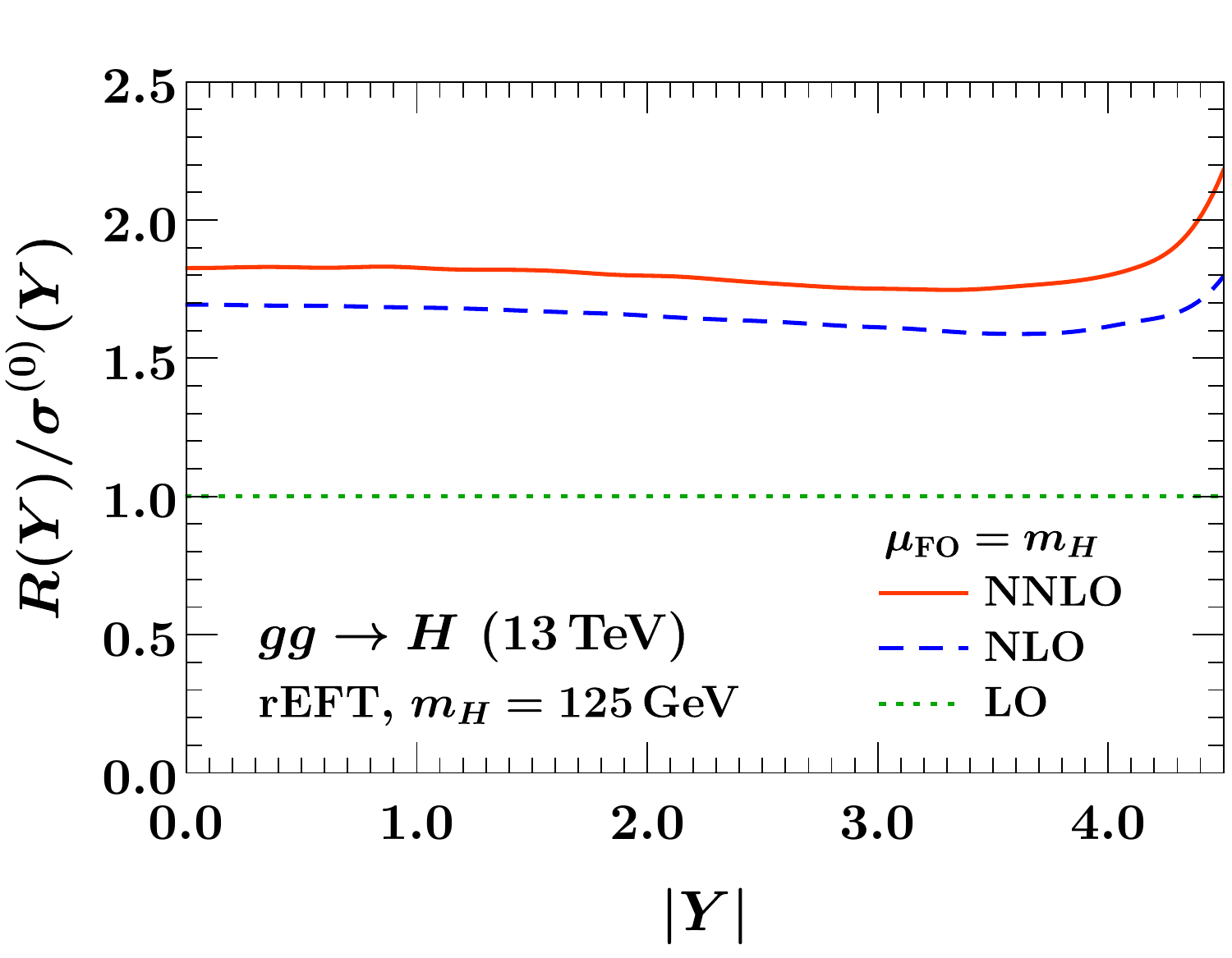}%
\hfill%
\includegraphics[width=\WidthTwoSubfigs]{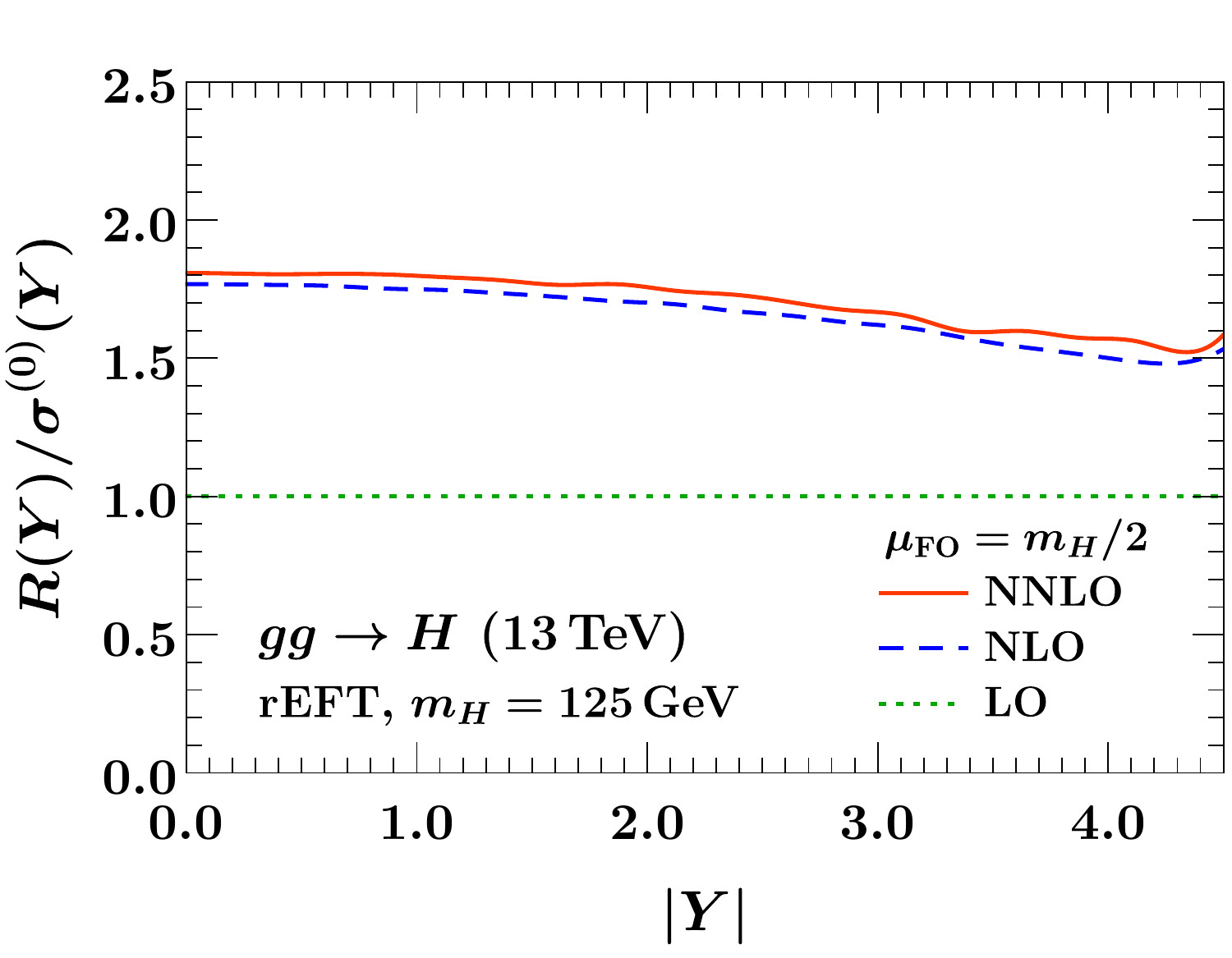}%
\caption{
The perturbative remainder $R(Y)/\sigma^{(0)}(Y)$ as a function of the Higgs rapidity $Y$
normalized to the LO spectrum $\sigma^{(0)}(Y) \equiv \df \sigma^{(0)} / \df \rap$ in the rEFT limit
for $\mu_\FO = m_H$ (left) and $\mu_\FO = m_H/2$ (right).}
\label{fig:ggH:Spectrum:Remainder}
\end{figure*}

\begin{figure*}
\centering
\includegraphics[width=\WidthTwoSubfigs]{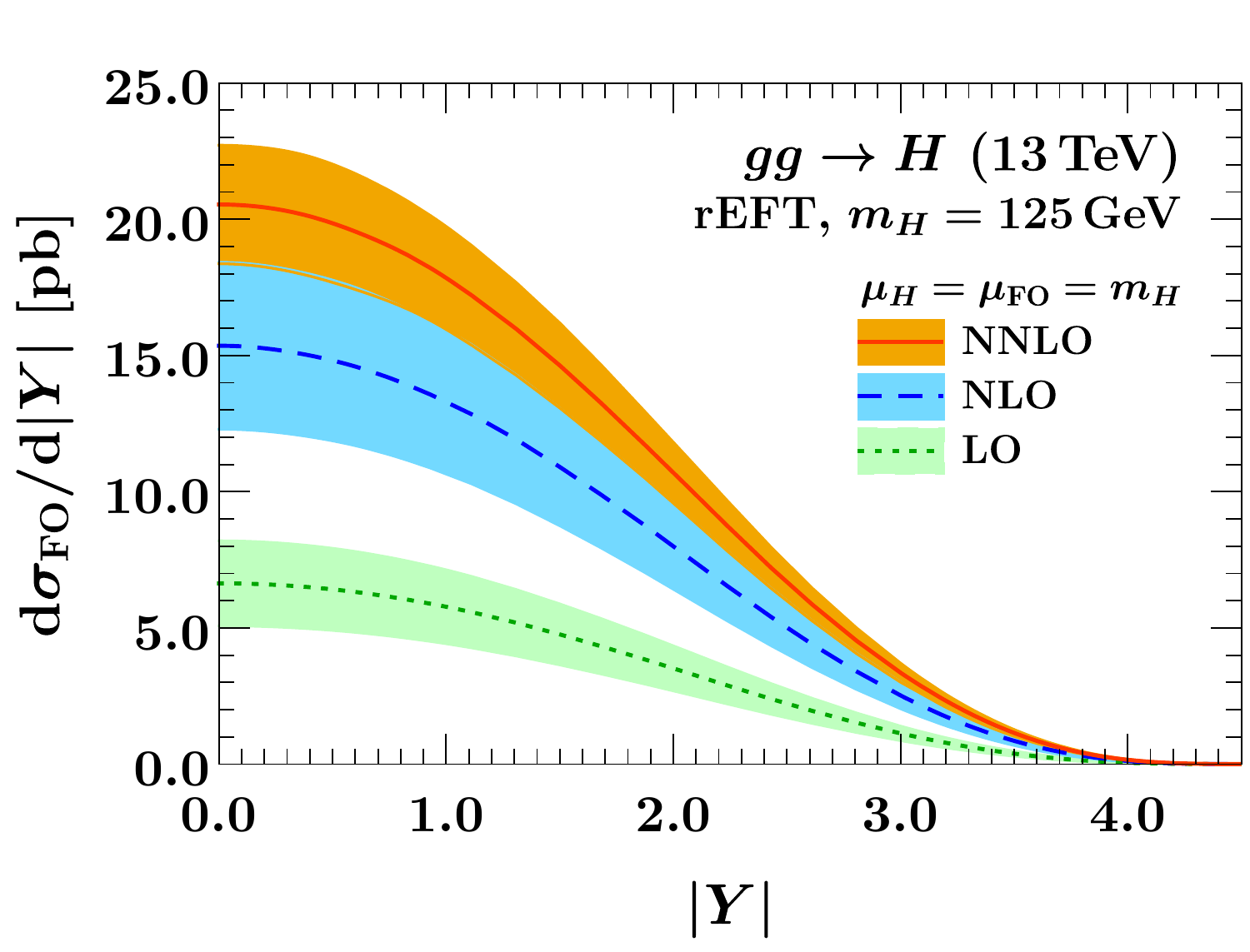}%
\hfill%
\includegraphics[width=\WidthTwoSubfigs]{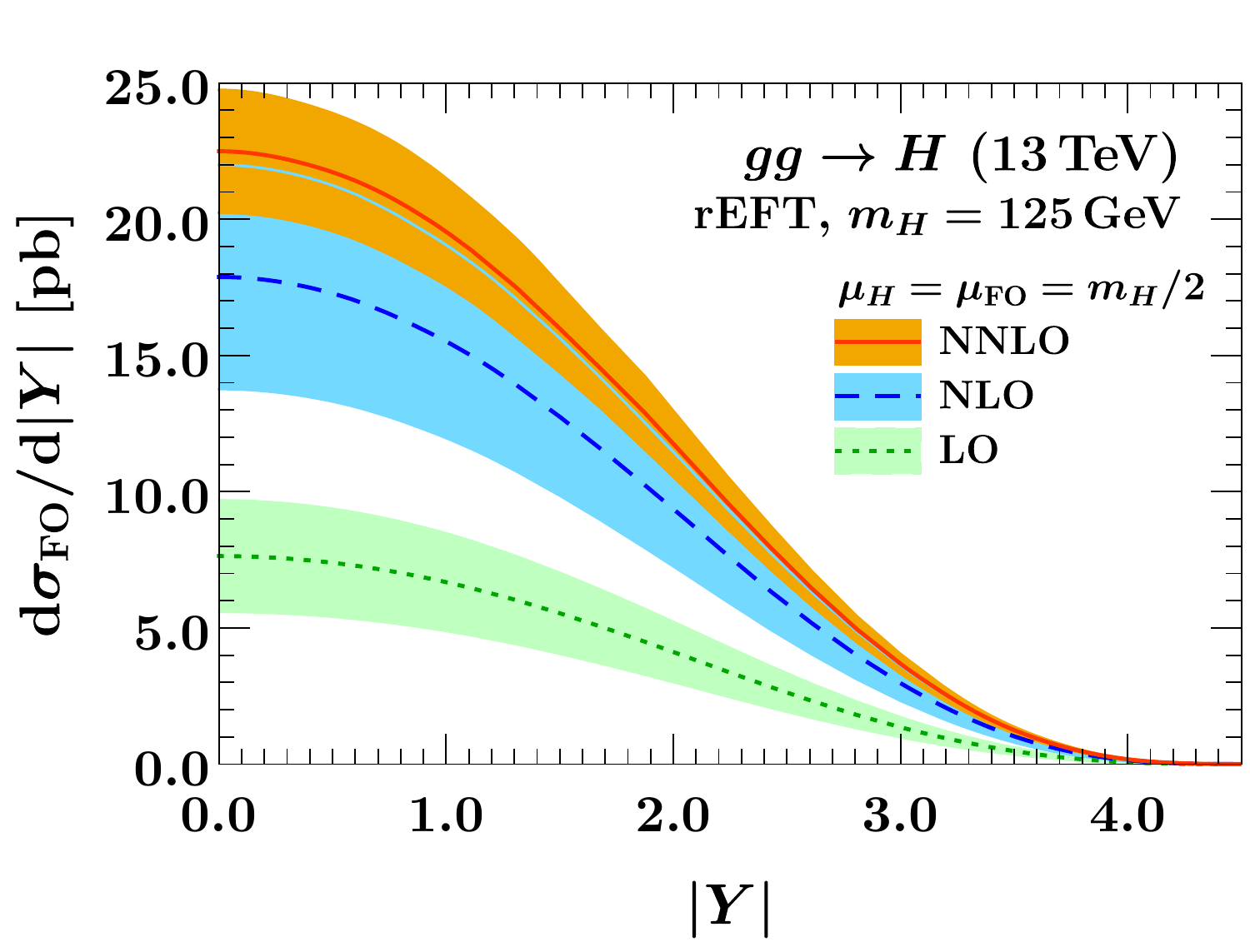}%
\\
\includegraphics[width=\WidthTwoSubfigs]{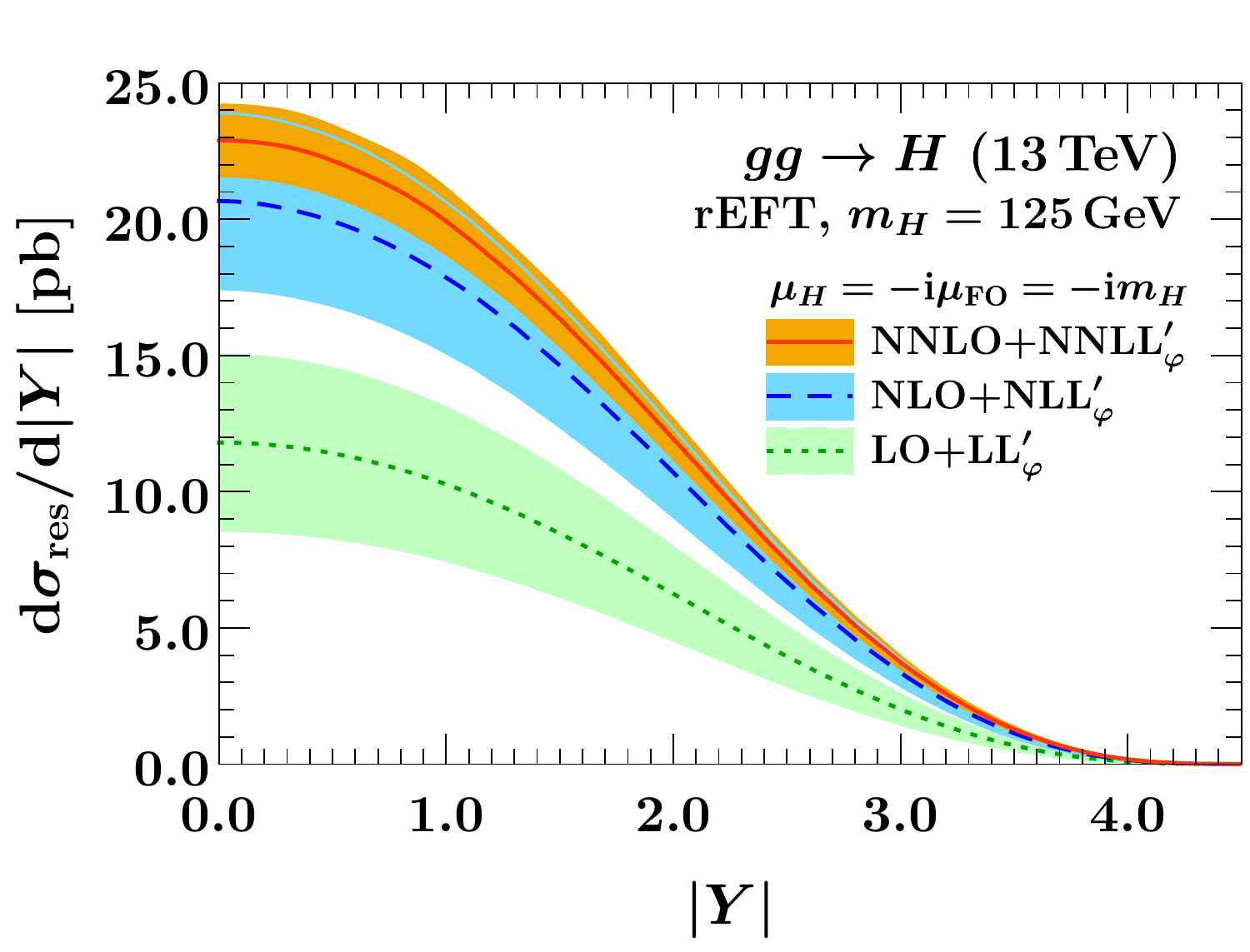}%
\caption{
The rapidity distribution for \ggH* at $\sqrts = 13 \TeV$ and $m_H = 125\GeV$ in the rEFT limit.
The fixed-order results are shown in the top row at $\mu_\FO = m_H$ (left) and
$\mu_\FO = m_H/2$ (right). The resummed result is shown on the bottom.
The uncertainty bands indicate $\DeltaFO$ and $\DeltaFO \oplus \Delta_\varphi$, respectively.
}
\label{fig:ggH:Spectrum}
\end{figure*}

The rapidity spectrum for gluon-fusion Higgs production is known to \nnlo*
\cite{Anastasiou:2004xq, Anastasiou:2005qj, Anastasiou:2007mz, Catani:2007vq, Grazzini:2008tf},
while the \nnnlo* corrections are available in the threshold limit~\cite{Ravindran:2006bu,Ahmed:2014uya}.
The resummation in the small-$x$ limit is also known~\cite{Caola:2010kv}.
We obtain the fixed-order bin-integrated rapidity distribution for $gg\to H$ to
\nnlo* with \texttt{HNNLO 2.0}~\cite{Catani:2007vq, Grazzini:2008tf}.
We use a binsize of $\Delta \rap = 0.25$ and for clarity in all plots interpolate
the binned results.

We first consider the rEFT limit and exclude additional quark mass effects.
In \fig{ggH:Spectrum:Remainder}, we display the perturbative remainder $R(\rap)$
as a function of $Y$. Although it has some intrinsic nontrivial rapidity dependence, the overall
behavior is as for the total cross section, namely it exhibits a noticeably
better convergence than the full fixed-order spectrum. Hence, we expect a similar
improvement from applying the resummation to the rapidity spectrum as for the
total cross section.

The upper panels of \fig{ggH:Spectrum} show the fixed-order results at $\mufo = m_H$
and $\mufo = m_H/2$, with the bands showing $\DeltaFO$.
The overall $K$ factor at \nlo* and \nnlo* is roughly constant in the
central rapidity range and similar to that of the total cross section. This is
consistent with the fact that a large part of the $K$ factor stems from
the timelike logarithms in the gluon form factor, which is independent of the rapidity.

The resummed result including fixed-order and resummation uncertainties,
$\DeltaFO \oplus \Delta_\varphi$, is shown in the bottom panel of \fig{ggH:Spectrum}.
Clearly, resumming the timelike logarithms improves the perturbative convergence
across the spectrum as it did for the total cross section.
The \nnllmatched* result has perturbative uncertainties that are almost a
factor of two smaller than at NNLO. At the same time, the
\nnllmatched* result is well covered by the lower-order \nllmatched* uncertainty band,
which is not the case at fixed order. Judging from the results for the total cross
section, for which the full N$^3$LO is known, we expect the precision of the
\nnllmatched* to be roughly comparable to what can be expected for the full \nnnlo* result,
and we can be confident that the corresponding \nnnllmatched* result will
have further reduced uncertainties and will be well contained within the current
\nnllmatched* uncertainties.

Next, we consider the cross section with a rapidity cut,
\begin{equation}
\sigma(\rap_\mathrm{cut})
= \int_0^{\rap_\mathrm{cut}} \! \df \vert \rap \vert \, \frac{\df \sigma}{\df \vert \rap \vert}
\,.\end{equation}
We now also include the exact $m_t$-dependence at NLO using \texttt{HNNLO 2.0}~\cite{Grazzini:2013mca},
from which we extract the rapidity-dependent analogue of $\delta\sigma_\nlo*^{t}$ in \eq{def delta sigma}.
The latter is included in the resummed results as discussed in the previous subsection:
It is either added to the resummed result based on $H_{gg}$ giving $\df (\sigma_\res + \delta\sigma^t ) / \df \rap$.
Alternatively, we can apply the resummation to the full $m_t$-exact spectrum
$\df (\sigma^\rEFT + \delta\sigma^t )_\res / \df \rap$ using $H_{gg}^t$.
The obtained differential rapidity spectra look extremely similar to those in \fig{ggH:Spectrum},
since the finite-$m_t$ correction yields a small negative shift $-0.2\pb < \df (\delta \sigma^t) / \df \rap < 0$ throughout.

\begin{table}
\centering
\begin{small}
\begin{tabular}{l|ll|ll}
\hline\hline
\multicolumn{5}{c}{$\sigma(\rap_\mathrm{cut} = 2.5)\,[\!\pb]$ for $gg\to H$, $\sqrts = 13 \TeV$, $m_H = 125\GeV$}
\\
\hline\hline
& \multicolumn{2}{c|}{$\sigma^\rEFT_\FO \!+ \delta\sigma^t_\nlo*$}
& \multicolumn{1}{c}{$\sigma^\rEFT_\res \!+ \delta\sigma^t_\nlo*$}
& \multicolumn{1}{c}{$(\sigma^\rEFT + \delta\sigma^t_\nlo*)_\res$}
\\
$n$
& \multicolumn{1}{c}{$\nNlo* ,\,\mufo = m_H$} & \multicolumn{1}{c|}{$\nNlo* ,\,\mufo = \frac{m_H}{2}$}
& \multicolumn{1}{c}{$\nNllmatched*\,\,(H_{gg})$} & \multicolumn{1}{c}{$\nNllmatched*\,\, (H^t_{gg})$}
\\
\hline
$0$
&$12.5\!\pm\! 2.9_\mu\,(23\%)$
&$14.5\!\pm\! 3.9_\mu\,(27\%)$
&$22.2\!\pm\! 5.2_\mu\!\pm\! 3.2_\varphi\,(27\%)$
&$21.1\!\pm\! 4.6_\mu\!\pm\! 3.1_\varphi\,(26\%)$
\\
$1$
&$28.5\!\pm\! 5.6_\mu\,(20\%)$
&$33.3\!\pm\! 7.5_\mu\,(23\%)$
&$38.4\!\pm\! 5.4_\mu\!\pm\! 2.4_\varphi\,(15\%)$
&$38.0\!\pm\! 5.2_\mu\!\pm\! 2.6_\varphi\,(15\%)$
\\
$2$
&$38.3\!\pm\! 4.0_\mu\,(11\%)$
&$41.9\!\pm\! 4.2_\mu\,(10\%)$
&$42.8\!\pm\! 2.3_\mu\!\pm\! 0.9_\varphi\,(5.8\%)$
&$42.9\!\pm\! 2.4_\mu\!\pm\! 0.9_\varphi\,(6.0\%)$
\\
\hline
$3$
& $\approx 42.0$
& $\approx 43.2$
& $\approx 43.2$
& $\approx 43.3$
\\
\hline\hline
\end{tabular}
\end{small}
\caption{
Cross section for \ggH* with a rapidity cut of $\rap_\mathrm{cut} = 2.5$.
All results include the exact $m_t$ dependence $\delta\sigma^t(\rap_\mathrm{cut})$ at \nlo*.
The percent uncertainties for the resummed results correspond to the total uncertainty
$\DeltaFO \oplus \Delta_\varphi$.
The approximate $n=3$ results are constructed as explained in the text and are given
for illustration only.}
\label{table:ggH:RapidityCumulant}
\end{table}

In \tab{ggH:RapidityCumulant}, we provide benchmark results for $\rap_\mathrm{cut} = 2.5$ at fixed order
and including resummation. As expected, the resummed results show the same
improved convergence and perturbative uncertainties as the rapidity spectrum and the
total cross section.

Since the full \nnnlo* rapidity spectrum is not yet known, one might think about approximating it by
rescaling the NNLO spectrum to the inclusive \nnnlo* cross section, e.g., by taking
$\sigma_\nnnlo*(\rap_\mathrm{cut})
\approx (\sigma_\nnnlo*/\sigma_\nnlo*) \times \sigma_\nnlo*(\rap_\mathrm{cut})$.
Although this likely improves the central value by moving it closer to the correct result,
it is unclear what uncertainties one should assign.
In fact, the resummed \nnllmatched* result for the rapidity spectrum provides
a clean way to essentially achieve this goal,
because it includes the dominant part of the inclusive \nnnlo* $K$ factor, and it does so
with the correct rapidity dependence and reliable uncertainties, which are already
substantially reduced compared to NNLO.

To illustrate this, first note that the entire rapidity dependence is contained in the
remainder $R$. Looking at \fig{ggH:Spectrum:Remainder}, one might assume that the \nnnlo*
contribution is roughly flat in rapidity such that taking
\begin{equation} \label{eq:R3Yflatapprox}
R^{(3)}(\rap)
\approx \frac{R^{(3)}}{\sigma^{(0)}}\, \sigma^{(0)}(\rap)
\,,\qquad
R^{(3)}(\rap_\mathrm{cut})
\approx \frac{R^{(3)}}{\sigma^{(0)}}\, \sigma^{(0)}(\rap_\mathrm{cut})
\,,\end{equation}
provides a reasonable approximation,
where $R^{(3)}/\sigma^{(0)}$ is the \nnnlo* correction for the total cross section.
Using \eq{R3Yflatapprox} as input and combining it with the known $H^{(3)}$ and lower-order results
we obtain the approximate numbers at \nnnlo* and \nnnllmatched* shown in the last line of
\tab{ggH:RapidityCumulant}. As expected, the \nnllmatched* results indeed capture
most of the expected shift from the NNLO result. Using the same procedure to evaluate
the scale variations, we can project that the relative uncertainties will be similar
to the inclusive cross section, namely around 4\% at \nnnlo* and around 2\% at \nnnllmatched*.
However, we caution again that it is debatable whether it is justified
to assign these without knowing the exact result.

\section{Quark annihilation}
\label{sec:QuarkAnnihilation}

We now turn to $q\bar q$ annihilation processes, for which the
perturbative corrections are typically much smaller than for gluon fusion.
For these, timelike Sudakov logarithms still arise in the corresponding quark
form factor at timelike momentum transfer and can be resummed to all orders.
We apply the resummation to Higgs production through bottom quark annihilation
in \sec{bbH} and to the Drell-Yan process in \sec{DY}.

\subsection{Higgs production through bottom-quark annihilation}
\label{sec:bbH}

The cross section for Higgs production through bottom-quark annihilation, $b\bar b\to H$,
is much smaller than for $gg\to H$, but is important phenomenologically as it provides
direct access to the bottom-quark Yukawa coupling and can be enhanced in theories
beyond the Standard Model.

The hard function for $b\bar b\to H$ corresponds to the quark scalar form factor
and is obtained from the SCET matching coefficient $C_{q\bar q}^S(Q^2, \mu)$ for
the scalar current with two massless quarks, see \eq{currents}. The scalar form
factor naturally arises in the five-flavor scheme calculation in which the $b$ quark
is treated as a massless quark at the hard matching scale, except for its Yukawa
coupling $y(\mu)$ to the Higgs, which always has its physical value corresponding to
$\overline{m}_b(\mu)$. We exclude the Yukawa coupling from the hard matching coefficient,
just as we excluded the overall $\vert \as C_X \vert^2$ for gluon fusion.

To the best of our knowledge, $C_{q\bar q}^S$ is not yet directly available in the literature.
We extract it in \app{Cqq} to \nnnlo* from the three-loop massless quark scalar form factor calculated in
\refcite{Gehrmann:2014vha} (see also \refscite{Harlander:2003ai, Ravindran:2006cg, Anastasiou:2011qx}
for the \nnlo* form factor, and \refcite{Bernreuther:2005gw} for the \nnlo* form factor
including the full mass dependence). The RGE of $C_{q\bar q}^S$ is given by
\begin{align} \label{eq:RGECqqS}
\mu \frac{\df}{\df \mu} C_{q\bar q}^S (m_H^2, \mu)
&= \gamma_{q\bar q}^{S}(m_H^2, \mu) \, C_{q\bar q}^S (m_H^2, \mu)
\,, \nn \\
\gamma_{q\bar q}^{S}(m_H^2, \mu)
&= \Gamma^q_\mathrm{cusp}[\as(\mu)] \ln \frac{- m_H^2-\img 0}{\mu^2}
   + 2 \gamma_C^q [\as(\mu)]  - \gamma_m[\as(\mu)]
\,. \end{align}
where $\Gamma^q_\mathrm{cusp}$ is the quark cusp anomalous dimension, and the last two terms are
the total noncusp contribution, where $2\gamma_C^q$ is the usual hard quark noncusp anomalous dimension
(also appearing for the vector current) and $\gamma_m$ is the mass anomalous dimension of the
Yukawa coupling.

We obtain the $b\bar{b} \to H$ total cross section to NNLO in the five-flavor
scheme~\cite{Barnett:1987jw, Dicus:1988cx} from \texttt{SusHi 1.6.0}~\cite{Chetyrkin:2000yt,
Harlander:2002wh, Harlander:2003ai, Harlander:2012pb, Harlander:2016hcx}. The
Yukawa coupling is evaluated at $\mufo$ and obtained by evolving from
$\overline{m}_b(\overline{m}_b) = 4.18\GeV$ using three-loop running.
We use the reevolved five-flavor \texttt{PDF4LHC} PDF sets from \refscite{Bonvini:2015pxa, Bonvini:2016fgf},
which use a $b$-quark pole mass for the $b$ PDF that is consistent with our choice of
$\overline{m}_b(\overline{m}_b)$ (namely the one-loop pole mass $m_b = 4.58\GeV$).
These PDF sets also distinguish the $b$-quark matching scale $\mu_b$
from the physical mass parameter $m_b$, allowing to estimate the perturbative
uncertainty $\Delta_b$ associated with the low-scale matching onto the $b$-quark PDFs
by varying $\mu_b$ (see \refcite{Bonvini:2016fgf} for details).
For the central value we use $\mu_b = m_b$.

\begin{figure*}
\includegraphics[width=\WidthTwoSubfigs]{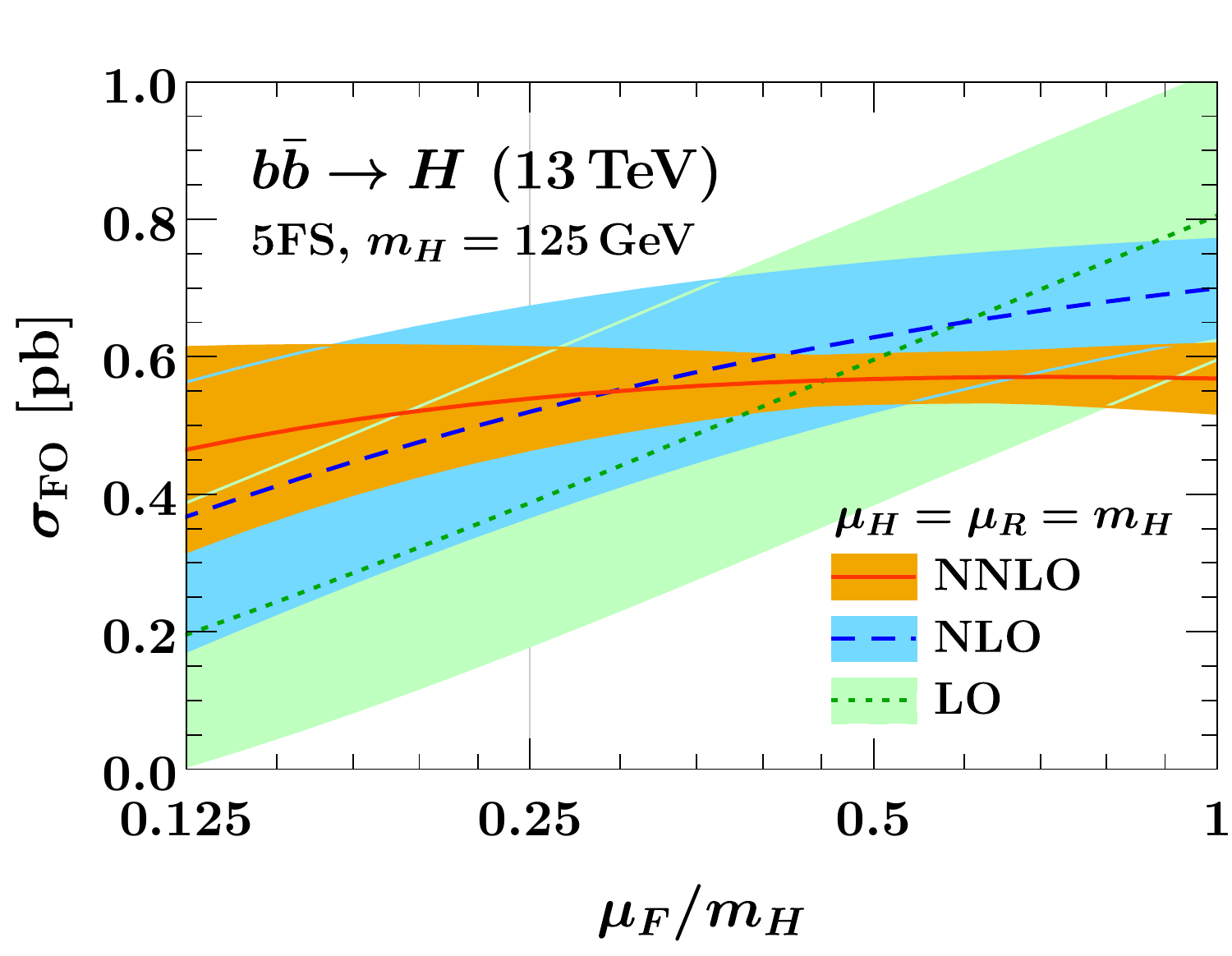}%
\hfill%
\includegraphics[width=\WidthTwoSubfigs]{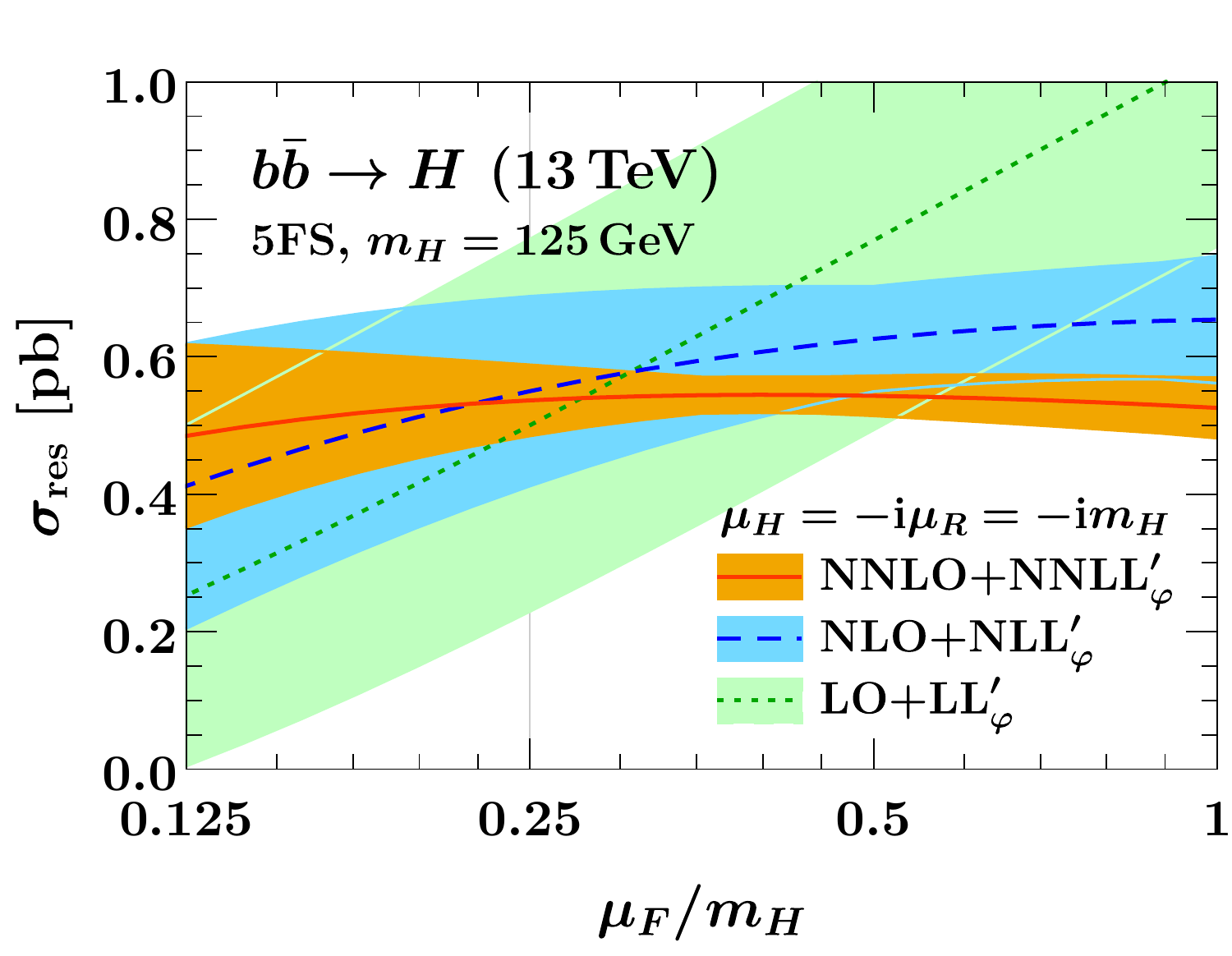}%
\caption{
The total $b \bar b \to H$ cross section at $\mufo = \mu_R = m_H$ as a function
of the central choice for $\kappa_F = \mu_F/m_H$.
The fixed-order results are shown on the left and the resummed results on the right.
The uncertainty bands show $\DeltaFO$ (fixed order) and $\DeltaFO \oplus \Delta_\varphi$ (resummed).
}
\label{fig:bbH:muF}
\end{figure*}

In contrast to the gluon-initiated case, here the $\mu_F$ dependence of the cross section
plays the dominant role, while the $\mu_R$ dependence is much less important.
For our central scales we use $\mufo = m_H$ and $\kappa_F = 1/4$ corresponding to
$(\mu_R, \mu_F) = (m_H, m_H/4)$, as typically adopted in the five-flavor scheme
calculation. This low central value for $\mu_F$ is motivated by the observation
that it roughly minimizes the collinear logarithms in the partonic cross section
and leads to more stable predictions, see e.g.\ \refscite{Plehn:2002vy,
Harlander:2003ai, Maltoni:2003pn, Maltoni:2005wd, Maltoni:2012pa, Bonvini:2015pxa, Harlander:2015xur}.
This is also seen in the left panel of \fig{bbH:muF}, which shows the fixed-order results
as a function of the central value used for $\kappa_F$ in the range $\kappa_F = \mu_F/m_H \in [1/8, 1]$
(always using $\mufo = \mu_R = m_H$ for the central value). The bands show the fixed-order
uncertainty $\Delta_\mu$, which is obtained from the $\mufo$
and $\kappa_F$ variations as discussed in \sec{Uncertainties}.

The perturbative series of the fixed-order cross section and its separation into $H_{q\bar q}^S$ and $R$
are given by
\begin{align} \label{eq:bbHxsecFO}
\sigma_\FO(\mu_R = m_H, \mu_F = m_H/4) &= (1 + 0.342 + 0.050) \times 0.387 \pb 
\,, \nn \\
H_{q\bar q}^S(m_H^2, \mu_H=m_H) &= \,\, 1 + 0.227 + 0.054
\,, \nn \\
R(\mu_R = m_H, \mu_F = m_H/4) &= (1 + 0.115 - 0.031) \times 0.387 \pb 
\,.\end{align}
Although the perturbative corrections to the cross section are more moderate
than for $gg\to H$, they are still clearly driven by the hard function,
while the corrections to the remainder at $\mu_F = m_H/4$ are relatively small.
The hard function itself shows a markedly improved convergence
when evaluated at $\mu_H = -\img m_H$. Up to \nnnlo*, we have
\begin{alignat}{3} \label{eq:BottomHardNumbers}
&H_{q\bar q}^S(m_H^2, \mu_H=m_H)          &&= 1 + 0.22742 + 0.05447 + 0.00507
\,, \nn \\
&H_{q\bar q}^S(m_H^2, \mu_H=-\img m_H)    &&= 1 - 0.00807 + 0.01202 + 0.00237
\,,\end{alignat}
showing that the corrections at real scales $\mu_H = m_H$ are mostly due to the
timelike logarithms, which are eliminated at $\mu_H = -\img m_H$.

The resummed results are shown in the right panel of \fig{bbH:muF}, again
as a function of the central choice for $\kappa_F$.
From the above observations, we expect the resummed results to have an improved
convergence at $\mu_F = m_H/4$. This is clearly seen, as the resummed predictions
at the different orders roughly intersect at $\mu_F = m_H/4$,
with the corrections beyond \llmatched* amounting to less than $10\%$.
The explicit numerical
results at each order including a breakdown of the perturbative uncertainties
are collected in \tab{bbH}.

Note that the remainder $R$ carries the full $\mu_F$ dependence of the cross section.
Evaluating it at $\kappa_F = 1$, corresponding to $\mu_F = m_H$,
its corrections are substantially larger than at $\mu_F = m_H/4$,
\begin{equation} \label{eq:bbHRemainder}
R(\mu_R = \mu_F = m_H) = \sigma^{(0)}(\mu_R = \mu_F = m_H) \times(1 - 0.359 - 0.136)
\,.\end{equation}
We also checked that in the considered range of $\mu_F$
the perturbative coefficients of the remainder are indeed minimized around $\mu_F = m_H/4$.
This explains why the resummed results in \fig{bbH:muF} do not improve
the fixed-order results above $\kappa_F = 1/4$, since there the cross section
becomes dominated by large negative corrections to the remainder and is also affected
by accidental numerical cancellations between $H$ and $R$.
Hence, the central choice $\kappa_F = 1/4$ is also quite optimal from this point of view,
as it leads to a clear ``division of labor'' between the remainder
and the hard function in capturing the perturbative corrections to \bbH*:
By evaluating the former at an appropriate $\mu_F$ the collinear logarithms arising from
initial-state $g\to b\bar b$ splittings are resummed into the $b$-quark PDFs,
while the latter captures the hard virtual contributions to the $b\bar{b}H$ vertex,
the bulk of which are enhanced at timelike kinematics and are resummed by setting $\mu_H = -\img m_H$.

The resummation also reduces the fixed-order uncertainty $\Delta_\mu$ by reducing the
$\mufo$ dependence and by eliminating large cross terms, which can also
reduce the impact of the large $\mu_F$ dependence from lower-order contributions.
This is seen in the slightly reduced $\mu_F$ dependence in \fig{bbH:muF}
at \nllmatched* and \nnllmatched* compared to their fixed-order counterparts.
However, the reduction of $\Delta_\mu$ is not nearly as dramatic as for $gg\to H$,
because the $\mu_F$ dependence plays a much bigger role here.
(In principle, the $\mu_F$ dependence and resulting uncertainties can be reduced
further by reorganizing the perturbative series as discussed in \refcite{Bonvini:2015pxa},
which is however beyond our scope here.)
In \tab{bbH} we also include the $\Delta_b$ uncertainty from the low-scale $b$-quark PDF.
(The parametric uncertainty in $m_b$ is much smaller than $\Delta_b$ and not considered.)
The resummation uncertainty $\Delta_\varphi$ is completely negligible compared to
$\Delta_\mu$, thanks to the very stable resummed hard function.

\begin{table}
\centering
\begin{tabular}{l|l|l}
\hline\hline
\multicolumn{3}{c}{$\sigma\,[\!\pb]$ for $b\bar b\to H$, $\sqrts = 13 \TeV$, $m_H = 125\GeV$}
\\
\hline\hline
$n$ & \multicolumn{1}{c|}{$\sigma_\FO$ at \nNlo*} & \multicolumn{1}{c}{$\sigma_\res$ at \nNllmatched*}
\\
\hline
$0$
&$0.387\!\pm\! 0.208_\mu\!\pm\! 0.020_b\,(54\%)$
&$0.500\!\pm\! 0.269_\mu\!\pm\! 0.026_b\!\pm\! 0.033_\varphi\,(54\%)$
\\
$1$
&$0.520\!\pm\! 0.153_\mu\!\pm\! 0.027_b\,(30\%)$
&$0.550\!\pm\! 0.138_\mu\!\pm\! 0.028_b\!\pm\! 0.006_\varphi\,(26\%)$
\\
$2$
&$0.539\!\pm\! 0.074_\mu\!\pm\! 0.028_b\,(15\%)$
&$0.537\!\pm\! 0.052_\mu\!\pm\! 0.028_b\!\pm\! 0.002_\varphi\,(11\%)$
\\
\hline\hline
\end{tabular}
\caption{
Total \bbH* cross section at $m_H = 125\GeV$ at the LHC with $\sqrts = 13\TeV$.
The central scales are $\mufo = m_H$, $\kappa_F = 1/4$, $\mu_b = m_b$.
The percent uncertainties correspond to the quadratic sum of all uncertainties.
}
\label{table:bbH}
\end{table}

Overall, we find that the \nnlo* and \nnllmatched* results are very similar, which
is reassuring, and that the resummation of timelike logarithms provides a useful tool
to accelerate the convergence of the \bbH* cross section and to reduce
its perturbative uncertainties.

\subsection{Drell-Yan rapidity spectrum}
\label{sec:DY}
As final example we consider the rapidity spectrum of the Drell-Yan process, $pp \rightarrow Z/\gamma^\ast \rightarrow \ell^+ \ell^-$, which is known to \nnlo* \cite{Anastasiou:2003yy,Anastasiou:2003ds,Melnikov:2006di,Melnikov:2006kv,Catani:2009sm}, and at \nnnlo* in the threshold limit \cite{Ravindran:2006bu,Ravindran:2007sv,Ahmed:2014uya}.
(For the inclusion of threshold resummation effects see e.g.\ \refscite{Sterman:1986aj,Catani:1989ne,Forte:2002ni,Idilbi:2005ky,Mukherjee:2006uu,Bolzoni:2006ky,Becher:2007ty}.)
The necessary quark vector form factor is known to three loops~\cite{Kramer:1986sg, Matsuura:1987wt, Matsuura:1988sm, Gehrmann:2005pd, Moch:2005tm, Moch:2005id, Baikov:2009bg, Lee:2010cga, Gehrmann:2010ue}.
The corresponding hard matching coefficient $C_{q\bar q}^V$ for the quark vector current to \nnnlo* was obtained in \refscite{Abbate:2010xh,Gehrmann:2010ue}.
We also need the matching coefficient $C_{q\bar q}^A$ for the axial-vector current,
which is equal to the vector coefficient up to singlet corrections that start entering
at $\ord{\as^2}$~\cite{Stewart:2009yx}. At NNLO the axial-vector coefficient receives
a nonvanishing singlet contribution from the axial-vector anomaly due to the large bottom-top
mass splitting, but these contributions have been found to be small at cross section level~\cite{Dicus:1985wx, Hamberg:1990np}.
Since they are also not implemented in the program \texttt{Vrap 0.9}~\cite{Anastasiou:2003ds}, which we use for our fixed-order predictions, we set $C_{q\bar q}^A = C_{q\bar q}^V$, dropping any singlet terms, and take
$H_{q\bar q} \equiv \abs{C_{q\bar q}^V}^2$ for our analysis.
The RGE in either case is identical, since the axially anomalous terms are nonlogarithmic, and reads
\begin{align} 
   \mu \frac{\df}{\df \mu} C_{q\bar q}^{V,A} (Q^2, \mu)
   &= \gamma_{q\bar q}^{V,A}(Q, \mu) \, C_{q\bar q}^{V,A} (Q^2, \mu)
\,, \nn \\
   \gamma_{q\bar q}^{V,A}(Q^2, \mu)
   &= \Gamma^q_\mathrm{cusp} \bigl[ \as(\mu) \bigr] \ln \frac{- Q^2-\img 0}{\mu^2}
      + 2 \gamma_C^q \bigl[ \as(\mu) \bigr]
   \,. \label{eq:RGECqqVA}
\end{align}

We consider the double differential cross section $\df^2 \sigma/\df \rap\,\df Q$,
where $\rap$ and $Q$ are the rapidity and invariant mass of the produced lepton pair,
and for simplicity focus on the $Z$ pole, $Q = m_Z$.
Note that perturbative corrections to the quark current can only have a weak logarithmic dependence on $Q$, 
so our observations also hold away from the $Z$ pole.
We obtain the rapidity distribution from \texttt{Vrap 0.9}~\cite{Anastasiou:2003ds},
taking into account $V = Z, \gamma^\ast$ and their interference terms.

\begin{figure*}
\centering
\includegraphics[width=\WidthTwoSubfigs]{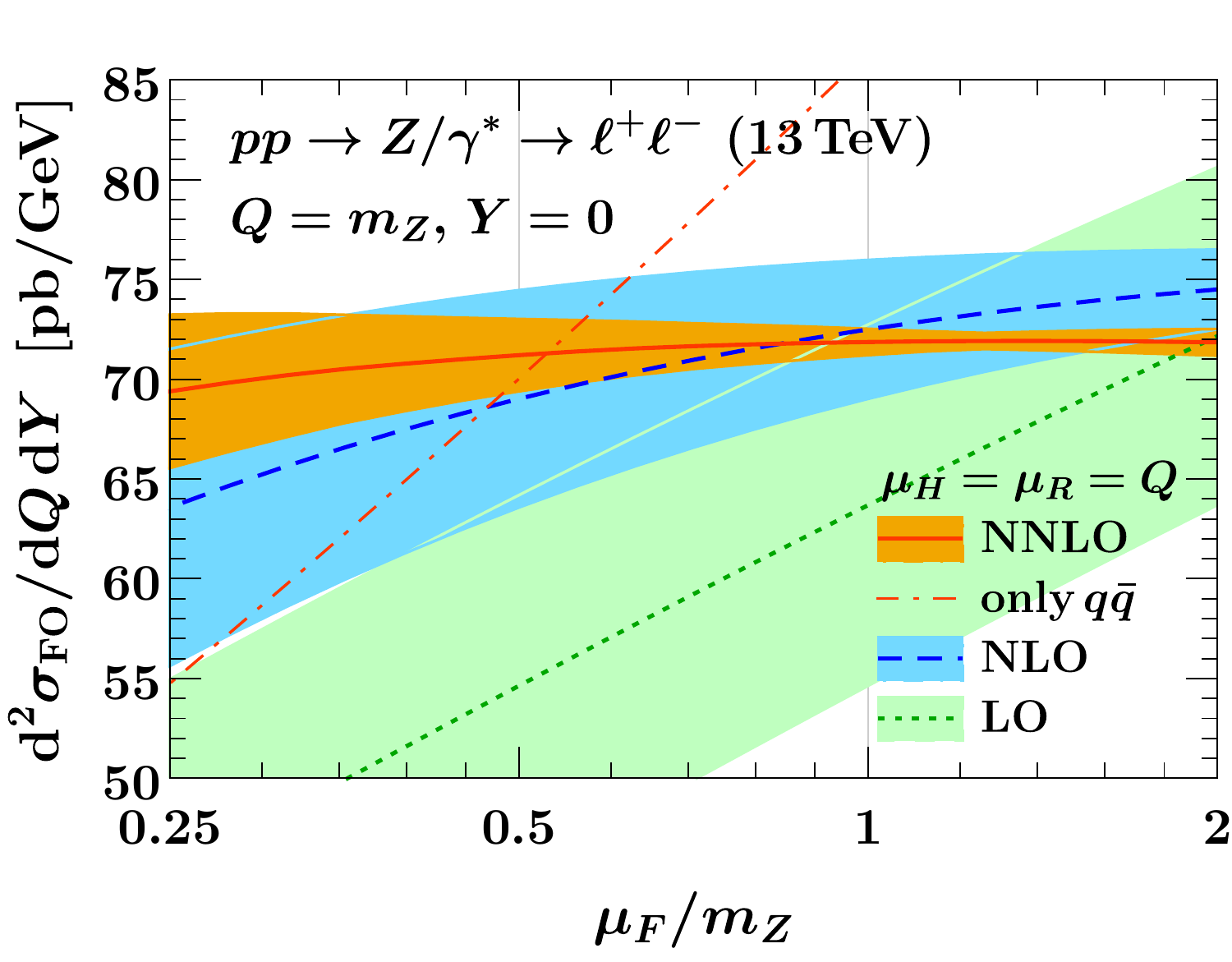}%
\hfill%
\includegraphics[width=\WidthTwoSubfigs]{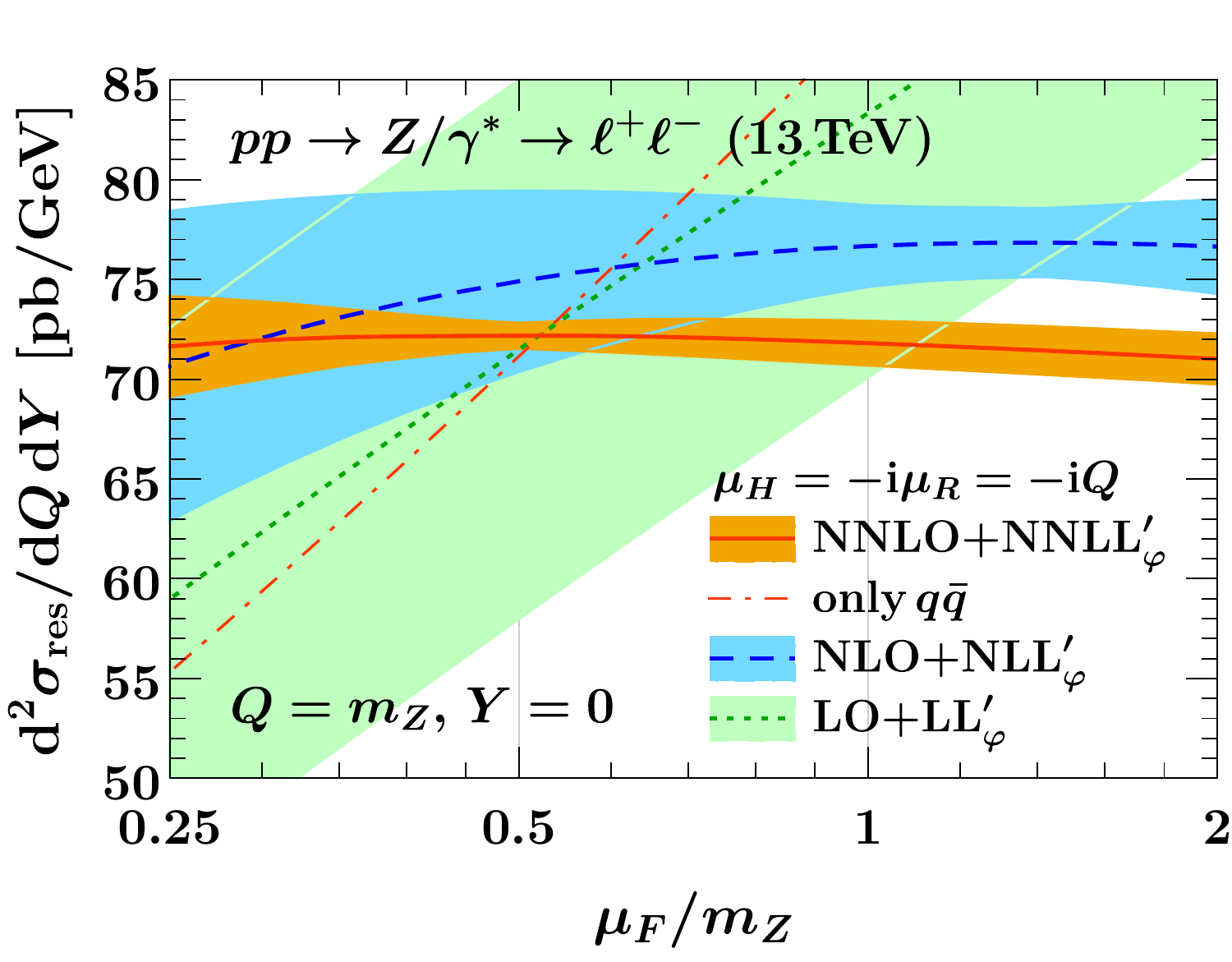}%
\\\vspace{2ex}
\includegraphics[width=\WidthTwoSubfigs]{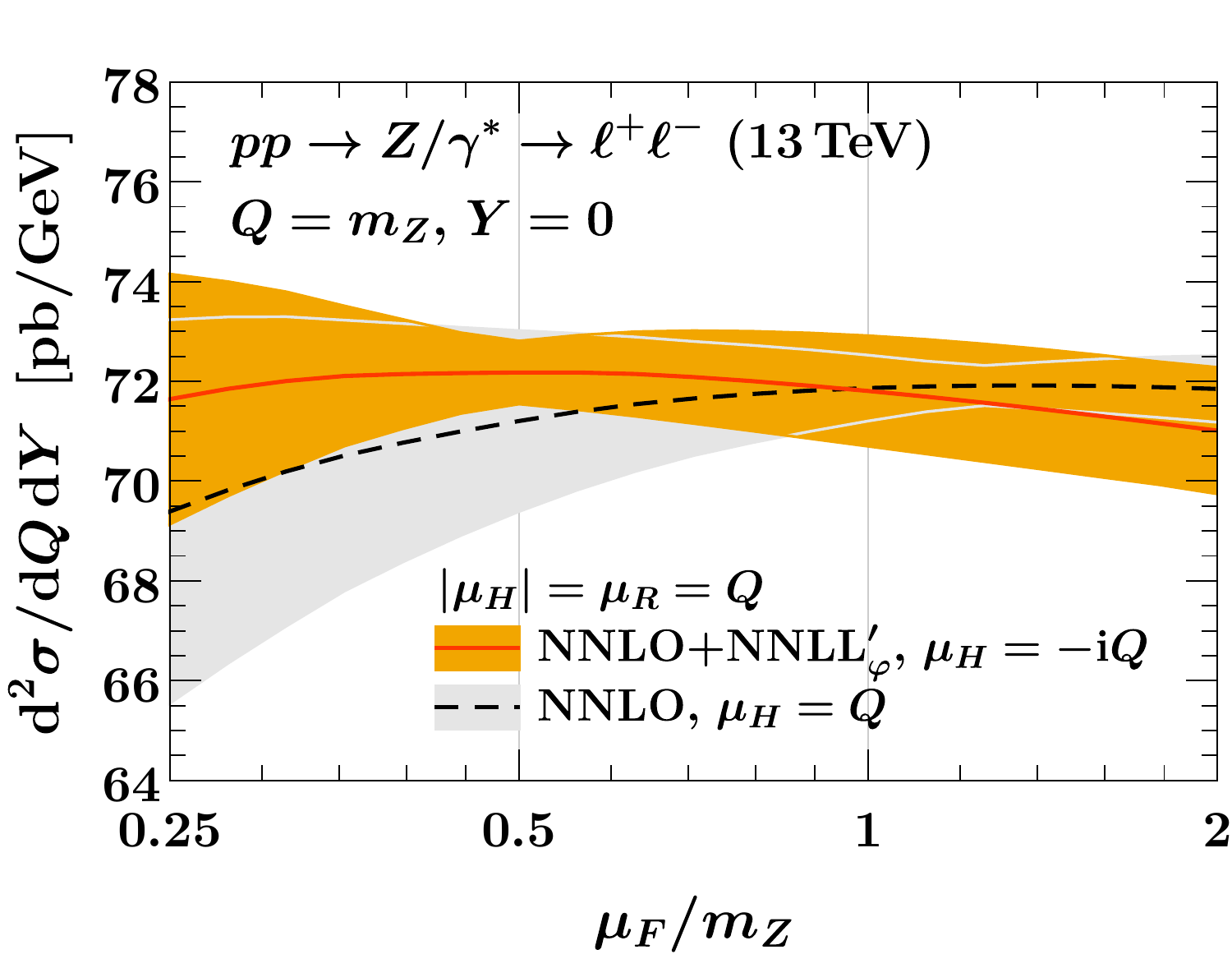}%
\caption{
Double-differential cross section $\df^2 \sigma / \df \rap\,\df Q$ for
$pp \to Z/\gamma^\ast \to \ell^+ \ell^-$ at $Q = m_Z$ and $\rap = 0$ and for
$\mufo = \mu_R = m_Z$
as a function of the central choice for $\kappa_F = \mu_F/m_Z$. The fixed-order results are
shown on the left and the resummed results on the right.
For illustration, the dot-dashed line shows the contribution from the $q\bar q$ channel at the highest order.
In the bottom panel, the highest-order results at \nnlo* and \nnllmatched* are directly compared
with a further zoomed-in $y$-axis.
The uncertainty bands show $\DeltaFO$ (fixed order) and $\DeltaFO \oplus \Delta_\varphi$ (resummed).}
\label{fig:DY muF dependence}
\end{figure*}

As for bottom-quark fusion, we can expect that for Drell-Yan
the choice of $\mu_F$ will be important to improve the predictions by minimizing
the perturbative corrections to the remainder.
We first consider the rapidity spectrum at fixed $Y = 0$.
(Alternatively, one could integrate over rapidity, which yields similar conclusions.)
Figure~\ref{fig:DY muF dependence} shows the dependence on the factorization scale $\mu_F$
at fixed order (top left panel) and including the resummation of timelike logarithms in the quark
form factor (top right panel), always using $\mufo = \mu_R = m_Z$.
In the bottom panel of \fig{DY muF dependence} we directly compare the \nnlo* (gray)
and \nnllmatched* (orange) results.
The fixed-order results show the best convergence around $\mu_F = m_Z$,
which is the typical choice adopted for Drell-Yan.
In contrast, the resummed results clearly favor lower values of $\mu_F$.
We therefore take $\mu_F = m_Z/2$ as our central choice, as in this region the lower-order
uncertainties provide the best coverage of the higher-order results, while
at the same time the $\mu_F$ dependence at the highest order is the most stable.
(In principle, one might even consider going as low as $\mu_F \approx m_Z/4$ close to
where the \nllmatched* and \nnllmatched* central values coincide. However, the $\mu_F$
dependence is much larger there, and we also prefer to stay within a factor of two of
the canonical value $\mu_F = m_Z$.)
The dot-dashed line in \fig{DY muF dependence} shows the results at the highest order
only including the $q\bar q$-initiated contributions.
At $\mu_F = m_Z/2$, the cross section is dominated by the $q\bar q$ contribution,
while at $\mu_F = m_Z$ the total NNLO corrections are small only due to substantial cancellations
between the $q\bar q$ and remaining channels. This supports our central choice of $\mu_F = m_Z/2$
in the resummed results as the quark form factor is associated primarily with the $q\bar q$ channel.
The explicit results at $\rap = 0$ are given in table \ref{table:DY:CentralRapidity},
including a breakdown of the uncertainties.

\begin{table}
\centering
\begin{small}
\renewcommand{\tabcolsep}{3ex}
\begin{tabular}{l|ll|l}
\hline\hline
\multicolumn{4}{c}{$\df^2 \sigma / \df \rap \, \df Q\,[\!\pb/\!\GeV]$ for $pp \to Z/\gamma^\ast \to \ell^+ \ell^-$,
$\sqrts = 13 \TeV$, $Q = m_Z$, $Y = 0$}
\\
\hline\hline
& \multicolumn{2}{c|}{$\df\sigma_\FO$ at \nNlo*} & \multicolumn{1}{c}{$\df\sigma_\res$ at \nNllmatched*}
\\
$n$ & \multicolumn{1}{c}{$\mu_F = m_Z$} & \multicolumn{1}{c|}{$\mu_F = m_Z/2$}  & \multicolumn{1}{c}{$\mu_F = m_Z/2$}
\\
\hline
$0$
&$63.7\!\pm\! 9.1_\mu\,(14\%)$
&$54.6\!\pm\! 9.6_\mu\,(17\%)$
&$71.5\!\pm\! 12.6_\mu\!\pm\! 5.0_\varphi\,(19\%)$
\\
$1$
&$72.5\!\pm\! 3.5_\mu\,(4.8\%)$
&$69.0\!\pm\! 5.4_\mu\,(7.9\%)$
&$74.9\!\pm\! 4.2_\mu\!\pm\! 1.6_\varphi\,(6.1\%)$
\\
$2$
&$71.9\!\pm\! 0.7_\mu\,(0.9\%)$
&$71.2\!\pm\! 1.8_\mu\,(2.5\%)$
&$72.2\!\pm\! 0.6_\mu\!\pm\! 0.2_\varphi\,(0.9\%)$
\\
\hline\hline
\end{tabular}
\end{small}
\caption{
Cross section for $pp \to Z/\gamma^\ast \to \ell^+ \ell^-$ at $Q = m_Z$ and $\rap = 0$ at the LHC with $\sqrts = 13 \TeV$.
The central scale is always $\mufo = \mu_R = m_Z$.
The percent uncertainties for the resummed results correspond to the total uncertainty $\DeltaFO \oplus \Delta_\varphi$.
}
\label{table:DY:CentralRapidity}
\end{table}

The perturbative series of the fixed-order cross section at $Q = m_Z$ and $\rap = 0$
and its decomposition into hard function $H_{q\bar q}$ and remainder $R$ at $\mu_F = m_Z$ and $\mu_F = m_Z/2$
is given by
\begin{align} \label{eq:DYxsecFO}
\df^2 \sigma_\FO(\mu_R = m_Z, \mu_F = m_Z)   &= (1 + 0.138 - 0.010) \times 63.7 \pb/\!\GeV
\,, \nn \\
H_{q\bar q}(m_Z^2, \mu_H=m_Z) &= \,\, 1 + 0.088 + 0.0317
\,, \nn \\
R(\mu_R = m_Z, \mu_F = m_Z)   &= (1 + 0.050 - 0.046) \times 63.7 \pb/\!\GeV
\,, \nn \\[1ex]
\df^2 \sigma_\FO(\mu_R = m_Z, \mu_F = m_Z/2) &= (1 + 0.263 + 0.040) \times 54.6 \pb/\!\GeV
\,, \nn \\
H_{q\bar q}(m_Z^2, \mu_H=m_Z) &= \,\,1 + 0.088 + 0.0317
\,, \nn \\
R(\mu_R = m_Z, \mu_F = m_Z/2) &= (1 + 0.175 - 0.007) \times 54.6 \pb/\!\GeV
\,.\end{align}
The corrections to the fixed-order cross section are smaller at $\mu_F = m_Z$.
However, its small NNLO contribution stems from a numerical cancellation between
$H_{q\bar q}$ and $R$, and as discussed in \sec{Calculation}, there is a priori no reason to expect that
this continues to happen at higher orders. Also, the NLO and NNLO contributions for $R$
are of the same size, indicating that the NLO contribution is artificially small or the NNLO
contribution unusually large or a mixture of both.
For $\mu_F = m_Z/2$, the NNLO contribution to $R$ is very small and the NNLO
contribution to the cross section primarily comes from $H_{q\bar q}$.
It will be interesting to see how this pattern continues at N$^3$LO.

The hard function itself shows again a notably improved convergence at
$\mu_H = -\img m_Z$ compared to $\mu_H = m_Z$,
\begin{alignat}{3} \label{eq:DYHardNumbers}
  &H_{q\bar q}^V(m_Z, \mu_H=m_Z)         &&= 1 + 0.08801 + 0.03169 + 0.00745
\,, \nn \\
  &H_{q\bar q}^V(m_Z, \mu_H=-\img m_Z)   &&= 1 - 0.15048 - 0.00126 - 0.00101
\,.\end{alignat}
As for $b\bar b H$, the improvement is not as dramatic as for gluon fusion due to the
reduced color factor for quarks vs. gluons. Nevertheless, by eliminating the timelike logarithms
in $H_{q \bar q}$, the higher-order contributions are substantially reduced, except
that the NLO contribution actually gets larger (in contrast to $b\bar b H$).
The reason for this is an accidental numerical cancellation in the one-loop matching coefficient
\begin{equation}
 C_{q\bar q}^V(Q,\mu) = 1 + \frac{\as(\mu)}{4\pi} C_F \left[ - \ln^2\left(\frac{-Q^2-\img 0}{\mu^2}\right) + 3\ln\left(\frac{-Q^2-\img 0}{\mu^2}\right) - 8 + \frac{\pi^2}{6} \right] + \ORd{\as^2(\mu)}
\,,\end{equation}
where the rather large nonlogarithmic constant term of $-8+\pi^2/6$ partially cancels the
$-\ln^2(-1)=\pi^2$ when $H_{q\bar q}$ is evaluated at $\mu_H = m_Z$. As discussed in
\sec{Calculation}, the separation of the nonlogarithmic constant terms between $H$ and $R$
amounts to a scheme choice and only their sum is ultimately relevant. Hence,
this large NLO constant term is a scheme-dependent artifact and in fact cancels
most of the equally large NLO contribution in $R(\mu_F = m_Z/2)$. This also means that the
$+0.263$ NLO contribution in the cross section at $\mu_F = m_Z/2$ in \eq{DYxsecFO}
does in fact primarily come from the NLO timelike logarithm in $H_{q\bar q}$, which gives a contribution
of $+0.247$ to it, even though this is not immediately obvious from \eq{DYxsecFO}.
This explains the much improved convergence of the resummed results at $\mu_F = m_Z/2$
compared to the fixed-order results at the same scale.

It was already noted in \refcite{Magnea:1990zb} that the constant terms in $C_{q\bar q}^V$ are scheme dependent
and hence not physical, unlike the ratio of form factors. Since the constant terms in $H$ and $R$ are
evaluated at different scales, there is a residual scheme dependence, which is analogous to a scale
choice in that it affects the numerical results but is formally of higher order [see \eq{finite terms cancel}].
To check this, we can consider an alternative renormalization scheme for the Wilson coefficient
$\tilde{C}_{q\bar q}^V$, for which all constant
terms exactly vanish. That is, the corresponding hard function solely consists of timelike logarithms, $\tilde{H}_{q\bar q}^V(m_Z^2, \mu_H = m_Z) = 1 + 0.247 + 0.073 + 0.013$, while $\tilde{H}_{q\bar q}^V(m_Z^2, \mu_H = -\img m_Z) = 1 + 0 + 0 + 0$.
Hence, the constant terms are moved entirely into the remainder. In this scheme, the resummed result at \nnllmatched*
is $\df^2 {\tilde{\sigma}}_\res(\rap = 0, Q = m_Z, \mu_R = m_Z, \mu_F = m_Z/2) = (71.4\!\pm\! 0.7_\mu\!\pm\! 0.2_\varphi) \pb/\!\GeV \,(1.0\%)$. Compared to the last line of \tab{DY:CentralRapidity}, the difference of $\pm 0.8\pb$ is of the same
size as the uncertainties and thus of the typical size we expect for an $\ord{\as^3}$ effect.

\begin{figure*}
\centering
\includegraphics[width=\WidthTwoSubfigs]{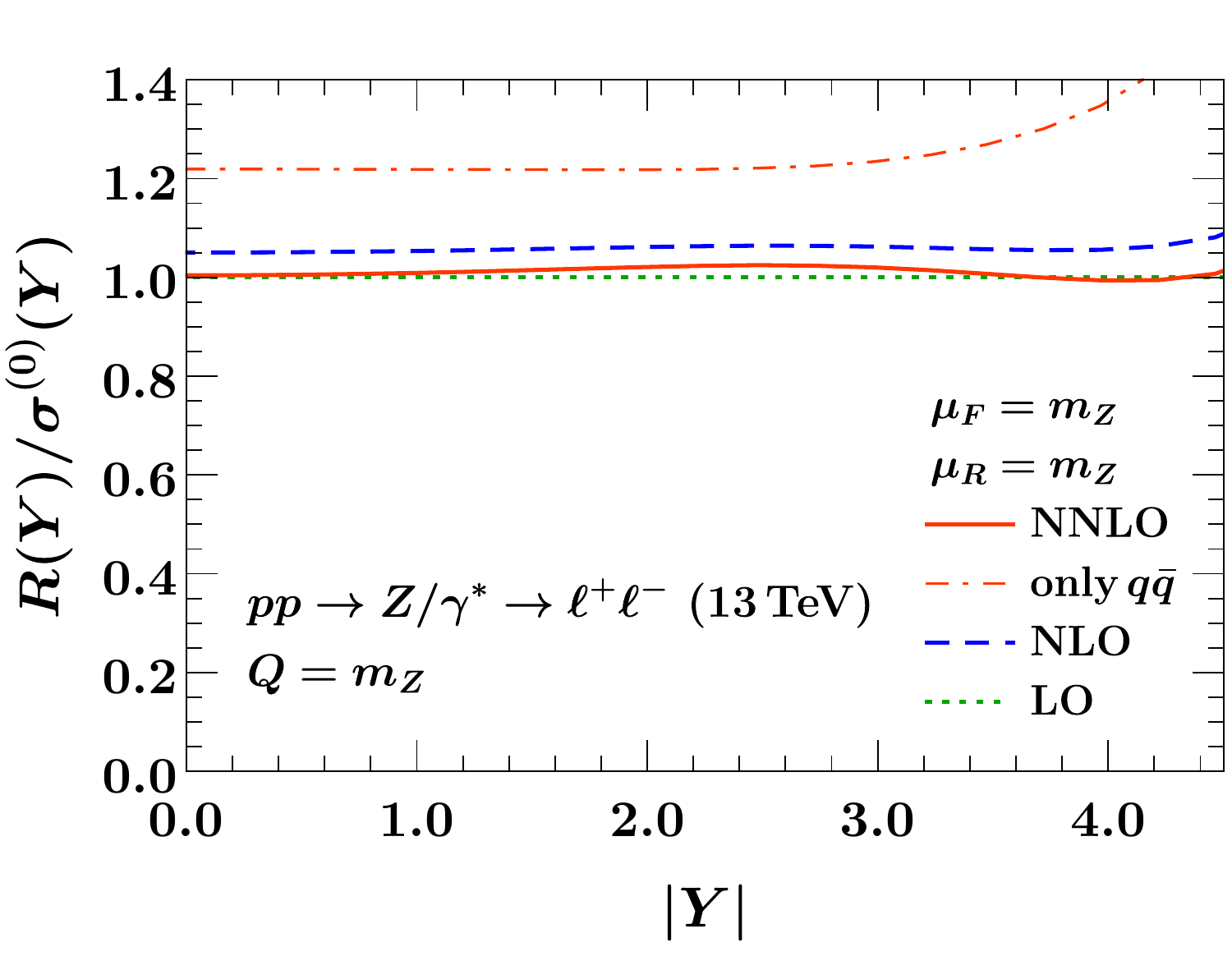}%
\hfill%
\includegraphics[width=\WidthTwoSubfigs]{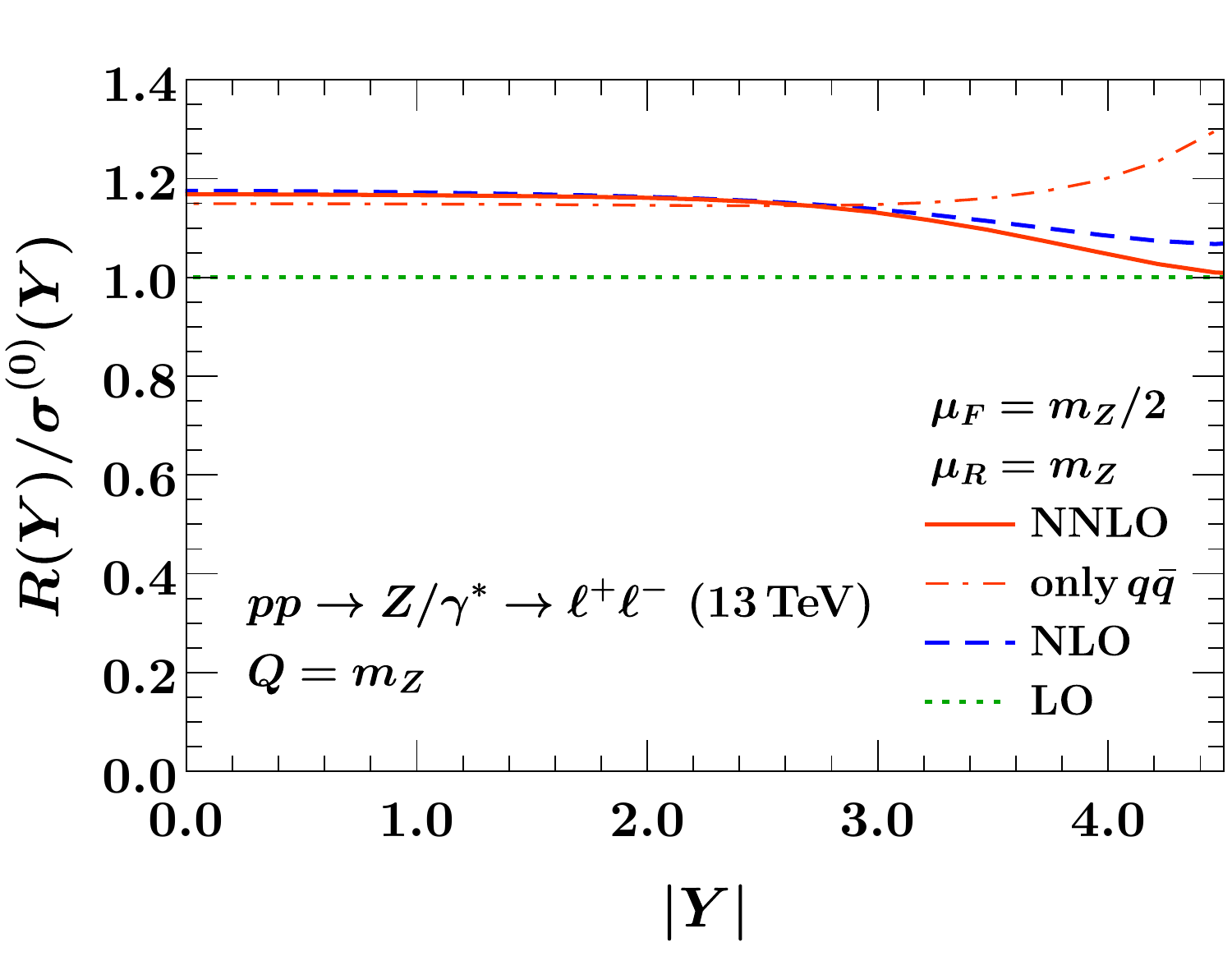}%
\caption{
The perturbative remainder $R(Y)$ for $pp \to Z/\gamma^\ast \to \ell^+ \ell^-$,
normalized to $\sigma^{(0)}(Y) \equiv \df^2\sigma^{(0)} / \df Q \,\df \rap$ at $Q=m_Z$
for $\mu_F = m_Z$ (left) and $\mu_F = m_Z/2$ (right).
The dot-dashed line shows the result including only the $q\bar q$ channel.}
\label{fig:DY remainder}
\end{figure*}

\begin{figure*}
\centering
\includegraphics[width=0.6\textwidth]{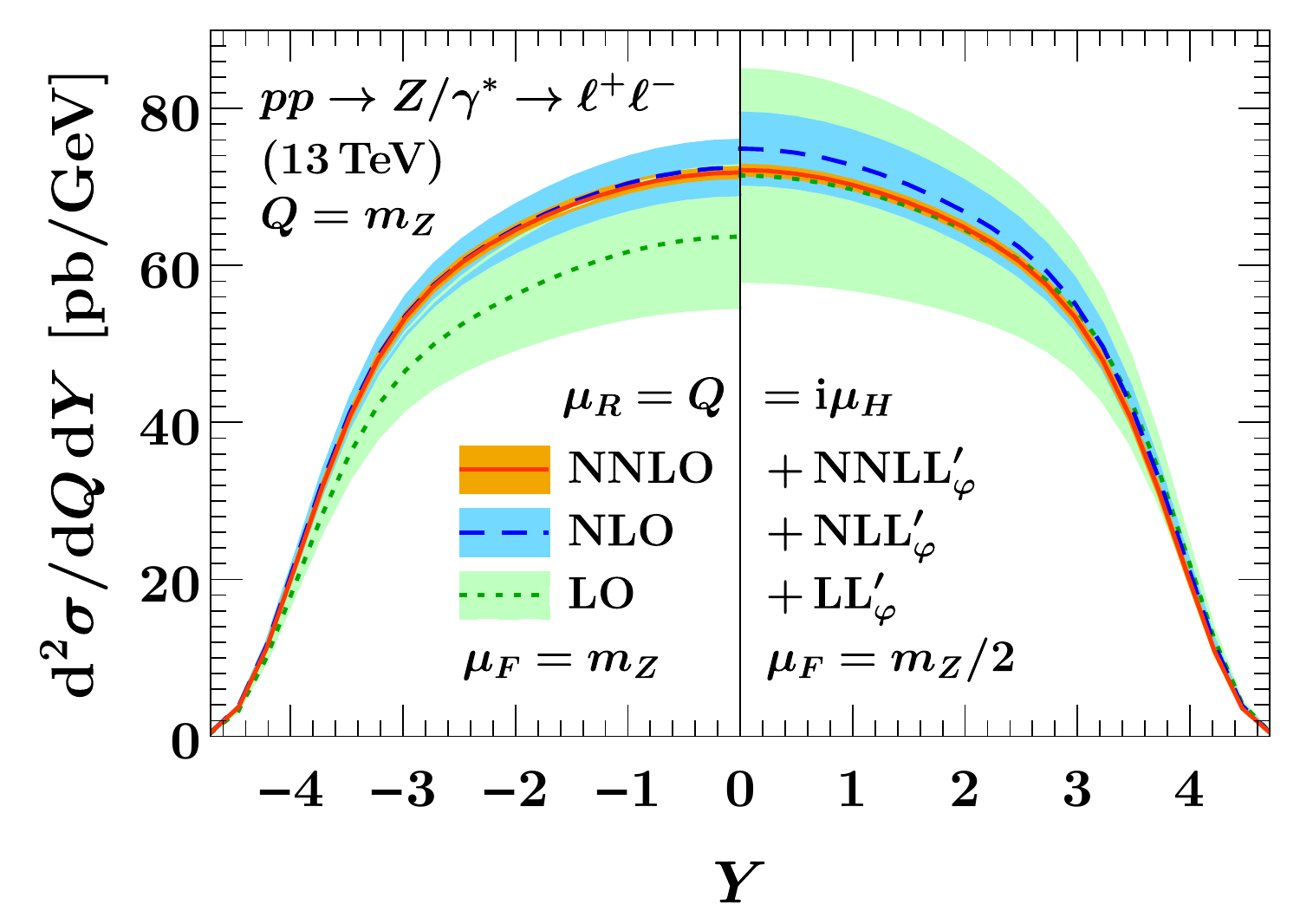}%
\caption{Rapidity spectrum for $pp \to Z/\gamma^\ast \to \ell^+ \ell^-$
at $Q = m_Z$. The fixed-order results are shown for $\rap < 0$ and the resummed
results for $\rap > 0$. For the central scale we use $\mufo = m_Z$ in both cases,
while $\kappa_F = 1$ (fixed order) and $\kappa_F = 1/2$ (resummed).
The uncertainty bands indicate $\DeltaFO$ (fixed order) and $\DeltaFO \oplus \Delta_\varphi$ (resummed).}
\label{fig:DY}
\end{figure*}

We now discuss the effect of the resummation on the rapidity spectrum.
In \fig{DY remainder}, we show the remainder $R(Y)$ normalized to the Born cross section as a function of $Y$
for $\mu_F = m_Z$ (left) and $\mu_F = m_Z/2$ (right). The behavior discussed for $Y = 0$ above is
similar throughout most of the spectrum. For $\mu_F = m_Z$, the \nnlo* corrections are comparably
large and of the same size and opposite sign as the \nlo* contributions,
while the \nnlo* corrections are almost negligible at $\mu_F = m_Z/2$ over most of the rapidity
range. We see again that at $\mu_F = m_Z$ the NNLO corrections involve substantial cancellations
between the $q\bar q$ and non-$q\bar q$ channels, which individually are very large. In contrast,
at $\mu_F = m_Z/2$ also the individual corrections to the remainder are very small, again
supporting this central choice when including the resummation.
(A large part of the rapidity-independent constant shift at \nlo* will again be canceled by the
constant term in $H_{q\bar q}$.)
In fig.~\ref{fig:DY}, we compare the rapidity spectrum at $Q=m_Z$,
where the fixed-order calculation at its optimal $\mu_F=m_Z$ is shown for $\rap<0$
and the resummed result at $\mu_F = m_Z/2$ for $\rap>0$.

Overall, we find that the \nnlo* and \nnllmatched* predictions provide very similar results.
On the one hand, this is reassuring, as it shows that the good convergence of the
fixed-order series is not spoiled by the resummation. On the other hand,
given the extreme reduction of the perturbative uncertainties in the fixed-order results
at the conventional choice of $\mu_F = m_Z$ by a factor of five when going from NLO to NNLO and the substantially
larger uncertainties at $\mu_F = m_Z/2$, one might perhaps be worried that the fixed-order
uncertainties are somewhat underestimated, in part due to the accidentally small NNLO
contribution. In this respect, the resummed results provide a useful confirmation and
increased confidence in the very small perturbative uncertainties in the Drell-Yan
predictions.

\section{Conclusion}
\label{sec:Conclusion}

We have investigated in detail the resummation of timelike logarithms $\ln^2(-1) = -\pi^2$
that arise to all orders in perturbation theory and are an important source of perturbative
corrections in $s$-channel color-singlet production processes, which involve a timelike hard momentum transfer.
These logarithms can be resummed to all orders using the RG evolution of the corresponding
quark or gluon form factors from spacelike to timelike scales.

We have shown how to incorporate the resummed form factor in a completely straightforward
manner into predictions for generic inclusive cross sections with arbitrary dependence or cuts on the Born kinematics.
We have verified that this does not spoil the perturbative series in all considered cases.
We have also discussed the assessment of the uncertainties intrinsic to the resummation.

We first revisited the resummation for the total gluon-fusion cross section, for which it
has been discussed before, considering both the production of a generic scalar
as well as the SM Higgs boson in the $m_t\to\infty$ limit up to \nnnllmatched*.
For the latter we have also shown how to incorporate quark-mass and electroweak effects.
We confirm that the resummation significantly improves the
perturbative series, and find that it reduces the perturbative uncertainties at the
highest orders by about a factor of two.

For the Higgs rapidity spectrum as well as the cross section with a cut on the Higgs rapidity
we obtain results at \nnllmatched*, which provide the currently most precise predictions
with central values close to what might be expected at \nnnlo*, and perturbative uncertainties
of $\sim 6\%$, which are almost a factor of two smaller than at NNLO.
Once \nnnlo* results for the rapidity dependence become available,
we project that the corresponding resummation at \nnnllmatched* will provide a similar
improvement.

We also studied the resummation of timelike logarithms for quark-induced processes,
namely Higgs production through bottom-quark annihilation and the Drell-Yan rapidity spectrum.
For the former, the resummation provides a small improvement in the perturbative convergence
and resulting uncertainties. For Drell-Yan production, the resummation provides no clear improvement
but also no worsening of the predictions, due to the already fast convergence of the fixed-order
perturbative series. In this case it provides a useful confirmation of the very small
residual perturbative uncertainties.

We conclude that utilizing the resummed timelike quark and gluon form factors is viable
and beneficial for obtaining precise and reliable predictions for $s$-channel
color-singlet production processes.

\acknowledgments
We like to thank Stefan Liebler for his support with \texttt{SusHi} and for comments on the manuscript as well as Dirk Rathlev for his expertise on \texttt{HNNLO}. We thank the anonymous referee for suggesting
to also study the individual partonic channels for the Drell-Yan process.
This work was supported by the DFG Emmy-Noether Grant No. TA 867/1-1 and the PIER Helmholtz Graduate school.
J.\,M. thanks DESY for hospitality
and gratefully acknowledges support by M{\"u}nster University funds designated for student research.

\appendix
\section{Perturbative ingredients}
\label{app:ingredients}

\subsection{Master formula for hard Wilson coefficients to three loops}
\label{app:GenericWilson}

The hard matching coefficients $C$ satisfy an RGE of the form
\begin{equation} \label{eq:generic C RGE}
\mu \frac{\df}{\df \mu} C(q^2,\mu)
= \Bigl\{\Gamma_\cusp[\as(\mu)] \ln\frac{-q^2-\img0}{\mu^2} + \gamma[\as(\mu)] \Bigr\} C(q^2,\mu)
\,,\end{equation}
which allows us to completely predict the logarithmic structure in terms of the cusp
and noncusp anomalous dimension coefficients.
We write the perturbative expansion of the hard coefficient as
\begin{equation}
C(q^2,\mu) = \sum_{n = 0}^\infty C^{(n)}(L) \biggl[\frac{\as(\mu)}{4\pi} \biggr]^{n}
\,,\qquad
L = \ln\frac{-q^2 - \img 0}{\mu^2}
\,,\qquad
C_n = C^{(n)}(0)
\,.\end{equation}
Normalizing $C$ such that the tree-level result is $C_0 = 1$, the perturbative solution of \eq{generic C RGE} to \nnnlo*
is given by
\begin{align} \label{eq:master formula}
C^{(0)} &= 1
\,,\nn \\
C^{(1)}(L) &= - \frac{L^2}{4}\Gamma_0 - \frac{L}{2}\gamma_0 + C_1
\,,\nn \\
C^{(2)}(L)
&= \frac{L^4}{32} \Gamma_0^2
   + \frac{L^3}{24} \Gamma_0 ( 2\beta_0 + 3\gamma_0 )
   + \frac{L^2}{8} ( 2\beta_0 \gamma_0 + \gamma_0^2 - 2C_1 \Gamma_0 - 2\Gamma_1 )
\nn \\ & \quad
   - \frac{L}{2} (2C_1 \beta_0 + C_1\gamma_0 + \gamma_1 )
   + C_2
\,,\nn \\
C^{(3)}(L)
&= - \frac{L^6}{384} \Gamma_0^3
   - \frac{L^5}{192} \Gamma_0^2 (4 \beta_0 + 3\gamma_0)
   + \frac{L^4}{96} \Gamma_0 \bigl( - 4 \beta_0^2 - 10 \beta_0 \gamma_0 - 3 \gamma_0^2 + 3C_1 \Gamma_0 + 6 \Gamma_1 \bigr)
\nn \\ & \quad
   + \frac{L^3}{48}\Bigl[ -8\beta_0^2 \gamma_0 - 6\beta_0 \gamma_0^2 - \gamma_0^3
     + \Gamma_0(16 C_1 \beta_0 + 6 C_1 \gamma_0 + 6 \gamma_1 + 4\beta_1)
     + \Gamma_1 (8 \beta_0 + 6 \gamma_0) \Bigr]
\nn \\ & \quad
   + \frac{L^2}{8}\Bigl[C_1(8 \beta_0^2 + 6 \beta_0 \gamma_0 + \gamma_0^2 - 2\Gamma_1 ) + 2\beta_1 \gamma_0 + 4 \beta_0 \gamma_1
     + 2\gamma_0 \gamma_1 - 2 C_2 \Gamma_0 - 2\Gamma_2 \Bigr]
\nn \\ & \quad
   - \frac{L}{2} \bigl(4 C_2 \beta_0 + 2 C_1 \beta_1 + C_2 \gamma_0 + C_1 \gamma_1 + \gamma_2 \bigr)
   + C_3
\,.
\end{align}
Here, $\beta_n$ are the beta-function coefficients, $\Gamma_n \equiv \Gamma_n^i$
the appropriate quark or gluon cusp anomalous dimensions coefficients,
and $\gamma_n$ are the coefficients of the total noncusp
anomalous dimension $\gamma$ in \eq{generic C RGE} as appropriate for the
hard coefficient of interest.
All required anomalous dimension coefficients are given below in \app{AnomDim}.
The results for the nonlogarithmic constant terms $C_n$ for the different Wilson coefficients
are given below in \app{finite}

The full expression for the hard function is obtained by squaring $C$, accounting for cross terms.
In the case of $H_{gg}^t$ defined in \eq{GluonHardFull} the product of $C_t\,C_{gg}$ is reexpanded.

\subsection{Anomalous dimensions}
\label{app:AnomDim}

We expand the $\beta$ function of QCD as
\begin{equation} \label{eq:beta expand}
\mu \frac{\df \as(\mu)}{\df \mu}
= \beta[\as(\mu)]
\,, \qquad
\beta(\as) = -2\as \sum_{n = 0}^\infty \beta_n \left( \frac{\as}{4\pi} \right)^{n+1}
\,.\end{equation}
The coefficients up to four loops in the \MSbar scheme are~\cite{Tarasov:1980au, Larin:1993tp, vanRitbergen:1997va, Czakon:2004bu}
\begin{align} \label{eq:beta coeff}
\beta_0 &= \frac{11}{3}\,C_A -\frac{4}{3}\,T_F\,n_f
\,, \nn \\
\beta_1 &= \frac{34}{3}\,C_A^2  - \Bigl(\frac{20}{3}\,C_A\, + 4 C_F\Bigr)\, T_F\,n_f
\,, \\
\beta_2 &=
\frac{2857}{54}\,C_A^3 + \Bigl(C_F^2 - \frac{205}{18}\,C_F C_A
- \frac{1415}{54}\,C_A^2 \Bigr)\, 2T_F\,n_f
+ \Bigl(\frac{11}{9}\, C_F + \frac{79}{54}\, C_A \Bigr)\, 4T_F^2\,n_f^2
\,, \nn \\
\beta_3 &=  \Bigl( \frac{149753}{6} + 3564 \zeta_3 \Bigr)
- \Bigl( \frac{1078361}{162} + \frac{6508}{27} \zeta_3 \Bigr) n_f
+ \Bigl( \frac{50065}{162} + \frac{6472}{81} \zeta_3 \Bigr) n_f^2
+  \frac{1093}{729} n_f^3
\nn\,,\end{align}
where for $\beta_3$ we specified to $\mathrm{SU}(3)$ for brevity.
Throughout our analysis, we consider $n_f = 5$ active flavors.

The cusp and noncusp anomalous dimensions are expanded as
\begin{equation} \label{eq:anom dim expand}
\Gamma^i_\cusp(\as) = \sum_{n = 0}^\infty \Gamma^i_n \Bigl( \frac{\as}{4\pi} \Bigr)^{n+1}
\,,\qquad
\gamma(\as) = \sum_{n = 0}^\infty \gamma_n \Bigl( \frac{\as}{4\pi} \Bigr)^{n+1}
\,.\end{equation}
The coefficients of the $\overline{\mathrm{MS}}$ cusp anomalous dimension to three loops are~\cite{Korchemsky:1987wg, Moch:2004pa, Vogt:2004mw}
\begin{align} \label{eq:Gcusp coeff}
\Gamma^q_n &= C_F \Gamma_n
\,,\qquad
\Gamma^g_n = C_A \Gamma_n
\,,\qquad \text{(for $n = 0,1,2$)}
\,,\nn\\[1ex]
\Gamma_0 &= 4
\,,\nn\\
\Gamma_1
&= 4 \Bigl[ C_A \Bigl( \frac{67}{9} - \frac{\pi^2}{3} \Bigr)  - \frac{20}{9}\,T_F\, n_f \Bigr]
= \frac{4}{3} \bigl[ (4 - \pi^2) C_A + 5 \beta_0 \bigr]
\,,\nn\\
\Gamma_2
&= 4 \Bigl[
   C_A^2 \Bigl(\frac{245}{6} -\frac{134 \pi^2}{27} + \frac{11 \pi ^4}{45} + \frac{22 \zeta_3}{3}\Bigr)
   +  C_A\, T_F\,n_f \Bigl(- \frac{418}{27} + \frac{40 \pi^2}{27}  - \frac{56 \zeta_3}{3} \Bigr)
   \nn \\ & \qquad
   +  C_F\, T_F\,n_f \Bigl(- \frac{55}{3} + 16 \zeta_3 \Bigr)
   - \frac{16}{27}\,T_F^2\, n_f^2
   \Bigr]
\,.\end{align}
The resummation at N$^3$LL formally also requires the yet unknown four-loop coefficient $\Gamma_3^i$,
which we estimate as usual by the Pad{\'e} approximation
\begin{equation}
\Gamma^i_{3,\,\text{Pad{\'e}}} = \frac{(\Gamma^i_2)^2}{\Gamma_1^i}
\,,\end{equation}
and explicitly verify that a variation $\pm 200 \%$ only affects the hard evolution kernel
$U_H$ (and thus the resummed cross section) at the sub-permille level. We therefore neglect this source of theory uncertainty.

The gluon noncusp anomalous dimension $\gamma_C^g$ enters the RGE for the gluon-to-scalar matching coefficients $C_{gg}$
and $C_{gg}'$ in \eqs{RGECgg}{RGECggPrime}. The coefficients in \MSbar up to three loops are~\cite{Moch:2005tm, Idilbi:2005ni, Idilbi:2006dg}
\begin{align}
\gamma_{C\,0}^g &= -\beta_0
\,,\nn\\
\gamma_{C\,1}^g
&= C_A \biggl[C_A \Bigl(-\frac{59}{9} + 2\zeta_3\Bigr)
   + \beta_0 \Bigl(-\frac{19}{9}+\frac{\pi^2}{6} \Bigr) \biggr] - \beta_1
\,,\nn\\
\gamma_{C\,2}^g
&= \frac{C_A}{2} \biggl[ C_A^2 \Bigl(-\frac{60875}{162} + \frac{634\pi^2}{81} +\frac{8\pi^4}{5}+\frac{1972\zeta_3}{9}
   - \frac{40\pi^2 \zeta_3}{9} - 32\zeta_5 \Bigr)
\nn\\ & \qquad\quad
   + C_A \beta_0 \Bigl(\frac{7649}{54}+\frac{134\pi^2}{81} - \frac{61\pi^4}{45} - \frac{500\zeta_3}{9}\Bigr)
   + \beta_0^2 \Bigl(\frac{466}{81}+\frac{5\pi^2}{9}-\frac{28 \zeta_3}{3}\Bigr)
\nn\\ & \qquad\quad
   + \beta_1 \Bigl(-\frac{1819}{54} + \frac{\pi^2}{3} + \frac{4\pi^4}{45} + \frac{152\zeta_3}{9}\Bigr)
\biggr]
   - \beta_2
\,.\end{align}

The evolution of $C_{gg}^t$ in the one-step matching also requires the anomalous dimension $\gamma_t$ of the Wilson coefficient $C_t$ arising from integrating out the top quark. It is given by
\begin{equation}
\gamma_t(\as) = \as^2 \frac{\df}{\df\as} \frac{\beta(\as)}{\as^2}
\,,\qquad
\gamma_{t\,n} = - 2n \cdot \beta_n
\,.\end{equation}

The quark noncusp anomalous dimension $\gamma_C^q$ enters the RG \eqs{RGECqqS}{RGECqqVA} for both quark-induced processes
we consider. The coefficients in \MSbar up to three loops are~\cite{Moch:2005id, Moch:2005tm, Idilbi:2006dg, Becher:2006mr}
\begin{align} \label{eq:gaCq}
\gamma^q_{C\,0} &= -3 C_F
\,,\nn\\
\gamma^q_{C\,1}
&= - C_F \biggl[
   C_A \Bigl(\frac{41}{9} - 26\zeta_3\Bigr)
   + C_F \Bigl(\frac{3}{2} - 2 \pi^2 + 24 \zeta_3\Bigr)
   + \beta_0 \Bigl(\frac{65}{18} + \frac{\pi^2}{2} \Bigr) \biggr]
\,,\nn\\
\gamma^q_{C\,2}
&= -C_F \biggl[
   C_A^2 \Bigl(\frac{66167}{324} - \frac{686 \pi^2}{81} - \frac{302 \pi^4}{135} - \frac{782 \zeta_3}{9} + \frac{44\pi^2 \zeta_3}{9} + 136 \zeta_5\Bigr)
\nn\\ & \qquad\qquad
   + C_F C_A \Bigl(\frac{151}{4} - \frac{205 \pi^2}{9} - \frac{247 \pi^4}{135} + \frac{844 \zeta_3}{3} + \frac{8 \pi^2 \zeta_3}{3} + 120 \zeta_5\Bigr)
\nn\\ & \qquad\qquad
   + C_F^2 \Bigl(\frac{29}{2} + 3 \pi^2 + \frac{8\pi^4}{5} + 68 \zeta_3 - \frac{16\pi^2 \zeta_3}{3} - 240 \zeta_5\Bigr)
\nn\\ & \qquad\qquad
   + C_A \beta_0 \Bigl(-\frac{10781}{108} + \frac{446 \pi^2}{81} + \frac{449 \pi^4}{270} - \frac{1166 \zeta_3}{9} \Bigr)
\nn\\ & \qquad\qquad
   + \beta_1 \Bigl(\frac{2953}{108} - \frac{13 \pi^2}{18} - \frac{7 \pi^4 }{27} + \frac{128 \zeta_3}{9}\Bigr)
   + \beta_0^2 \Bigl(-\frac{2417}{324} + \frac{5 \pi^2}{6} + \frac{2 \zeta_3}{3}\Bigr)
\biggr]
\,.\end{align}

The evolution of $C_{q\bar q}^S$ also requires the anomalous dimension of the quark Yukawa coupling, which is equivalent to the quark mass anomalous dimension $\gamma_m$,
\begin{equation}
\mu \frac{\df}{\df \mu} y(\mu) = \gamma_m[\as(\mu)] \, y(\mu)
\,.\end{equation}
It is known to five loops~\cite{Tarrach:1980up,Tarasov:1982gk,Larin:1993tq,Vermaseren:1997fq,Chetyrkin:1997dh,Baikov:2014qja,Luthe:2016xec}. For our main analysis at NNLL we only require the two-loop result, while the three-loop coefficient $\gamma_{m\,2}$ serves to verify our \nnnlo* result for $C_{q\bar q}^S$. The results are
\begin{align} \label{eq:g quark mass coeff}
\gamma_{m\,0} &= -6 C_F
\,, \nn \\
\gamma_{m\,1}
&= -2 C_F \Bigl(
    \frac{3}{2}\,C_F + \frac{97}{6} \,C_A - \frac{10}{3}\, T_F\, n_f
    \Bigr)
\,, \nn \\
\gamma_{m\,2}
& = -2 C_F \biggl[
    \frac{11413}{108}\, C_A^2
    - \frac{129}{4}\,C_F C_A
    + \frac{129}{2}\, C_F^2
    + C_A\, T_F\, n_f\, \Bigl( -\frac{556}{27}-48\zeta_3 \Bigr)
\nn \\ &\qquad\qquad
    +C_F\, T_F\, n_f\, (-46+48\zeta_3)
    - \frac{140}{27} T_F^2\, n_f^2
    \biggr]
\,.\end{align}

\subsection{Constant terms to three loops}
\label{app:finite}

In the following, we provide the process-specific nonlogarithmic constant terms $C_n$ for the various hard matching coefficients.
For $C_{gg}$, $C_t$, and $C_{q\bar q}^V$, we can collect the results from the literature.
The result for $C_{q\bar q}^S$ we have extracted from the three-loop scalar quark form factor.
By convention, we normalize all coefficients to unity at \lo*,
\begin{equation}
C_{gg\,0} = C_{t\,0} = C_{q\bar q\,0}^V = C_{gg\,0}^S = 1
\,.\end{equation}
Note that for all coefficients quoted here, we closely follow the notation from the original publications.
For this reason, we set $T_F = 1/2$ in the following results.

\subsubsection{Gluon matching coefficient}

The finite terms of $C_{gg}$ can be read off from the full result given in \refcite{Gehrmann:2010ue},
\begin{align}
C_{gg\,1} &= C_A\, \zeta_2
\,,\nn\\
C_{gg\,2}
&= C_A^2 \Bigl(\frac{5105}{162}-\frac{143}{9}\zeta_3+\frac{67}{6}\zeta_2+\frac{1}{2}\zeta_2^2 \Bigr)
   + C_A\,n_f \Bigl(-\frac{916}{81}-\frac{46}{9}\zeta_3-\frac{5}{3}\zeta_2 \Bigr)
\nn \\ & \quad
   + C_F\,n_f \Bigl(-\frac{67}{6}+8\zeta_3 \Bigr)
\,,\nn\\
C_{gg\,3}
&= C_A^3 \Bigl(
      +\frac{29639273}{26244}-\frac{1939}{270}\zeta_2^2+\frac{2222}{9}\zeta_5+\frac{105617}{729}\zeta_2
      -\frac{24389}{1890}\zeta_2^3-\frac{152716}{243}\zeta_3
\nn \\  &\qquad\quad
      -\frac{605}{9}\zeta_2\zeta_3 - \frac{104}{9}\zeta_3^2 \Bigr)
\nn \\ & \quad
   + C_A^2\,n_f \Bigl(
      -\frac{3765007}{6561}+\frac{428}{9}\zeta_5-\frac{460}{81}\zeta_3-\frac{14189}{729}\zeta_2-\frac{82}{9}\zeta_2\zeta_3+\frac{73}{45}\zeta_2^2 \Bigr)
\nn \\ & \quad
   + C_A C_F\,n_f \Bigl(
      -\frac{341219}{972}+\frac{608}{9}\zeta_5+\frac{14564}{81}\zeta_3-\frac{68}{9}\zeta_2+\frac{64}{3}\zeta_2\zeta_3-\frac{64}{45}\zeta_2^2 \Bigr)
\nn \\ & \quad
   + C_F^2\,n_f\, \Bigl(\frac{304}{9}-160\zeta_5+\frac{296}{3}\zeta_3 \Bigr)
\nn \\ & \quad
   + C_A\,n_f^2 \Bigl(
       \frac{611401}{13122}+\frac{4576}{243}\zeta_3+\frac{4}{9}\zeta_2+\frac{4}{27}\zeta_2^2 \Bigr)
\nn \\ & \quad
   + C_F\,n_f^2 \Bigl(
       \frac{4481}{81}-\frac{112}{3}\zeta_3-\frac{20}{9}\zeta_2-\frac{16}{45}\zeta_2^2 \Bigr)
\,.\end{align}

\subsubsection{\texorpdfstring{$C_t$}{Ct} coefficient for Higgs production in the EFT limit}

The general expression for $C_t(m_t, \mu)$ up to $\ord{\as^3}$ is given by
\begin{align}
C_t(m_t, \mu)
&= 1
   + C_{t\,1} \frac{\alpha_s(\mu)}{4\pi}
   + \biggl[ (\beta_1 - \beta_0 C_{t\, 1}) \ln\frac{m_t^2}{\mu^2} + C_{t\,2} \biggr] \frac{\alpha_s^2(\mu)}{(4\pi)^2}
\nn \\ & \quad
   + \biggl[-\beta_0(\beta_1 - \beta_0 C_{t\,1} ) \ln^2\frac{m_t^2}{\mu^2}
      + 2(\beta_2 - \beta_0 C_{t\,2}) \ln\frac{m_t^2}{\mu^2}
      + C_{t\,3} \biggr] \frac{\alpha_s^3(\mu)}{(4\pi)^3}
\,.\end{align}
The constant terms are given by
\begin{align}
C_{t\,1} &= 5 C_A - 3 C_F
\,, \nn \\
C_{t\,2}
&=
\frac{91}{6} C_A^2 - \frac{100}{3} C_A C_F + \frac{27}{2} C_F^2
+  C_A \Bigl(\frac{47}{12}\beta_0 - \frac{5}{6} T_F \Bigr)
- C_F T_F \Bigl(\frac{4}{3} + 5 n_f\Bigr)
\,, \nn \\
C_{t\,3}
&=
- \frac{2761331}{648} + \frac{897943}{144}\zeta_3
+  n_f \Bigl(\frac{58723}{324} - \frac{110779}{216} \zeta_3 \Bigr)
- n_f^2 \, \frac{6865}{486}
\,,\end{align}
where $C_{t\,3}$ is taken from \refcite{Anastasiou:2016cez}, where all color factors are already evaluated for $N_c=3$.

The dependence of $H_{gg}^t$ on $\rho = m_H^2 / (4 m_t^2)$ at \nlo* is given by~\cite{Berger:2010xi}
\begin{align} \label{eq:Ftop}
F_1(\rho) &=
C_A \Bigl(5 - \frac{38}{45}\, \rho - \frac{1289}{4725}\, \rho^2 - \frac{155}{1134}\, \rho^3 - \frac{5385047}{65488500}\, \rho^4\Bigr)
\nn\\ & \quad
+ C_F \Bigl(-3 + \frac{307}{90}\, \rho + \frac{25813}{18900}\, \rho^2 + \frac{3055907}{3969000}\, \rho^3 +
   \frac{659504801}{1309770000}\, \rho^4 \Bigr) + \ord{\rho^5}
\,,\end{align}
where $F_1(0) = C_{t\,1}$.
The exact $\rho$ dependence of $F_1(\rho)$ in terms of harmonic polylogarithms is
known~\cite{Harlander:2005rq, Anastasiou:2006hc, Aglietti:2006tp}.
We use the results expanded in $\rho$, which are completely sufficient
for practical purposes because the corrections are small and the expansion in $\rho \simeq 0.13$ converges very quickly.

\subsubsection{Quark vector-current matching coefficient}

The finite terms of $C_{q\bar q}^V$ to three loops can be read off from \refcite{Gehrmann:2010ue},
\begin{align}
C^V_{q\bar q\,1} &= C_F (- 8 + \zeta_2)
\,, \nn \\
C^V_{q\bar q\,2}
&= C_F \biggl[
   C_A \Bigl(-\frac{51157}{648}+\frac{313}{9}\zeta_3-\frac{337}{18}\zeta_2+\frac{44}{5}\zeta_2^2 \Bigr)
   + C_F \Bigl(\frac{255}{8}-30\zeta_3+21\zeta_2-\frac{83}{10}\zeta_2^2 \Bigr)
\nn \\ & \qquad\quad
   + n_f \Bigl(\frac{4085}{324}+\frac{2}{9}\zeta_3+\frac{23}{9}\zeta_2 \Bigr)
   \biggr]
\,, \nn \\
C^V_{q\bar q\,3}
&= C_F \biggl[
   C_A^2 \Bigl(
      -\frac{51082685}{52488}-\frac{434}{9}\zeta_5+\frac{505087}{486}\zeta_3-\frac{1136}{9}\zeta_3^2-\frac{412315}{729}\zeta_2+\frac{416}{3}\zeta_2\zeta_3
\nn \\ & \qquad\qquad\quad
      + \frac{22157}{270}\zeta_2^2-\frac{6152}{189}\zeta_2^3 \Bigr)
\nn \\ & \qquad\quad
   + C_A C_F \Bigl(
      \frac{415025}{648}-\frac{2756}{9}\zeta_5-\frac{18770}{27}\zeta_3+\frac{296}{3}\zeta_3^2+\frac{538835}{648}\zeta_2-\frac{3751}{9}\zeta_2\zeta_3
\nn \\ & \qquad\qquad\quad
      -\frac{4943}{270}\zeta_2^2-\frac{12676}{315}\zeta_2^3 \Bigr)
\nn \\ & \qquad\quad
   + C_F^2 \Bigl(
      -\frac{2539}{12}-\frac{413}{5}\zeta_2^2+664\zeta_5-\frac{6451}{24}\zeta_2+\frac{37729}{630}\zeta_2^3-470\zeta_3+250\zeta_2\zeta_3+16\zeta_3^2 \Bigr)
\nn \\ & \qquad\quad
   + C_A\,n_f \Bigl(
      \frac{1700171}{6561}-\frac{4}{3}\zeta_5-\frac{4288}{27}\zeta_3+\frac{115555}{729}\zeta_2+\frac{4}{3}\zeta_2\zeta_3+\frac{2}{27}\zeta_2^2 \Bigr)
\nn \\ & \qquad\quad
   + C_F\,n_f \Bigl(
      \frac{41077}{972}-\frac{416}{9}\zeta_5+\frac{13184}{81}\zeta_3-\frac{31729}{324}\zeta_2-\frac{38}{9}\zeta_2\zeta_3-\frac{331}{27}\zeta_2^2 \Bigr)
\nn \\ & \qquad\quad
   + n_f^2\, \Bigl(
      -\frac{190931}{13122}-\frac{416}{243}\zeta_3-\frac{824}{81}\zeta_2-\frac{188}{135}\zeta_2^2 \Bigr)
\nn \\ & \qquad\quad
   + N_{F,V} \biggl(\frac{N_c^2-4}{N_c}\biggr)
   \Bigl(4-\frac{80}{3}\zeta_5+\frac{14}{3}\zeta_3+10\zeta_2-\frac{2}{5}\zeta_2^2 \Bigr)
   \biggr]
\,.\end{align}
The last term is the three-loop contribution from diagrams where the initial-state quarks do not couple directly to the vector boson. Here, $N_c = 3$ is the number of colors and we refer to \refcite{Gehrmann:2010ue} for details of $N_{F,V}$.
Since the full Drell-Yan fixed-order cross section is only available to \nnlo*,
the three-loop coefficient never enters our resummed predictions.
For the illustrative values of $H_{q\bar q}^V$ given in \eq{DYHardNumbers} we set $N_{F,V} = 0$ for the sake of comparison.

The explicit three-loop results for $C_{q\bar q}^V$ were also extracted in \refcite{Abbate:2010xh} from the three-loop form factor in \refcite{Lee:2010cga}. We verified that the above results agree with the numerical results for the vector hard function $H_{q\bar q}^V$ given in \refcite{Abbate:2010xh} after setting $N_{F,V} = 0$.

\subsubsection{Quark scalar-current matching coefficient}
\label{app:Cqq}

As far as we are aware, a result for $C_{q\bar q}^S$ has not been given explicitly
in the literature so far.
The quark scalar form factor $F$ in QCD has been computed to $\ord{\as^3}$ in \refcite{Gehrmann:2014vha},
from which we can extract $C_{q\bar q}^S$.
A slight difficulty arises as $F$ is only given at timelike kinematics and fixed $\mu = Q$ in \refcite{Gehrmann:2014vha}.
To obtain the full dependence on $L = 2\ln(-\img Q/\mu)$, we start from the bare form factor
$\mathcal{F}$ given in \refcite{Gehrmann:2014vha} and perform its UV-renormalization at an arbitrary \MSbar renormalization point $\mu$.
We explicitly checked that the ratio of the timelike to spacelike form factor is IR-finite as required.
We then proceed by subtracting the IR poles in $F$ in \MSbar by a multiplicative renormalization factor,
\begin{equation} \label{eq:subtract IR bbH}
C_{q\bar q}^S(\mu) = \frac{1}{y(\mu)} \lim_{\epsilon \rightarrow 0} Z^{-1}(\epsilon, \mu)\, F(\epsilon,\mu)
\,.\end{equation}
In SCET with pure dimensional regularization, the $1/\epsilon$ IR poles in $F(\epsilon, \mu)$
are the UV poles of the bare Wilson coefficient, so \eq{subtract IR bbH} is equivalent to the
\MSbar renormalization of $C_{q\bar q}^S$.

Here we have made explicit that the renormalized quark Yukawa coupling $y(\mu)$ is excluded from $C_{q\bar q}^S$.
We have also verified that the obtained renormalization factor $Z$ reproduces the correct anomalous dimension for $C_{q\bar q}^S$, i.e.\ that it satisfies
\begin{equation} \label{eq:mu dependence Z bbH}
-\frac{\df \ln Z(\mu)}{\df \ln \mu} = \Gamma^q_\cusp[\as(\mu)]\, L + 2\gamma_C^q[\as(\mu)]
\end{equation}
order-by-order in $\as$, which provides a strong check on the pole structure of $F$.
Equivalently, we also checked that the full result for $C_{q\bar q}^S(\mu)$ obtained from \eq{subtract IR bbH}
agrees with \eq{master formula} (with $\Gamma_\cusp \equiv \Gamma^q_\cusp$ and $\gamma = 2\gamma_C^q - \gamma_m$).
For the nonlogarithmic constant terms of $C_{q\bar q}^S$ we obtain
\begin{align} \label{eq:bbH finite terms}
C^S_{q\bar q\,1} &= C_F (-2 + \zeta_2)
\,, \nn \\
C^S_{q\bar q\,2}
&= C_F \biggl[C_F \Bigl(6  + 14 \zeta_2 - \frac{83}{10} \zeta_2^2 - 30 \zeta_3 \Bigr)
   + C_A \Bigl(-\frac{467}{81} -\frac{103}{18} \zeta_2 + \frac{44 }{5}\zeta_2^2 + \frac{151 }{9}\zeta_3 \Bigr)
\nn \\ & \qquad\quad
   + n_f \Bigl(\frac{200}{81} + \frac{5}{9} \zeta_2+\frac{2}{9} \zeta_3 \Bigr)
   \biggr]
\,, \nn \\
C^S_{q\bar q\,3}
&= C_F \biggl[
   C_A^2 \Bigl(
      -\frac{6152}{189} \zeta_2^3+\frac{10093}{135} \zeta_2^2+\frac{326}{3} \zeta_2 \zeta_3-\frac{264515}{1458} \zeta_2
      - \frac{1136}{9} \zeta_3^2+\frac{107648}{243} \zeta_3
\nn \\ & \qquad\qquad\quad
      +\frac{106}{9}\zeta_5+\frac{5964431}{26244} \Bigr)
\nn \\ & \qquad\quad
   + C_F C_A \Bigl(
      -\frac{12676}{315}\zeta_2^3-\frac{893}{270} \zeta_2^2-\frac{3049 }{9}\zeta_2 \zeta_3+\frac{31819 }{81}\zeta_2
      +\frac{296}{3}\zeta_3^2-\frac{4820}{27} \zeta_3
\nn \\ & \qquad\qquad\quad
      -\frac{1676}{9} \zeta_5-\frac{9335}{81} \Bigr)
\nn \\ & \qquad\quad
   + C_F^2 \Bigl(
    \frac{37729}{630}\zeta_2^3 - 77 \zeta_2^2 + 178 \zeta_2 \zeta_3 - \frac{353}{3} \zeta_2 + 16\zeta_3^2 - 654\zeta_3 + 424 \zeta_5 + \frac{575}{3}
   \Bigr)
\nn \\ & \qquad\quad
   + C_A\,n_f \Bigl(
      -\frac{476}{135} \zeta_2^2+\frac{4}{3} \zeta_2 \zeta_3+\frac{33259
      \zeta_2}{729}-\frac{2860}{27} \zeta_3-\frac{4}{3} \zeta_5-\frac{521975}{13122} \Bigr)
\nn \\ & \qquad\quad
   + C_F\,n_f \Bigl(
      -\frac{61}{27} \zeta_2^2-\frac{38}{9} \zeta_2 \zeta_3-\frac{6131}{162} \zeta_2+\frac{11996}{81} \zeta_3-\frac{416}{9} \zeta_5+\frac{35875}{972} \Bigr)
\nn \\ & \qquad\quad
   + n_f^2 \Bigl(
      -\frac{188}{135} \zeta_2^2-\frac{212}{81} \zeta_2-\frac{200}{243} \zeta_3+\frac{2072}{6561} \Bigr)
\biggr]
\,.\end{align}

\subsection{Renormalization group evolution}
\label{app:RGE}

For reference we collect the explicit expressions needed for the RG evolution of the hard functions.
The evolution factor $U_H$ is defined by \eq{def UH}. It is given explicitly by
\begin{equation}
U_H(Q, \mu_0, \mu)
= \biggl\lvert
\,\exp \biggl[
       2 \eta^i_\Gamma(\mu_0, \mu) \ln\Bigl( \frac{-\img Q}{\mu_0} \Bigr)
       - 2K^i_\Gamma(\mu_0, \mu)
       + K_\gamma(\mu_0, \mu)
\biggr] \biggr\rvert^2
\,,\end{equation}
where
\begin{align} \label{eq:Keta_def}
K^i_\Gamma(\mu_0, \mu)
& = \int_{\alpha_s(\mu_0)}^{\alpha_s(\mu)}\!\frac{\df\alpha_s}{\beta(\alpha_s)}\,
\Gamma^i_\cusp(\alpha_s) \int_{\alpha_s(\mu_0)}^{\alpha_s} \frac{\df \alpha_s'}{\beta(\alpha_s')}
\,,\quad
\eta^i_\Gamma(\mu_0, \mu)
= \int_{\alpha_s(\mu_0)}^{\alpha_s(\mu)}\!\frac{\df\alpha_s}{\beta(\alpha_s)}\, \Gamma^i_\cusp(\alpha_s)
\,,\nn \\
K_\gamma(\mu_0, \mu)
& = \int_{\alpha_s(\mu_0)}^{\alpha_s(\mu)}\!\frac{\df\alpha_s}{\beta(\alpha_s)}\, \gamma(\alpha_s)
\,.\end{align}
Here $\Gamma_\cusp(\as)$ is the relevant quark or gluon cusp anomalous dimension and $\gamma(\as)$
the appropriate noncusp anomalous dimension of the relevant hard matching coefficient.

Their explicit expressions at \nnllog* are
\begin{align} \label{eq:Keta}
K_\Gamma(\mu_0, \mu) &= -\frac{\Gamma_0}{4\beta_0^2}\,
\biggl\{ \frac{4\pi}{\alpha_s(\mu_0)}\, \Bigl(1 - \frac{1}{r} - \ln r\Bigr)
   + \biggl(\frac{\Gamma_1 }{\Gamma_0 } - \frac{\beta_1}{\beta_0}\biggr) (1-r+\ln r)
   + \frac{\beta_1}{2\beta_0} \ln^2 r
\nn\\ & \hspace{10ex}
+ \frac{\alpha_s(\mu_0)}{4\pi}\, \biggl[
  \biggl(\frac{\beta_1^2}{\beta_0^2} - \frac{\beta_2}{\beta_0} \biggr) \Bigl(\frac{1 - r^2}{2} + \ln r\Bigr)
  + \biggl(\frac{\beta_1\Gamma_1 }{\beta_0 \Gamma_0 } - \frac{\beta_1^2}{\beta_0^2} \biggr) (1- r+ r\ln r)
\nn\\ & \hspace{10ex}
  - \biggl(\frac{\Gamma_2 }{\Gamma_0} - \frac{\beta_1\Gamma_1}{\beta_0\Gamma_0} \biggr) \frac{(1- r)^2}{2}
     \biggr] \biggr\}
\,, \nn\\
\eta_\Gamma(\mu_0, \mu) &=
 - \frac{\Gamma_0}{2\beta_0}\, \biggl[ \ln r
 + \frac{\alpha_s(\mu_0)}{4\pi}\, \biggl(\frac{\Gamma_1 }{\Gamma_0 }
 - \frac{\beta_1}{\beta_0}\biggr)(r-1)
\nn \\ & \hspace{10ex}
 + \frac{\alpha_s^2(\mu_0)}{16\pi^2} \biggl(
    \frac{\Gamma_2 }{\Gamma_0 } - \frac{\beta_1\Gamma_1 }{\beta_0 \Gamma_0 }
      + \frac{\beta_1^2}{\beta_0^2} -\frac{\beta_2}{\beta_0} \biggr) \frac{r^2-1}{2}
    \biggr]
\,, \nn\\
K_\gamma(\mu_0, \mu) &=
 - \frac{\gamma_0}{2\beta_0}\, \biggl[ \ln r
 + \frac{\alpha_s(\mu_0)}{4\pi}\, \biggl(\frac{\gamma_1 }{\gamma_0 }
 - \frac{\beta_1}{\beta_0}\biggr)(r-1) \biggr]
\,,\end{align}
where $r = \alpha_s(\mu)/\alpha_s(\mu_0)$ and the running coupling is given by the three-loop expression
\begin{equation} \label{eq:alphas}
\frac{1}{\alpha_s(\mu)} = \frac{X}{\alpha_s(\mu_0)}
  +\frac{\beta_1}{4\pi\beta_0}  \ln X
  + \frac{\alpha_s(\mu_0)}{16\pi^2} \biggr[
  \frac{\beta_2}{\beta_0} \Bigl(1-\frac{1}{X}\Bigr)
  + \frac{\beta_1^2}{\beta_0^2} \Bigl( \frac{\ln X}{X} +\frac{1}{X} -1\Bigr) \biggl]
\,,\end{equation}
with $X\equiv 1+\alpha_s(\mu_0)\beta_0 \ln(\mu/\mu_0)/(2\pi)$.
For the resummation at lower logarithmic accuracies, the expressions in \eq{Keta} are truncated accordingly.
The relevant expressions at \nnnllog*, used for the inclusive $gg\to H$ cross sections,
can be found in \refcite{Abbate:2010xh}.

\section{Fixed-order estimates from resummed timelike logarithms}
\label{app:estimates}

It is instructive to compare explicitly the fixed-order contributions induced purely
by the timelike logarithms in the form factor with the full fixed-order result to assess whether
they are indeed a dominant part of the perturbative corrections.
However, we also stress that this is not a good way for judging the usefulness of
the resummation as a whole, since it does not capture the full resummed result and
in particular does not take into account the improvements in perturbative convergence
and uncertainties.

In \refcite{Anastasiou:2016cez}, such an analysis was carried out for $gg\to H$
for the coefficient $C_\delta$ of the $\delta(1-z)$ term in the partonic cross
section. This coefficient is fully determined by the partonic threshold limit
$z = m_H^2 / \hat s \to 1$, where $\hat s$ is the partonic center-of-mass energy,
and factorizes as in \eq{exclusive} into the product of the gluon form factor and
purely soft contributions. Ref.~\cite{Anastasiou:2016cez} found
that $C_\delta$ is poorly predicted from the timelike logarithms alone.
However, this can be very misleading since the $\delta(1-z)$ coefficient is strongly
scheme dependent and not a physical quantity. Rather, this type of analysis should
be carried out at the level of the cross section, which is a scheme-independent
physical observable.
To illustrate this, we repeat the analysis of \refcite{Anastasiou:2016cez}
and compare it to a different convention for the soft function,
as well as considering the hadronic $K$ factor.

Following \refcite{Anastasiou:2016cez}, we choose $\mufo = m_H$
and work in the pure EFT limit of \sec{ggX} with $C_X(\mufo) = 1$.
(The result for Higgs production is easily restored by reexpanding against $\vert C_t\vert^2$.)
The relevant hard function is hence $H_{gg}$ in \eq{GluonHard}.
Given the exact hard function to $\ord{\as^n}$, which is fully included in the N$^n$LL$'$
resummed result, the $\ord{\as^{n+1}}$ contribution predicted by and included
in the resummation is given by
\begin{align}
H^{(n+1)}_\text{appr} &= 2 \Bigl(\frac{\as}{4\pi}\Bigr)^{n+1} \Re \Bigl[ C^{(n+1)}(L = -\img \pi) \Bigr]_{C_{n+1} = 0} + \text{cross terms}
\nn \\
&= H^{(n+1)} - 2 \Bigl(\frac{\as}{4\pi}\Bigr)^{n+1} C_{n+1}
\,,\end{align}
where all logarithmic terms in the $\ord{\as^{n+1}}$ Wilson coefficient $C^{(n+1)}$
are predicted by the RGE [see \eq{master formula}],
and the only missing ingredient compared to the full result for $H^{(n+1)}$ is the nonlogarithmic
$\ord{\as^{n+1}}$ term $C_{n+1}$.
Denoting the soft function contribution to the $\delta(1-z)$ coefficient by $S_\delta = 1 + S_\delta^{(1)} + \dotsb$,
the corresponding approximate result for $C_\delta$ at $\ord{\as^{n+1}}$ is
\begin{align} \label{eq:prediction}
C^{(n+1)}_{\delta\,\text{appr}} &= H^{(n+1)}_\text{appr} + H^{(n)}\, S_\delta^{(1)} + \cdots + H^{(1)}\, S_\delta^{(n)}
\,.\end{align}

The result for $S_\delta$ to $\ord{\as^3}$ can be obtained from \refcite{Li:2014afw}, which
writes the soft function in terms of the standard (plus) distributions
\begin{equation}
\delta(1-z)\,,\qquad \biggl[ \frac{\ln^n(1-z)}{1-z} \biggr]_+ \quad(n\ge0)
\,.\end{equation}
Applying \eq{prediction} at each successive order, we find
\begin{alignat}{9} \label{eq:CdeltaFO}
\text{\llmatched*:}\qquad
C_\delta
&= 1 && + 14.80 \Bigl(\frac{\alpha_s}{\pi}\Bigr) && + \dotsb
\,, \nn \\
\text{\nllmatched*:}\qquad
C_\delta &=
1 	&&  +9.87 \Bigl(\frac{\alpha_s}{\pi}\Bigr)
&&  +45.35 \Bigl(\frac{\alpha_s}{\pi}\Bigr)^2  && + \dotsb
\,, \nn \\
\text{\nnllmatched*:}\qquad
C_\delta &=
1 	&& +9.87 \Bigl(\frac{\alpha_s}{\pi}\Bigr)
&& +13.61 \Bigl(\frac{\alpha_s}{\pi}\Bigr)^2
&&  -644.26  \Bigl(\frac{\alpha_s}{\pi}\Bigr)^3 + \dotsb
\,, \nn \\
\text{\nnnlo*:}\qquad
C_\delta &=
1 	&& +9.87 \Bigl(\frac{\alpha_s}{\pi}\Bigr)
&& +13.61 \Bigl(\frac{\alpha_s}{\pi}\Bigr)^2
&& +1124.31 \Bigl(\frac{\alpha_s}{\pi}\Bigr)^3
\,.\end{alignat}
The last coefficients in the first three lines are those predicted by the resummation beyond
the included fixed-order accuracy. In the last line the
\nnnlo* result is given for comparison.
(These numbers agree with those given in \refcite{Anastasiou:2016cez} except for $C^{(3)}_{\delta\,\text{appr}}$,
where they find $-554.79$ rather than our $-644.26$. We were unable to resolve
this numerical difference, but it is immaterial for the present discussion.)

From \eq{CdeltaFO} it looks like the resummation does a poor job at
approximating the higher fixed-order result, which would be in stark contrast to
what we have seen in \sec{GluonFusion}.
The resolution lies in the cross terms with the soft function
in \eq{prediction}. The $S_\delta^{(n)}$ coefficients depend on the
(in principle arbitrary) boundary condition chosen for the plus distributions.
In other words, the distinction between the soft-function cross terms included in \eq{CdeltaFO}
and those between $H^{(n)}$ and the remaining soft-function terms is arbitrary.
To illustrate this, we can instead write the soft function in terms of the different
set of plus distributions
\begin{equation}
  \delta(1-z)\,,\qquad \biggl[ \frac{1}{1-z} \ln^n\frac{(1-z)^2}{z} \biggr]_+ \quad(n\ge0)
\,,\end{equation}
used e.g.\ in \refscite{Ahrens:2008nc, Bonvini:2014tea}, for which we get a different
$\tilde S_\delta$ coefficient%
\footnote{
Ref.~\cite{Bonvini:2014tea} seems to miss a minus sign in $e^{-2\gamma_E \eta}$.
Restoring this we find full agreement with \refcite{Li:2014afw} and with the two-loop result
in \refcite{Ahrens:2008nc}.}
and a corresponding different $\tilde C_\delta$ coefficient of $\delta(1-z)$,
\begin{alignat}{9} \label{eq:tildeCdeltaFO}
\text{\llmatched*:}\qquad
\tilde C_\delta
&= 1 && +14.80 \Bigl(\frac{\alpha_s}{\pi}\Bigr)  && + \dotsb
\,, \nn \\*
\text{\nllmatched*:}\qquad
\tilde C_\delta
&= 1 &&  +19.74 \Bigl(\frac{\alpha_s}{\pi}\Bigr)
&&  +215.82 \Bigl(\frac{\alpha_s}{\pi}\Bigr)^2  && + \dotsb
\,, \nn \\*
\text{\nnllmatched*:}\qquad
\tilde C_\delta
&= 1 && +19.74 \Bigl(\frac{\alpha_s}{\pi}\Bigr)
&& +210.07 \Bigl(\frac{\alpha_s}{\pi}\Bigr)^2
&&  +1484.58  \Bigl(\frac{\alpha_s}{\pi}\Bigr)^3 + \dotsb
\,, \nn \\*
\text{\nnnlo*:}\qquad
\tilde C_\delta
&= 1 && +19.74 \Bigl(\frac{\alpha_s}{\pi}\Bigr)
&& +210.07 \Bigl(\frac{\alpha_s}{\pi}\Bigr)^2
&& +1372.11 \Bigl(\frac{\alpha_s}{\pi}\Bigr)^3
\,.\end{alignat}
In this convention, the resummation approximates the higher fixed-order terms
of $\tilde C_\delta$ very well.
The strong scheme dependence of the $\delta(1-z)$ coefficient is obvious from the
completely different coefficients in the exact results for $C_\delta$ and $\tilde C_\delta$
in \eqs{CdeltaFO}{tildeCdeltaFO}.

Instead, it is much more meaningful to consider physical quantities such as the inclusive hadronic cross section.
The approximate result analogous to \eq{prediction} for the total $K$ factor is given by
\begin{equation} \label{eq:prediction K}
K^{(n+1)}_\text{appr} = H^{(n+1)}_\text{appr} + H^{(n)} R^{(1)} + \cdots + H^{(1)} R^{(n)}
\,,\end{equation}
where the $S_\delta$ coefficient is now replaced by the full perturbative remainder $R$ defined in \eq{Rdef}.
The scheme dependence in this case is how the nonlogarithmic constant terms are split between $H$ and $R$,
which as discussed in \sec{ResummationScheme} cancels in their product and by construction does not
enter $H^{(n+1)}_\text{appr}$.
The analogous fixed-order expansions of the resummed results for the $K$ factor are given by
\begin{alignat}{9} \label{eq:prediction hadronic+inclusive}
\text{\llmatched*:}\qquad
K_{\ggX*}
&= 1 && +14.80 \Bigl(\frac{\alpha_s}{\pi} \Bigr) && + \dotsb
\,, \nn \\
\text{\nllmatched*:}\qquad
K_{\ggX*}
&= 1 &&  +30.52 \Bigl(\frac{\alpha_s}{\pi}\Bigr)
&&  +402.00 \Bigl(\frac{\alpha_s}{\pi}\Bigr)^2  &&  + \dotsb
\,, \nn \\
\text{\nnllmatched*:}\qquad
K_{\ggX*}
&= 1 && +30.52 \Bigl(\frac{\alpha_s}{\pi}\Bigr)
&& +425.27 \Bigl(\frac{\alpha_s}{\pi}\Bigr)^2
&&  +3820.46  \Bigl(\frac{\alpha_s}{\pi}\Bigr)^3  + \dotsb
\,, \nn \\
\text{\nnnlo*:}\qquad
K_{\ggX*}
&= 1 && +30.52 \Bigl(\frac{\alpha_s}{\pi}\Bigr)
&& +425.27 \Bigl(\frac{\alpha_s}{\pi}\Bigr)^2
&& +3576.94 \Bigl(\frac{\alpha_s}{\pi}\Bigr)^3
\,.\end{alignat}
Evidently, the resummed results approximate the higher fixed-order terms in the
$K$ factor very well, except at NLO, where $H^{(1)}$ and $R^{(1)}$ each contribute about
half of the full $K$ factor. This is precisely equivalent to our discussion
in \sec{ggX} that the large corrections to the $K$ factor are primarily driven by
the timelike logarithms in $H$, while the nonlogarithmic constant terms,
$R^{(n)}$ and $H^{(n)}(\mu_H = -\img m_X)$, are much smaller.

\bibliographystyle{jhep}
\bibliography{refs_draft}

\end{document}